\documentclass[letterpaper,twocolumn,10pt]{article}
\usepackage{zhanggroup}

\usepackage{color}
\usepackage{xcolor}
\usepackage{xspace} 
\usepackage{amssymb}
\usepackage{amsmath, bbm}
\usepackage{amsthm}
\usepackage{mathrsfs}
\usepackage{cite}
\usepackage{url}
\usepackage{caption}
\usepackage{subcaption}
\usepackage{float}
\usepackage[absolute]{textpos}
\usepackage{balance}
\usepackage{graphicx}
\usepackage{textcomp}
\usepackage{balance}
\usepackage{booktabs}
\usepackage{tabularx}
\usepackage[pagebackref=true,breaklinks=true,colorlinks,bookmarks=false]{hyperref}
\hypersetup{
  colorlinks,
  linkcolor={blue!70!green},
  citecolor={green!70!blue},
  urlcolor={orange!70!red}
}

\newcommand{\dataset}{\mathcal{D}}
\newcommand{\train}{\it{train}}
\newcommand{\property}{\mathcal{P}}
\newcommand{\target}{\it{target}}
\newcommand{\shadow}{\it{shadow}}
\newcommand{\auxiliary}{\it{auxiliary}}
\newcommand{\classify}{\it{classifier}}
\newcommand{\generator}{\mathsf{G}}
\newcommand{\discriminator}{\mathsf{D}}
\newcommand{\real}{\it{real}}
\newcommand{\infer}{\it{infer}}

\newcommand{\samples}{\mathbf{X}}
\newcommand{\noise}{z}
\newcommand{\noises}{\noise}
\newcommand{\optnoises}{\noises^{*}}
\newcommand{\classifier}{f_{\property}}
\DeclareMathOperator*{\argmin}{arg\,min}
\newcommand{\loss}{\mathcal{L}}
\newcommand{\rand}{\circ}
\newcommand{\mypara}[1]{\smallskip\noindent{\bf {#1}.} \xspace}
\begin{document}
% ======================================================

\begin{textblock}{12}(2,1)
\centering
To Appear in the 29th Network and Distributed System Security Symposium, 27 February – 3 March, 2022.
\end{textblock}

\date{}
% ======================================================
\title{\Large \bf Property Inference Attacks Against GANs}
% ======================================================

% ======================================================
\author{
{\rm Junhao Zhou\textsuperscript{1}\thanks{The first two authors made equal contributions.}}\ \ \ \ \
{\rm Yufei Chen\textsuperscript{1}\textsuperscript{\textcolor{blue!60!green}{$\ast$}}}\ \ \
{\rm Chao Shen\textsuperscript{1}\thanks{Corresponding authors.}}\ \ \
{\rm Yang Zhang\textsuperscript{2}\textsuperscript{\textcolor{blue!60!green}{$\dagger$}}}\ \ \
\\
\textsuperscript{1}\textit{Faculty of Electronic and Information Engineering, Xi'an Jiaotong University}\ \ \
\\
\textsuperscript{2}\textit{CISPA Helmholtz Center for Information Security}
}
% ======================================================

\maketitle

% ======================================================
\begin{abstract}
% ======================================================

While machine learning (ML) has made tremendous progress during the past decade, recent research has shown that ML models are vulnerable to various security and privacy attacks.
So far, most of the attacks in this field focus on discriminative models, represented by classifiers.
Meanwhile, little attention has been paid to the security and privacy risks of generative models, such as generative adversarial networks (GANs).
In this paper, we propose the first set of training dataset property inference attacks against GANs.
Concretely, the adversary aims to infer the macro-level training dataset property, i.e., the proportion of samples used to train a target GAN with respect to a certain attribute.
A successful property inference attack can allow the adversary to gain extra knowledge of the target GAN's training dataset, thereby directly violating the intellectual property of the target model owner.
Also, it can be used as a fairness auditor to check whether the target GAN is trained with a biased dataset.
Besides, property inference can serve as a building block for other advanced attacks, such as membership inference.
We propose a general attack pipeline that can be tailored to two attack scenarios, including the full black-box setting and partial black-box setting.
For the latter, we introduce a novel optimization framework to increase the attack efficacy.
Extensive experiments over four representative GAN models on five property inference tasks show that our attacks achieve strong performance.
In addition, we show that our attacks can be used to enhance the performance of membership inference against GANs.\footnote{Our code is available at \url{https://github.com/Zhou-Junhao/PIA_GAN}.}

% ======================================================
\end{abstract}
% ======================================================

% ======================================================
\section{Introduction}
% ======================================================

Machine learning (ML) has progressed rapidly during the past decade, and ML models have been adopted in a wide range of applications, such as image recognition~\cite{HZRS16,KSH12}, speech recognition~\cite{GMH13,HDYDMJSVNSK12}, machine translation~\cite{BCB15}, etc.
Most of the current ML applications are based on discriminative models, represented by classifiers. 
Generative models, on the other hand, have attracted an increasing amount of attention recently.
The most representative generative model is generative adversarial networks (GANs)~\cite{GPMXWOCB14}.
Due to the ability to produce novel samples from high-dimensional data distributions, GANs are finding appealing application scenarios, such as image-to-image translation~\cite{ZPIE17,LBK17,WLZTKC18}, image inpainting~\cite{ISI17}, text generation~\cite{AGLMZ17,VTBE15}, and sound generation~\cite{MKGKJSCB17,ODZSVGKSK16}.

ML models have been shown to exhibit severe security and privacy vulnerabilities.
Existing attacks including adversarial examples~\cite{SZSBEGF14,EEFLRXPKS18}, membership inference~\cite{SSSS17,SZHBFB19}, and model stealing~\cite{TZJRR16,PMGJCS17,OSF19} mainly focus on discriminative models.
Recent research has also demonstrated the security and privacy risks of generative models.
In particular, Hayes et al.~\cite{HMDC19} show that an adversary can effectively determine whether a data sample is used to train a target GAN.
Chen et al.~\cite{CYZF20} further generalize this attack by proposing a taxonomy of membership inference scenarios against GANs.
While most of the attacks against generative models focus on membership inference, other vulnerabilities are left largely unexplored.

% ======================================================
\subsection{Our Contributions}
% ======================================================

\mypara{Motivation}
In this paper, we perform the first \emph{property inference attack} against GANs: an adversary aims to infer whether a target GAN's underlying training dataset exhibits a certain general property.
Here, the general property is related to the macro-level information of the target GAN's training dataset.
More importantly, the property is not related to the design purpose of the target GAN.
For instance, if a GAN is trained to generate human faces, the property can be the proportion of white people in its training dataset.
A successful property inference attack against a target GAN can lead to severe consequences.
For instance, learning the property of a GAN's training dataset gains an adversary extra information of the data, which directly violates the intellectual property (IP) of the model owner, as the dataset is often expensive to collect.
Also, an effective property inference attack can be used as a fairness auditing tool to make sure a GAN is not trained on biased data~\cite{BG18}.
Moreover, this attack can be further leveraged as a stepping stone to perform more advanced attacks, such as membership inference~\cite{SSSS17}.

\begin{figure*}[!t]
\centering
\begin{subfigure}{1.9\columnwidth}
\includegraphics[width=\columnwidth]{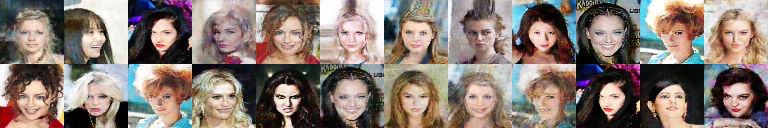}
\caption{All white females}
\label{figure:intuition_a}
\end{subfigure}
\begin{subfigure}{1.9\columnwidth}
\includegraphics[width=\columnwidth]{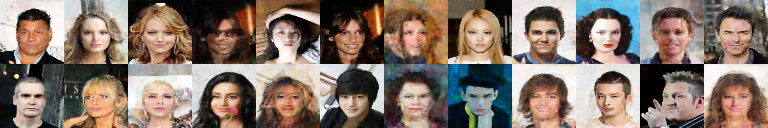}
\caption{Multiple groups}
\label{figure:intuition_b}
\end{subfigure}
\caption{Samples generated by WGAN trained on (a) 256 images of white females drawn from CelebA dataset (b) 256  images of uniformly distributed demographic background (gender and pale skin) from CelebA dataset.
We can see that almost all images in (a) are white females and images in (b) are rather diverse.}
\label{figure:intuition}
\end{figure*}

\mypara{Attack Methodology}
Our attack follows the intuition that the generated samples of a GAN can reflect its underlying training dataset's property.
For instance, in \autoref{figure:intuition_a}, we can see a WGAN~\cite{ACB17} trained with faces of only white females mainly generates images of white females, while in \autoref{figure:intuition_b}, a WGAN trained with images from people with a diverse demographic background can produce a diverse set of images.

We propose two attack scenarios, i.e., full black-box setting and partial black-box setting.
For the former, we assume the adversary can just get samples blindly from the target GAN's generator.
For the latter, the adversary can decide the input of the target GAN's generator, i.e., the latent code.
Note that for both attack scenarios, the adversary does not have access to the target GAN's parameters, which means we focus on the most difficult setting for the adversary~\cite{SSSS17}.

Both of our property inference attacks follow a general pipeline.
The adversary first queries the target GAN model to obtain a set of generated samples.
Then she relies on a property classifier to label these samples with respect to the target property.
For instance, if the target property is the gender distribution of the GAN's training dataset, the property classifier is a gender classifier for the generated samples.
In the end, she infers the target property by summarizing the results of the property classifier.
For the partial black-box setting, since the adversary can choose the input, i.e., the latent code, for the target GAN, we propose a novel optimization framework which allows us to generate a set of latent codes, namely optimized latent code set, to reduce the number of queries to the target model.

\mypara{Evaluation}
Extensive experiments over four GANs on five different property inference tasks show that our attacks achieve very effective performance.
In the gender prediction task on CelebA~\cite{LLWT15}, the average absolute difference between the inferred proportion and the ground truth proportion of males in the target training dataset is 2.4\% in the full black-box setting and 9.8\% in the partial black-box setting.
In the age prediction task (proportion of youth) on AFAD~\cite{NZWGH16}, the average absolute difference is 9.7\% and 10.1\% in the full and partial black-box settings respectively.
Meanwhile, in the income prediction task on the US Census dataset~\cite{UCIINCOME}, the average absolute difference is 2.9\% and 4.5\% in the full and partial black-box settings respectively.
We further compare the two methodologies in detail and conclude that the partial black-box attack behaves better when using a limited number of generated samples (around 100 to 125), while the full black-box attack results in a high accuracy using a large number of random samples.
We also observe that our full black-box attack works well even without any knowledge of the target training dataset.
Also, our partial black-box attack works robustly with respect to different optimization starting points as well as the number, the structure, and the dataset of the shadow models.

\mypara{Enhancing Membership Inference}
We further enhance the state-of-the-art membership inference attack against GANs~\cite{CYZF20} leveraging our proposed property inference attacks.
In detail, we calibrate a sample's membership prediction result based on its attributes and the property of the target GAN's training dataset obtained by our attack.
Experimental result shows that our enhanced methodology increases the membership inference's AUC from 0.52 to 0.61 on the CelebA dataset.
This further demonstrates the applicability of our proposed attacks.

In summary, we make the following contributions in this paper:
\begin{itemize}
\item We propose the first property inference attack against generative models.
\item We specify two attack scenarios and propose corresponding technical solutions relying on advanced ML techniques.
\item We perform extensive experiments and demonstrate the efficacy of our proposed attacks.
\item We show that our property inference attacks can serve as a building block to enhance other advanced attacks, in particular, membership inference against GANs.
\end{itemize}

% ======================================================
\subsection{Organization}
% ======================================================

The rest of the paper is organized as follows.
We introduce generative models, property inference attacks, and threat models used in this paper in \autoref{section:preliminaries}.
Then, \autoref{section:attack} presents our attack workflow and detailed methodologies.
The experiment setup and evaluation results are shown in \autoref{section:evaluation}.
We further show how to leverage our attack to enhance membership inference in \autoref{section:MIA}.
\autoref{section:related} discusses the related work and \autoref{section:conclusion} concludes this paper.

% ======================================================
\section{Preliminaries}
\label{section:preliminaries}
% ======================================================

In this section, we first introduce generative models.
Then, we present property inference attacks.
The threat models considered in this paper are discussed in the end.

% ======================================================
\subsection{Generative Models}
% ======================================================

Machine learning models can be categorized into generative models and discriminative models.
Discriminative models are mainly designed to solve classification problems, such as image recognition and text sentiment prediction.
On the other hand, generative models aim to learn the underlying training data distribution and generate new data based on it.
There exist various types of generative models, including Variational AutoEncoders (VAEs) and Generative Adversarial Networks (GANs).
In this paper, we focus on GANs as they are the most popular generative models nowadays.

A GAN is assembled with two neural networks, i.e., the generator and the discriminator.
The generator takes random noise (latent code) as input and generates samples,
while the discriminator performs adversarial training to distinguish the real and fake (generated) samples.
In the training stage, these two networks are updated in turns: the generator learns to generate samples as realistic as possible while the discriminator learns to better separate real and fake samples.

Mathematically, the loss function of a GAN is defined as the following.
\[
\mathbb{E}_{x \sim \dataset_{\train}}[\log \discriminator(x)] + \mathbb{E}_{z \sim \mathcal{Z}}[\log(1-\discriminator(\generator(z)))]
\]
Here, $\generator$ and $\discriminator$ represent the generator and the discriminator, respectively.
$ z \sim \mathcal{Z} $ denotes the latent code following a prior, normally multivariate Gaussian or uniform distribution.
The training dataset of the GAN is represented by $\dataset_{\train}$.
As $\generator$ is trained to minimize the loss and $\discriminator$ aims to maximize the loss, the optimization for GAN follows a two-player minimax game.

After being first introduced by Goodfellow et al.~\cite{GPMXWOCB14}, GANs have attracted a lot of attention.
Over the years, many works have been proposed to enhance the original GANs, such as WGAN~\cite{ACB17}, DCGAN~\cite{RMC15}, WGAPNGP~\cite{GAADC17}, TGAN~\cite{XV18}, PGGAN~\cite{KALL18}, and BigGAN~\cite{BDS19}.
In this paper, we focus on WGANGP, PGGAN, DCGAN, and TGAN as they have achieved strong performance in various settings empirically.
Note that our method is general and can be applied to other types of GANs as well.

% ======================================================
\subsection{Property Inference Attacks}
% ======================================================

Property inference attacks aim to extract some general properties of a target ML model's training dataset, which the target model owner does not intend to share.
Also, these general properties are not related to the target GAN's original design purposes.
For instance, if the target model is used to generate realistic human profile photos, the property can be the gender distribution of the samples in the training dataset.
A successful property inference attack allows an adversary to gain insights on the training data of the target model, which may violate the intellectual property of the model owner as high-quality data is normally expensive to collect. 
Also, property inference can serve as an important building block for other more advanced attacks, such as membership inference attacks~\cite{CYZF20} (see \autoref{section:MIA}).
Moreover, property inference attacks can serve as a fairness auditor for the target model's training dataset, e.g., whether the samples used to train a model are equally distributed among different genders~\cite{BG18}.

Recently, Ganju et al.~\cite{GWYGB18} propose the first property inference attack against discriminative models, in particular fully connected neural networks. 
In this setting, the adversary is assumed to have white-box access to the target model and uses a meta classifier to predict the property of the corresponding training dataset.
The features of the meta classifier are summarized over the parameters of the model.
The authors propose two approaches for feature summarization, including neuron sorting and DeepSets.
To train the meta classifier, the adversary needs to establish multiple shadow models.
Different from~\cite{GWYGB18}, our attack is set up in more practical cases, where the victim GAN is completely black-box or only part of the GAN parameters are accessible. 

So far, property inference attacks concentrate on discriminative models in the white-box setting.
In this paper, we propose the first set of property inference attacks against generative models, represented by GANs. 
More importantly, we focus on the black-box setting which is the most difficult and realistic scenario for the adversary~\cite{SSSS17,SBBFZ20}.

% ======================================================
\subsection{Threat Models}
\label{section:threat_model}
% ======================================================

Similar to the setting of discriminative models, the goal of the adversary here is to infer whether the target model's training dataset $\dataset_{\target}$ has a certain property $\property$.
We first assume the adversary has an auxiliary dataset $\dataset_{\auxiliary}$ that comes from the same distribution of $\dataset_{\target}$.
The adversary leverages this auxiliary dataset to build local shadow GANs and classifiers for her attacks, i.e., $\dataset_{\auxiliary}=\dataset_{\shadow}\cup \dataset_{\classify}$ (see~\autoref{section:attack} for more details).
This is also the common assumption used in most of the works in machine learning privacy~\cite{SSSS17,GWYGB18,SZHBFB19,SBBFZ20}.
We also assume the adversary only has access to the generator of the target model as the discriminator is normally disregarded after the training phase.
For simplicity, we use $\generator_{\target}$ to represent the target model in the rest of the paper.

We consider two scenarios for the adversary including \emph{full black-box setting} and \emph{partial black-box setting}.

\mypara{Full Black-box Setting}
This is the least knowledgeable setting for the adversary, where she can just get the generated samples blindly from the target black-box generator $\generator_{\target}$.
This attack scenario provides a good simulation of the online closed-source API~\cite{CYZF20}.

\mypara{Partial Black-box Setting}
In this scenario, the adversary also has no knowledge about the parameters of the target GAN but can construct the latent code $\noise$ to generate the corresponding sample from $\generator_{\target}$.
In this way, she can feed her chosen latent codes to obtain specific generated samples.
Formally, we use the latent code set $\{\noise_{i}\}_{i=1}^{\vert\samples\vert}$ to represent her chosen latent codes in our paper. 
Moreover, we assume the adversary knows the architecture of the target GAN, as well as the training algorithm.
Such information can be directly inferred by performing hyperparameter stealing attacks~\cite{OASF18,WG18}.

% ======================================================
\section{Attack Methodology}
\label{section:attack}
% ======================================================

In this section, we first introduce our general attack pipeline. 
Then, we present the details of our attacks under two different scenarios.

% ======================================================
\subsection{Attack Workflow}
% ======================================================

Our attacks are designed based on the intuition that the generated samples of a target GAN can reflect the underlying training dataset's property.
For instance, if a GAN is mainly trained with images of white males, we expect the generated images are more likely to be white males compared to other demographic backgrounds. 
\autoref{figure:intuition_a} shows a WGAN trained only on images of white female mainly generates samples that are recognized as white females.
Meanwhile, \autoref{figure:intuition_b} shows a WGAN trained with samples from a diverse demographic background generates a diverse set of samples.
Therefore, an adversary can estimate a target GAN's underlying property $\property_{\infer}$ by inspecting the corresponding property of its generated samples ($\property_{\real}$).

\begin{figure}[!t]
\centering
\includegraphics[width=0.9\columnwidth]{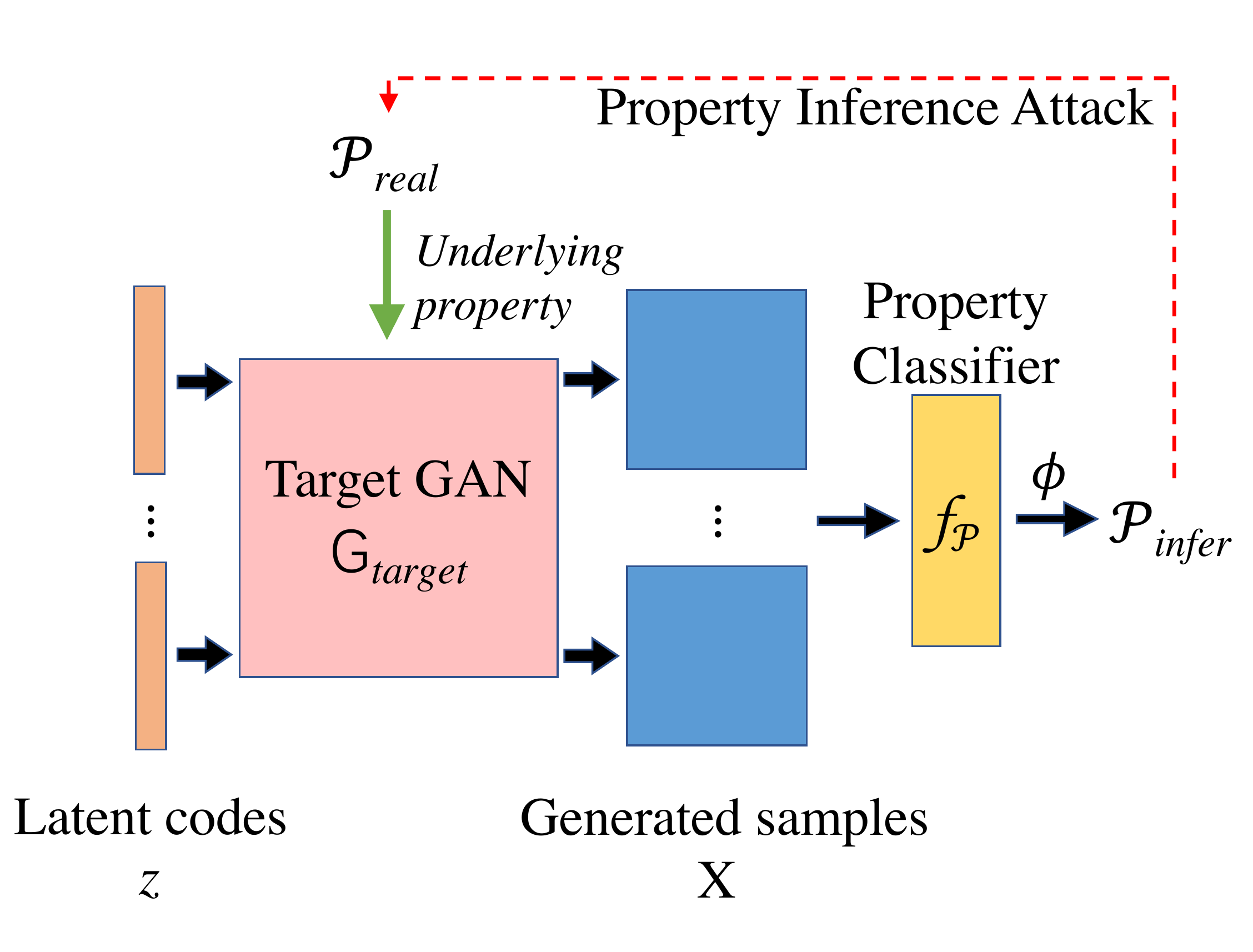}
\caption{Workflow of the general property inference attack strategy.
With the help of a property classifier $\classifier$, the adversary obtains $\property_{\infer}$ to inspect $\property_{\real}$, the target underlying property in the training dataset of $\generator_{\target}$.}
\label{figure:strategy}
\end{figure}

\autoref{figure:strategy} depicts the general attack workflow, which can be roughly categorized into three steps. 
In the first step, the adversary queries the target generator $\generator_{\target}$ to produce synthetic samples $\samples=\{\generator(\noise_1),\generator(\noise_2),\cdots,\generator(\noise_{\vert\samples\vert})\}$.
Here, $\generator(\noise_i)$ represents the $i$th generated sample from the target GAN with respect to the corresponding latent code $\noise_i$.
The concrete methods for generating samples, i.e., choosing latent codes, for both full and partial black-box settings, are presented later in \autoref{section:full_black-box} and \autoref{section:partial_black-box}, respectively. 

Next, the adversary constructs a \emph{property classifier} $\classifier$ tailored for classifying the previously generated samples with respect to the target property she is interested in.
For instance, if the target property is the gender distribution of the samples (profile images) in the underlying training dataset, the property classifier then predicts the gender of each sample.
The property classifier here is trained with part of the auxiliary dataset, i.e., $\dataset_{\classify}$ (as described in \autoref{section:preliminaries}) that is disjoint from the underlying training dataset of target GAN.

In the end, the adversary predicts $\property_{\infer}$ based on the output of the property classifier.
Concretely, she computes a function $\phi$ over the prediction of the property classifier, defined as:
\[
\phi\left(\{\classifier(\generator_{\target}(\noise_i))\}_{i=1}^{\vert\samples\vert}\right)
\]
In this paper, our attack focuses on inferring a general property as the distribution of a certain attribute, such as the gender distribution of the samples in the target underlying training dataset.
Therefore, $\phi$ is realized as a function to summarize the distribution of the target attribute.
However, we emphasize that our methodology is general and can be applied to infer other types of property.

\begin{figure*}[!t]
\centering
\includegraphics[width=1.55\columnwidth]{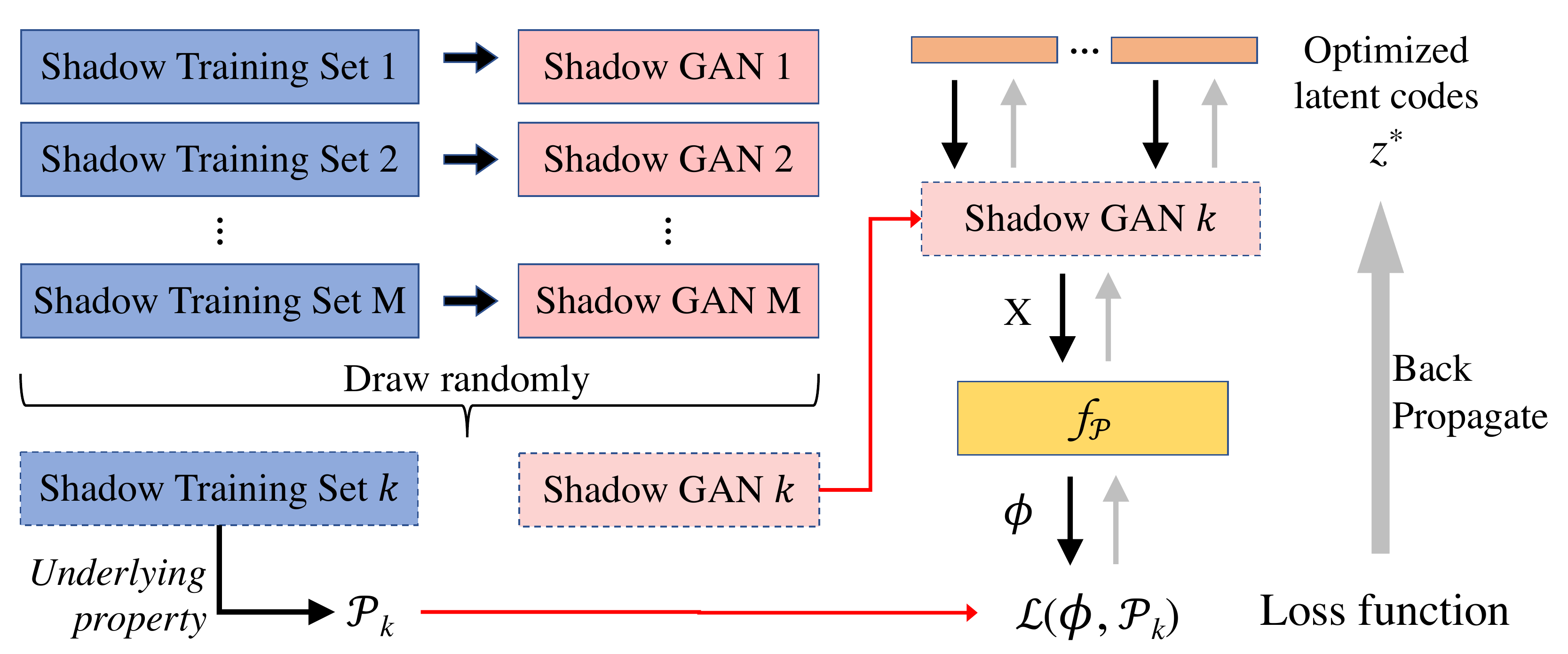}
\caption{Methodology of optimizing input latent codes $(\noises\leftarrow\noises^{*})$ with shadow models.}
\label{figure:opt_noise}
\end{figure*}

% ======================================================
\subsection{Full Black-box Adversary}
\label{section:full_black-box}
% ======================================================

For a full black-box adversary, she can only obtain generated samples blindly from the target GAN.
These acquired samples $\samples$, generated from a random latent code set, are consumed by $\classifier$, and then $\phi$ to get the attack result $\property_{\infer}$, just as presented in the basic attack strategy.

More formally, the property inference attack through full black-box GANs can be described as the following, where the latent codes are just drawn randomly from a prior.
\[
\phi\left(\{\classifier\left(\generator_{\target}\left(\noise_{i}^{\rand} \right)\right)\}_{i=1}^{\vert\samples\vert}\right)
\]
Here, each latent code is denoted by $\noise_{i}^{\rand}$, as a member of the latent code set.

% ======================================================
\subsection{Partial Black-box Adversary}
\label{section:partial_black-box}
% ======================================================

Different from the full black-box adversary, the partial black-box adversary can choose the latent code to feed into the target GAN.
Thus, she can construct/craft a specific latent code set to allow the target GAN to generate corresponding samples that can help her to achieve an effective property inference.
Crafting a latent code set is a training process. To this end, the adversary needs to establish a set of shadow models to simulate GANs trained with datasets of different properties.

The process to construct the latent code set with the help of shadow models can be divided into three stages. 
\autoref{figure:opt_noise} provides a schematic overview of it.

In the first stage, the adversary generates shadow training datasets for training shadow GANs.
More formally, she samples $M$ shadow training datasets $\{\dataset_1,\dataset_2,\cdots,\dataset_M\}$ from $\dataset_{shadow}$ (obtained from the local auxiliary dataset presented in \autoref{section:preliminaries}) corresponding to $M$ shadow models.
Each shadow dataset $\dataset_{k}$ is sampled to fulfill a certain property denoted by $\property_{k}$, and all the shadow training datasets' properties are uniformly distributed. 

In the next stage, the adversary trains each local shadow GAN $\generator_{k}$ with the corresponding shadow training dataset $\dataset_{k}$.
Note that each $\generator_{\shadow}$ has the same architecture with the target GAN $\generator_{\target}$ (see \autoref{section:preliminaries}).

Finally, the adversary crafts an optimized latent code set, denoted by $\{\optnoises_i\}_{i=1}^{\vert\samples\vert}$, over the $M$ shadow GANs.
Mathematically, the optimization is defined as the following.
\[
\argmin_{\{\optnoises_i\}_{i=1}^{\vert\samples\vert} } \sum_{k=1}^{M} \loss \left(\phi\left(\{\classifier(\generator_k(\optnoises_i))\}_{i=1}^{\vert\samples\vert}\right), \property_k\right)
\]
Here, $\loss$ represents the adopted loss function, and we utilize the stochastic gradient descent (SGD) method to minimize the loss function.
At the beginning of the optimization, we need to set a random starting of the latent code set.
Particularly, we present an experiment about how the optimization starting point affects the partial black-box attack in \autoref{section:evaluation_part_black-box}.

With the optimized latent code set, the adversary can infer the target property $\property_{\infer}$ similar to the full black-box adversary:
\[
\phi\left(\{\classifier\left(\generator_{\target}\left(\noise_i^{*}\right)\right)\}_{i=1}^{\vert\samples\vert}\right)
\]
Here, each latent code is denoted by ${\noise}_i^{*}$, as a member of the optimized latent code set.

% ======================================================
\section{Evaluation}
\label{section:evaluation}
% ======================================================

In this section, we first describe the datasets used in our experiments, followed by descriptions of the evaluated GAN models and detailed experimental setup.
We then present the results of our proposed attacks.

% ======================================================
\subsection{Dataset}
% ======================================================

GANs have been demonstrated to be successfully used in the image domain ~\cite{ISI17,LBK17,WLZTKC18}, and current attacks against GANs are also demonstrated with computer vision targets~\cite{CYZF20}. 
Therefore, we mainly focus on GANs generating image outputs in this paper. 
We also experiment on tabular data to prove that our attacks are general and can be applied to other domains.

\mypara{MNIST}
The MNIST database of handwritten digits~\cite{MNIST} is a commonly adopted benchmark repository for computer vision and machine learning projects.
It includes 70,000 handwritten digits labeled with corresponding digit numbers.
In this paper, we focus on inferring \emph{the proportion of 0s and 1s} used to train target GANs.
Concretely, we construct a subset MNIST$_{\rm{01}}$ with over 6.9K 0s and 7.8K 1s and evaluate our inference attack on the proportion of 0s.
We also show an extended experiment in \autoref{section:Multi} to evaluate our attack facing the property with multiple classes (digit 0\textasciitilde9) based on the whole MNIST dataset.

\mypara{CelebA}
CelebA~\cite{LLWT15} is a benchmark dataset for face-related problems.
This large-scale face attributes dataset contains more than 200K celebrity images, and each of them has 40 binary attributes. 
In this paper, we focus on the gender attribute, which not only is easy for a property classifier to discriminate but also has a relatively balanced proportion on females and males (around 4:6).
As a result, we intend to infer the \emph{gender distribution} of the samples used to train a target GAN.

\begin{table*}[!t]
\centering
\caption{The settings for each experiment, describing the dataset, the property classifier task, and the target property.}
\label{table:experiment_setting}
\scriptsize
\begin{tabular}{l | c | c | c | c | c | c | c }
\toprule
  Task & Dataset  & Property Classifier & Target Property & GAN structure & Size of $\dataset_{\target}$ & Size of $\dataset_{k}$ & Size of $\dataset_{\classify}$ \\
\midrule
$T_1$ & CelebA & Gender Classification & Proportion of Males & WGANGP & 40000 & 40000 & 82K \\
$T_2$ & AFAD$_{\rm{gender}}$ & Gender Classification & Proportion of Males & PGGAN & 10800 & 10800 & 92K \\
$T_3$ & AFAD$_{\rm{age}}$ & Age Classification & Proportion of Youth & PGGAN & 10800 & 10800 & 92K \\
$T_4$ & MNIST$_{\rm{01}}$ & Digit classification & Proportion of 0s & DCGAN & 3000 & 3000 & 8.8K \\
$T_5$ & Census Income & Income classification & Proportion of high-income & TGAN & 4200 & 4200 & 290k  \\
\bottomrule
\end{tabular}
\end{table*}

\mypara{AFAD}
The Asian Face Age Dataset (AFAD)~\cite{NZWGH16} is a dataset proposed mainly for age estimation tasks.
This dataset contains over 160K Asian faces, with the corresponding age and gender attributes.
In this paper, we take advantage of both attributes and focus on inferring the \emph{gender distribution} and the \emph{age distribution} of the images used to train target GANs.
Concretely, we construct two datasets from AFAD, i.e., AFAD$_{\rm{gender}}$ and AFAD$_{\rm{age}}$.
AFAD$_{\rm{gender}}$ contains the same number of images (160K) as the normal AFAD.
AFAD$_{\rm{age}}$, on the other hand, contains over 72K samples where $18 \le \rm{age} \le 20$ and $30 \le \rm{age} \le 39$ are chosen from AFAD.
In this way, the \emph{age distribution} is described as the proportion of youth ($18 \le \rm{age} \le 20$) in the underlying training dataset.

\mypara{US Census Income}
The US Census Income Dataset~\cite{UCIINCOME} is used to learn to predict whether a person earns over \$50K a year. 
It includes 299,285 instances and each of them has 41 demographic and employment related attributes, such as age, gender, education, and citizenship.
In this paper, we intend to infer the \emph{high-income distribution} (the proportion of records whose income is over \$50K) of the samples used to train a target GAN.

% ======================================================
\subsection{Models}
\label{section:target_models}
% ======================================================

We first introduce the four GAN models that we focus on in this paper, i.e., DCGAN~\cite{RMC15}, WGANGP~\cite{GAADC17}, PGGAN~\cite{KALL18}, and TGAN~\cite{XV18}, as they are typical and representative models used in multiple applications like image generation, image-to-image translation, and super-resolution.
Then, we describe the property classifier $\classifier$ which is used in both attack scenarios.

\mypara{DCGAN}
DCGAN bridges the gap between convolutional networks (CNNs) and unsupervised learning.
Thanks to the combination of CNNs, DCGANs are stable to train in most settings and are proved to learn good representations of images for supervised learning and generative modeling.
In this paper, the dimension of the latent code is chosen as 100, and the output size is set as 32$\times$32$\times$1, while our DCGANs are trained on MNIST.
The structure we use in this paper is shown in \autoref{table:DCGAN_structure}, and the detailed hyper-parameters are listed below.
The number of critic iterations per generator iteration $n_{critic}=1$, the batch size $m=100$, and parameters of the Adam optimizer $\alpha=0.0002$, $\beta_1=0.5$, $\beta_2=0.999$.

\mypara{WGANGP}
WGANGP is proposed to improve the training process of an ordinary Wasserstein GAN.
With the help of an addition called gradient penalty, their proposed method enables stable training of a wide variety of GAN architectures.
In this paper, we choose the dimension of each latent code as 100, and the pixel size of output images is 64$\times$64$\times$3.
The structure of our WGANGP is shown in \autoref{table:WGANGP_structure}.
Moreover, the hyper-parameters in the training process are configured as the following:
the gradient penalty coefficient $\lambda=10$, the number of critic iterations per generator iteration $n_{critic}=3$, the batch size $m=100$, and parameters of the Adam optimizer $\alpha=0.0002$, $\beta_1=0.9$, $\beta_2=0.999$.

\mypara{PGGAN}
The key idea of PGGAN is to grow both generator and discriminator progressively with the resolution becoming higher.
This training method has a better performance in training high-quality images.
In this paper, we set the image size as 64$\times$64$\times$3, with consideration of the training cost.
The remaining settings are the same as those provided in~\cite{KALL18}, including the dimension of input latent code is 512, the usage of WGANGP loss, and leaky ReLU with leakiness 0.2.
The structure of PGGAN we use is shown in \autoref{table:PGGAN_structure}, and the detailed hyper-parameters are shown as below:
the gradient penalty coefficient $\lambda=10$, the number of critic iterations per generator iteration $n_{critic}=1$, the batch size $m=36$, and parameters of the Adam optimizer $\alpha=0.001$, $\beta_1=0$, $\beta_2=0.99$.

\mypara{TGAN}
Tabular GAN is designed to generate tabular data like medical or educational records.
With the help of mode-specific normalization for numerical variables and smoothing for categorical variables, TGAN can simultaneously generate discrete and continuous variables.
In this paper, we set the dimension of the latent code to 200 and the size of the synthetic output to 503 (based on the total size of the one-hot vectors).
All of the settings of the TGAN are the same as those provided in~\cite{XV18}, while the output and hidden state size of LSTM are 100, the size of the hidden vector is 100, the batch size is 200, and the parameters of the Adam optimizer are set as $\alpha=0.001$, $\beta_1=0.5$, $\beta_2=0.99$.

\mypara{Property Classifier}
For each inference task, we construct a property classifier (binary in our case) with the corresponding generated image size.
For the gender classifiers, our neural networks start with two convolutional layers and follow with two fully connected layers, both obtaining over 95\% testing accuracy.
And the age classifier starts with three convolutional layers followed with three fully connected layers, getting an accuracy of around 80\%.
To discriminate the digit 0 and 1, our neural network has the same structure as the age classifier, with a testing accuracy of over 98\%.
Additionally, we build up a four-layer fully connected classifier to recognize the high-income census with over 86\% testing accuracy.
We also show extended experiments later in \autoref{section:property_classifier} to investigate the influence of the property classifier on the attack performance, i.e., the needlessness of the target model's training dataset and the irrelevance of architecture of the property classifier.
For the former, we adopt two off-the-shelf gender classifiers (based on IMDB-WIKI dataset and the Audience dataset) and a locally trained model (based on EMNIST dataset).
For the latter, we train five well-behaved classifiers with different architectures.

% ======================================================
\subsection{Experiment Setup}
% ======================================================

\begin{figure*}[!t]
\centering
\begin{subfigure}{0.4\columnwidth}
\includegraphics[width=\columnwidth]{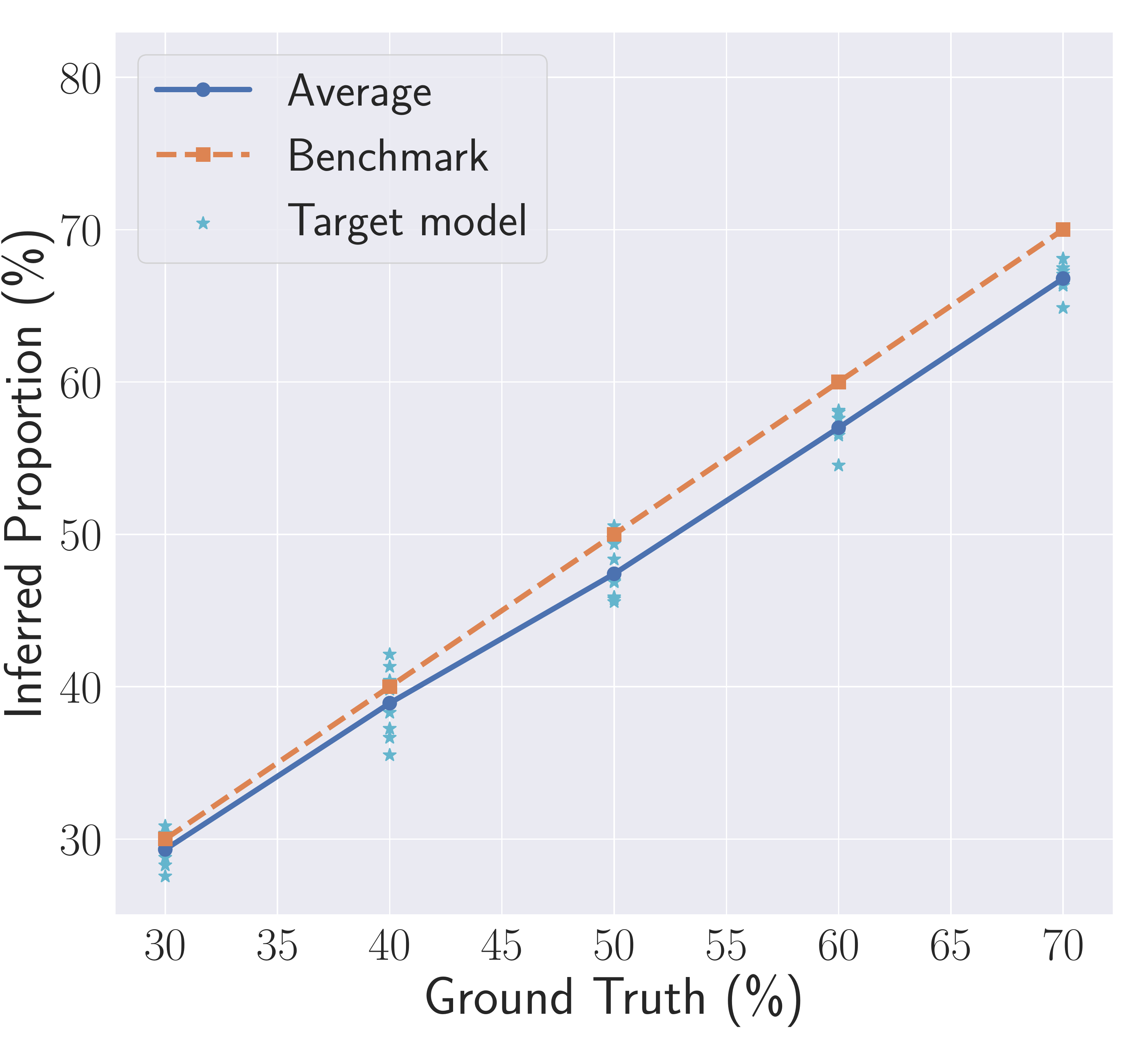}
\caption{Evaluation on $T_1$}
\label{figure:fullBBsight1}
\end{subfigure}
\begin{subfigure}{0.4\columnwidth}
\includegraphics[width=\columnwidth]{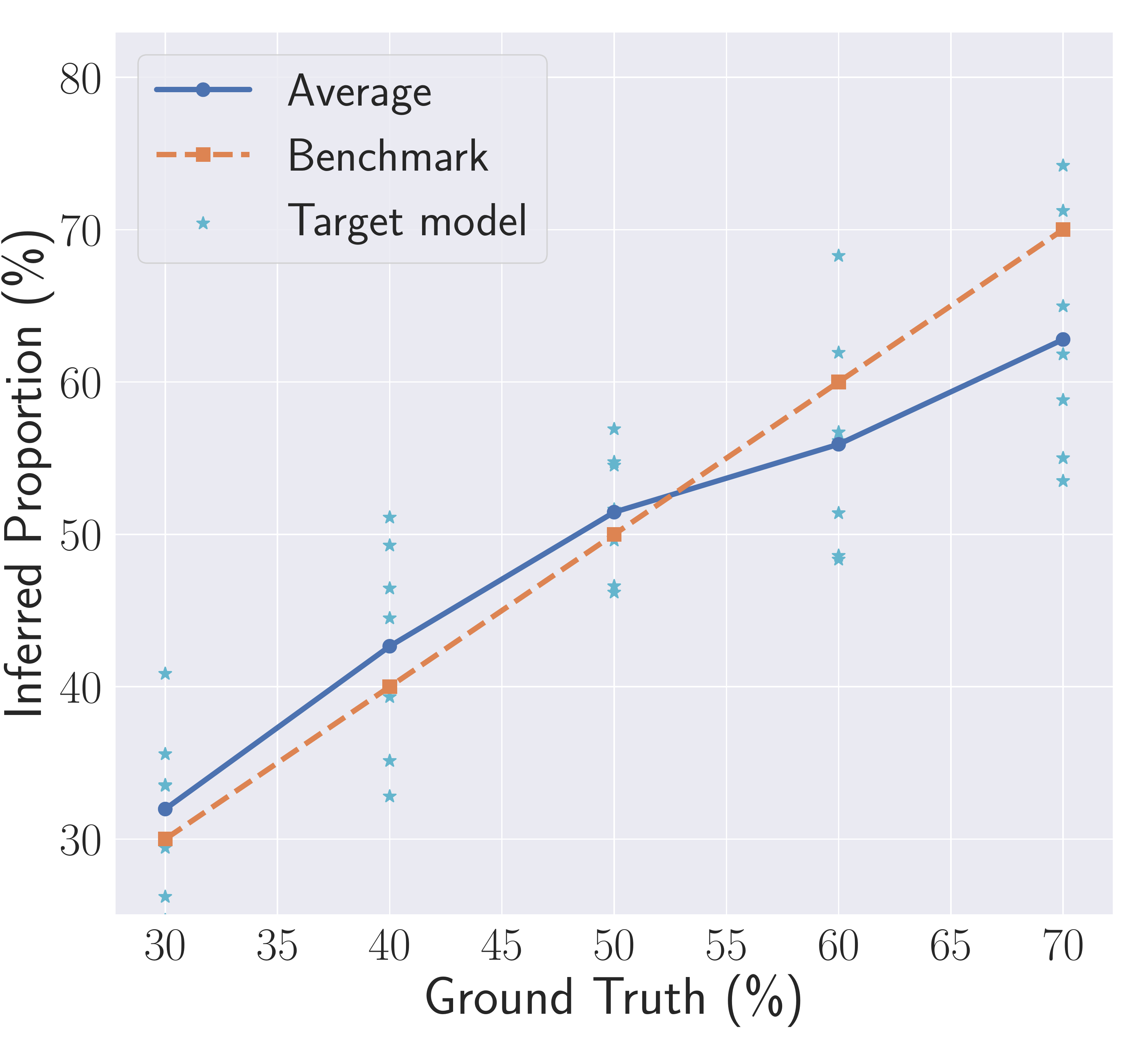}
\caption{Evaluation on $T_2$}
\label{figure:fullBBsight2}
\end{subfigure}
\begin{subfigure}{0.4\columnwidth}
\includegraphics[width=\columnwidth]{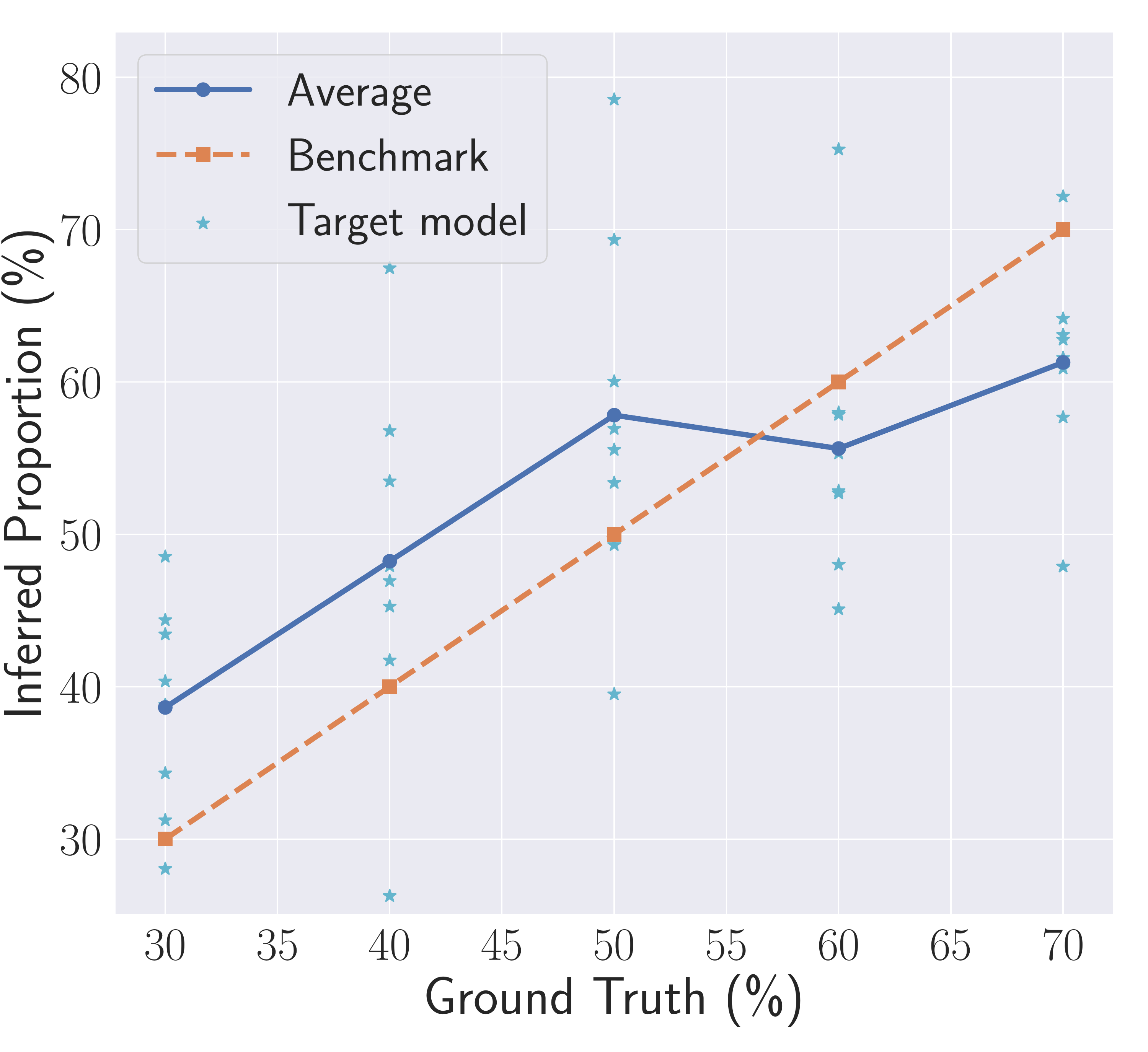}
\caption{Evaluation on $T_3$}
\label{figure:fullBBsight3}
\end{subfigure}
\begin{subfigure}{0.4\columnwidth}
\includegraphics[width=\columnwidth]{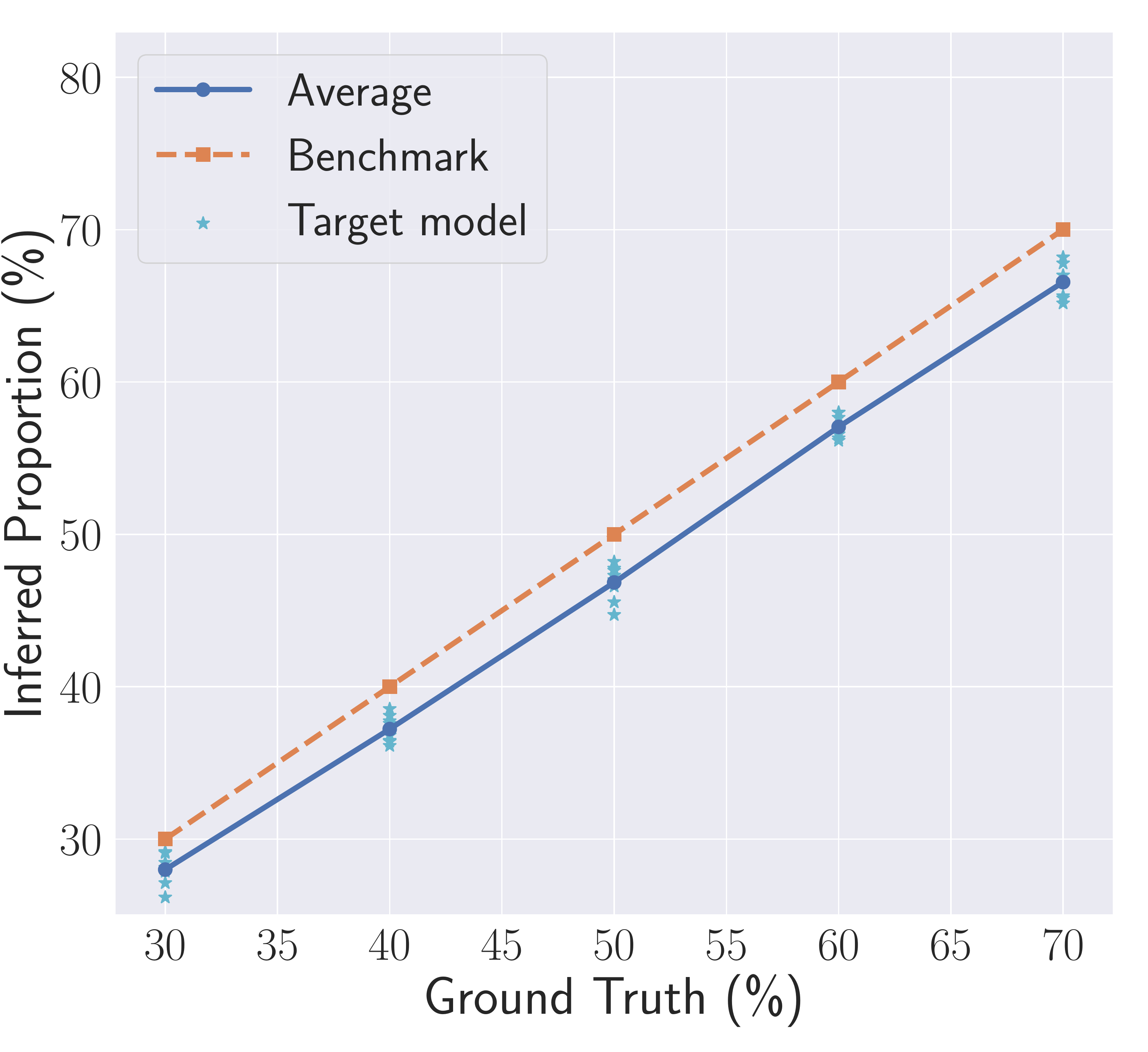}
\caption{Evaluation on $T_4$}
\label{figure:fullBBsight4}
\end{subfigure}
\begin{subfigure}{0.4\columnwidth}
\includegraphics[width=\columnwidth]{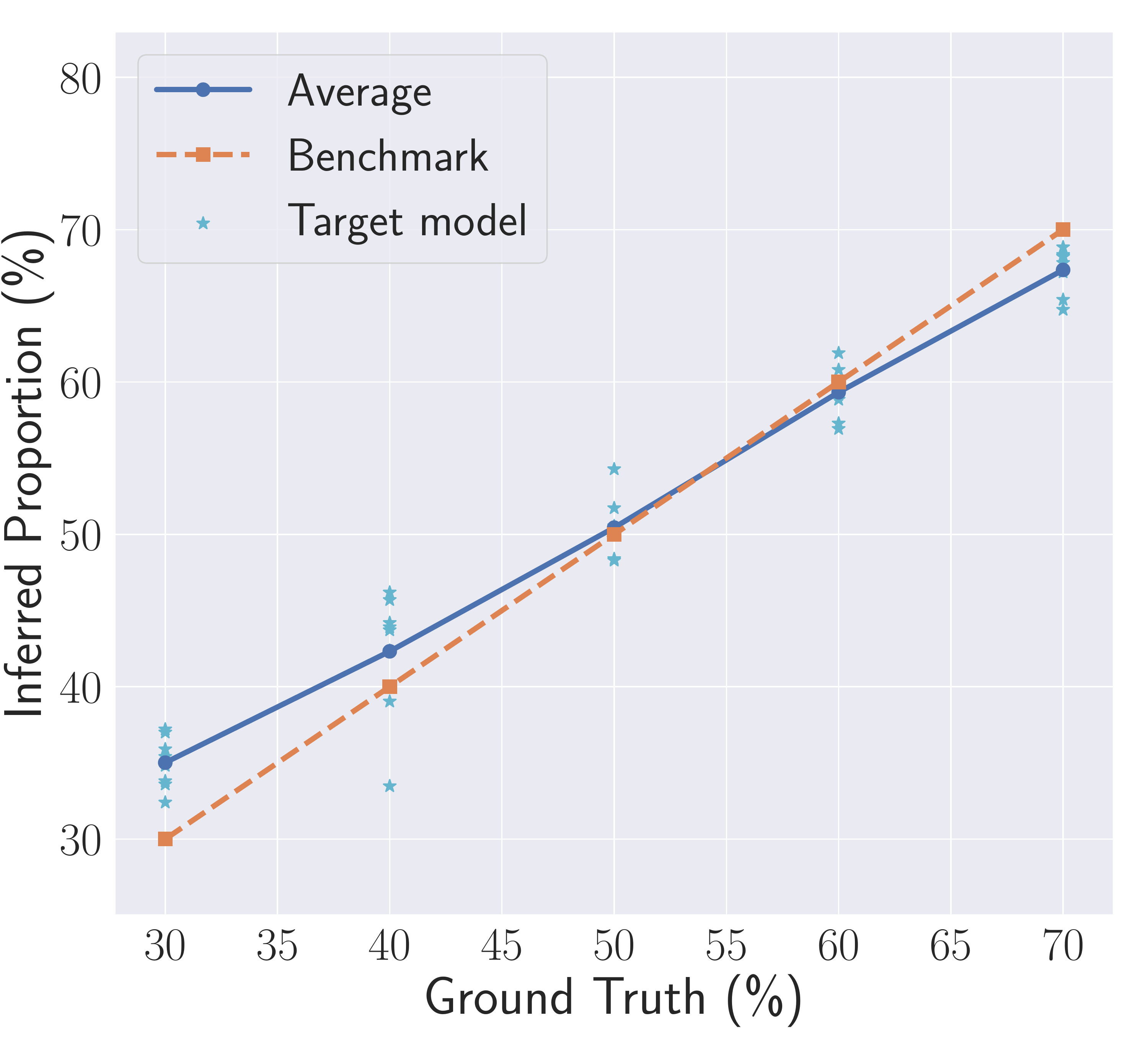}
\caption{Evaluation on $T_5$}
\label{figure:fullBBsight5}
\end{subfigure}
\caption{Full black-box attack performance.
Each point depicts a target model with corresponding underlying property (Ground Truth) and inferred property (Inferred Proportion).
The average curve gives an average attack result for target models with the same underlying property.
The benchmark line refers to the best attack result, where the inferred proportion is exactly equal to the ground truth.
The inferred result for each target model is obtained using 20K randomly generated samples.}
\label{figure:fullBBsight}
\end{figure*}

\begin{figure*}[!t]
\centering
\begin{subfigure}{0.4\columnwidth}
\includegraphics[width=\columnwidth]{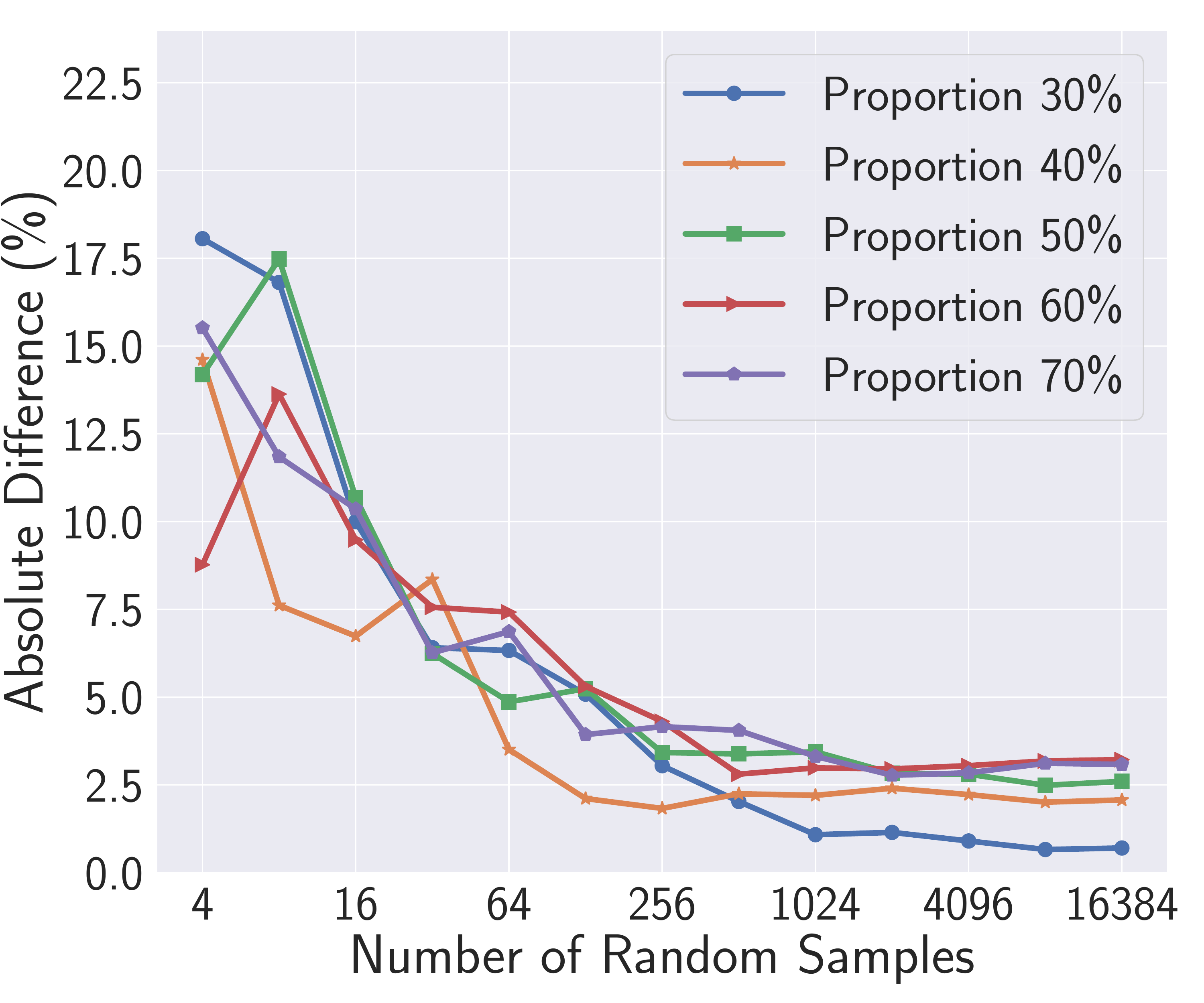}
\caption{Evaluation on $T_1$}
\label{figure:fullBBnumEUC1}
\end{subfigure}
\begin{subfigure}{0.4\columnwidth}
\includegraphics[width=\columnwidth]{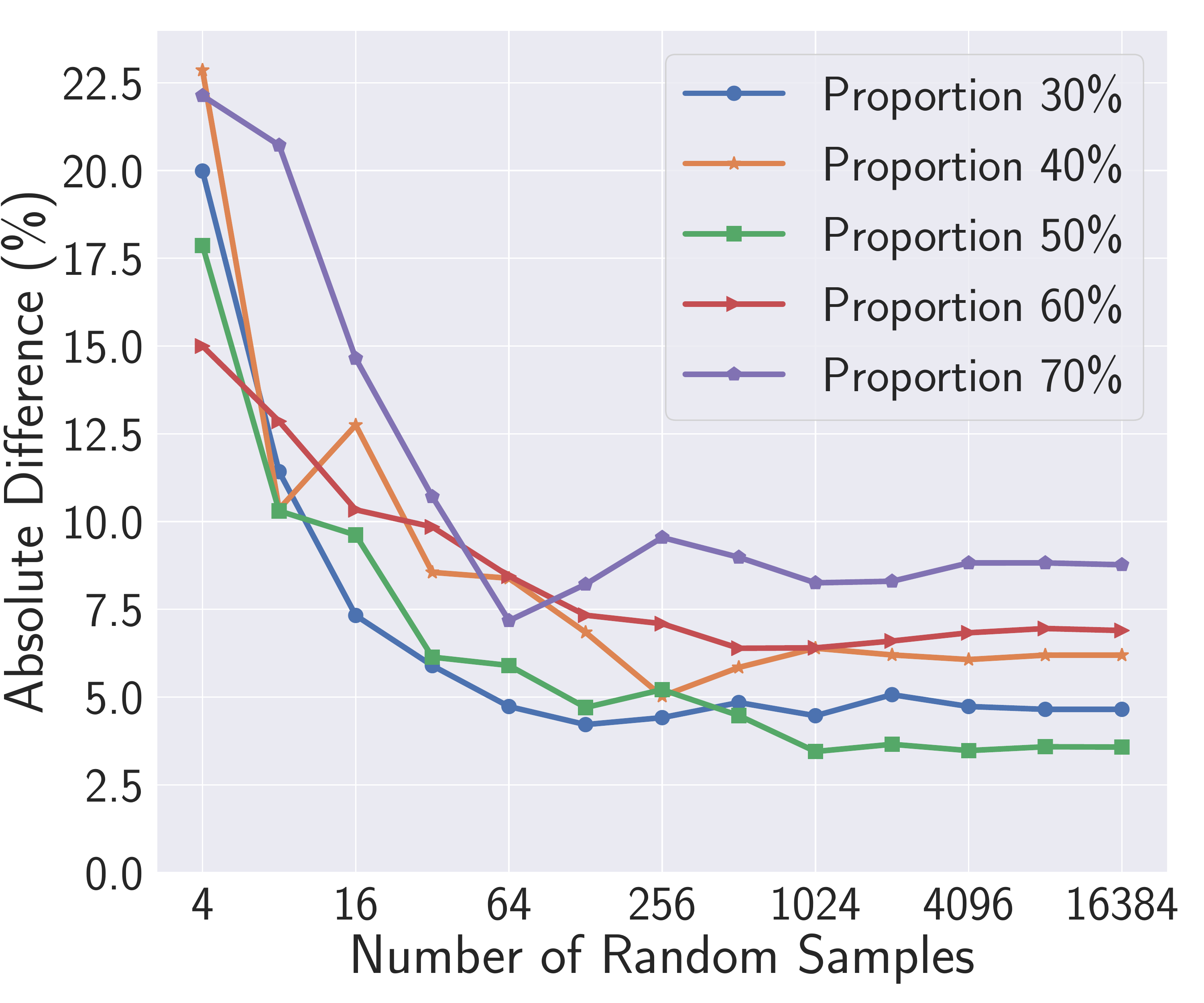}
\caption{Evaluation on $T_2$}
\label{figure:fullBBnumEUC2}
\end{subfigure}
\begin{subfigure}{0.4\columnwidth}
\includegraphics[width=\columnwidth]{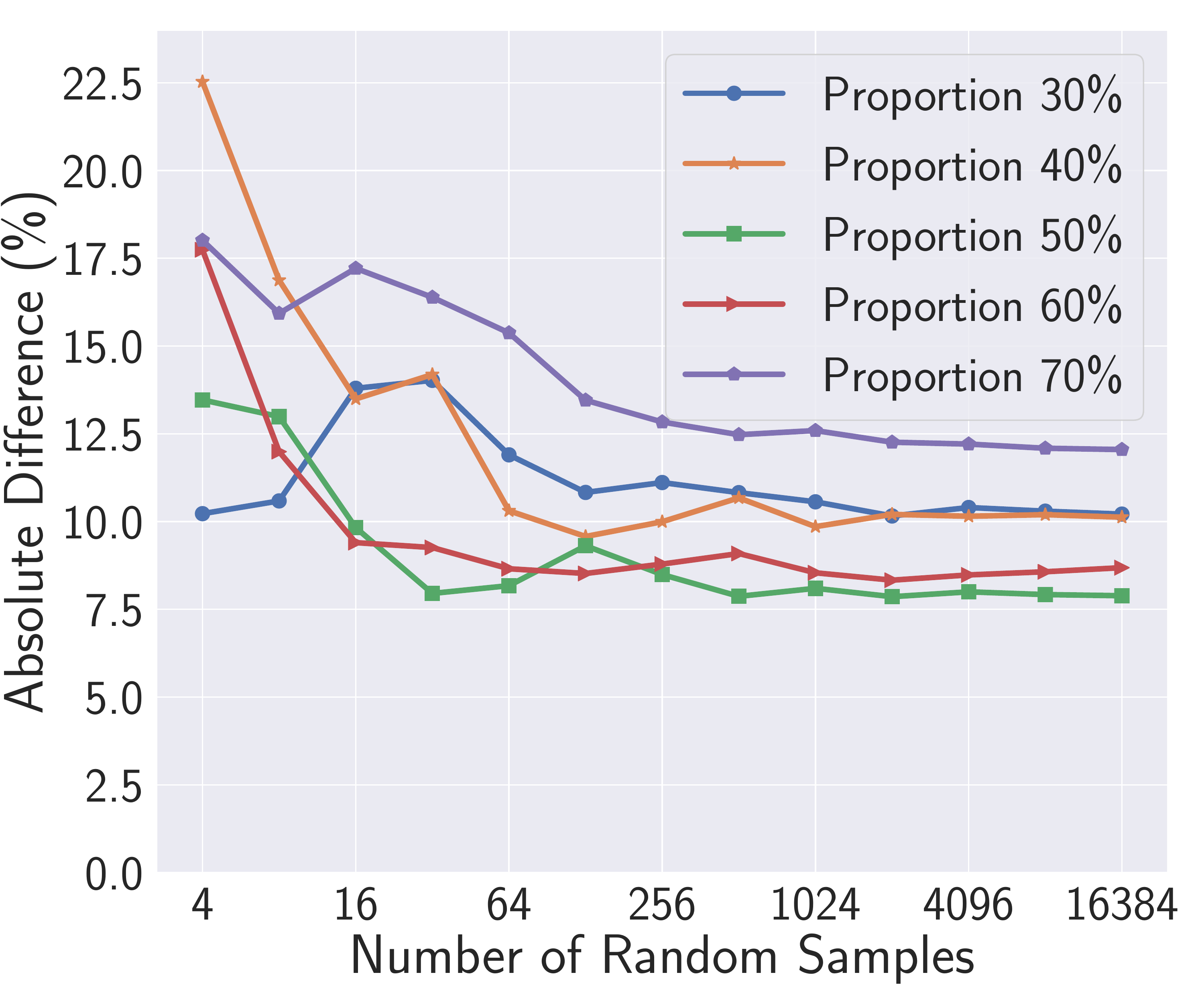}
\caption{Evaluation on $T_3$}
\label{figure:fullBBnumEUC3}
\end{subfigure}
\begin{subfigure}{0.4\columnwidth}
\includegraphics[width=\columnwidth]{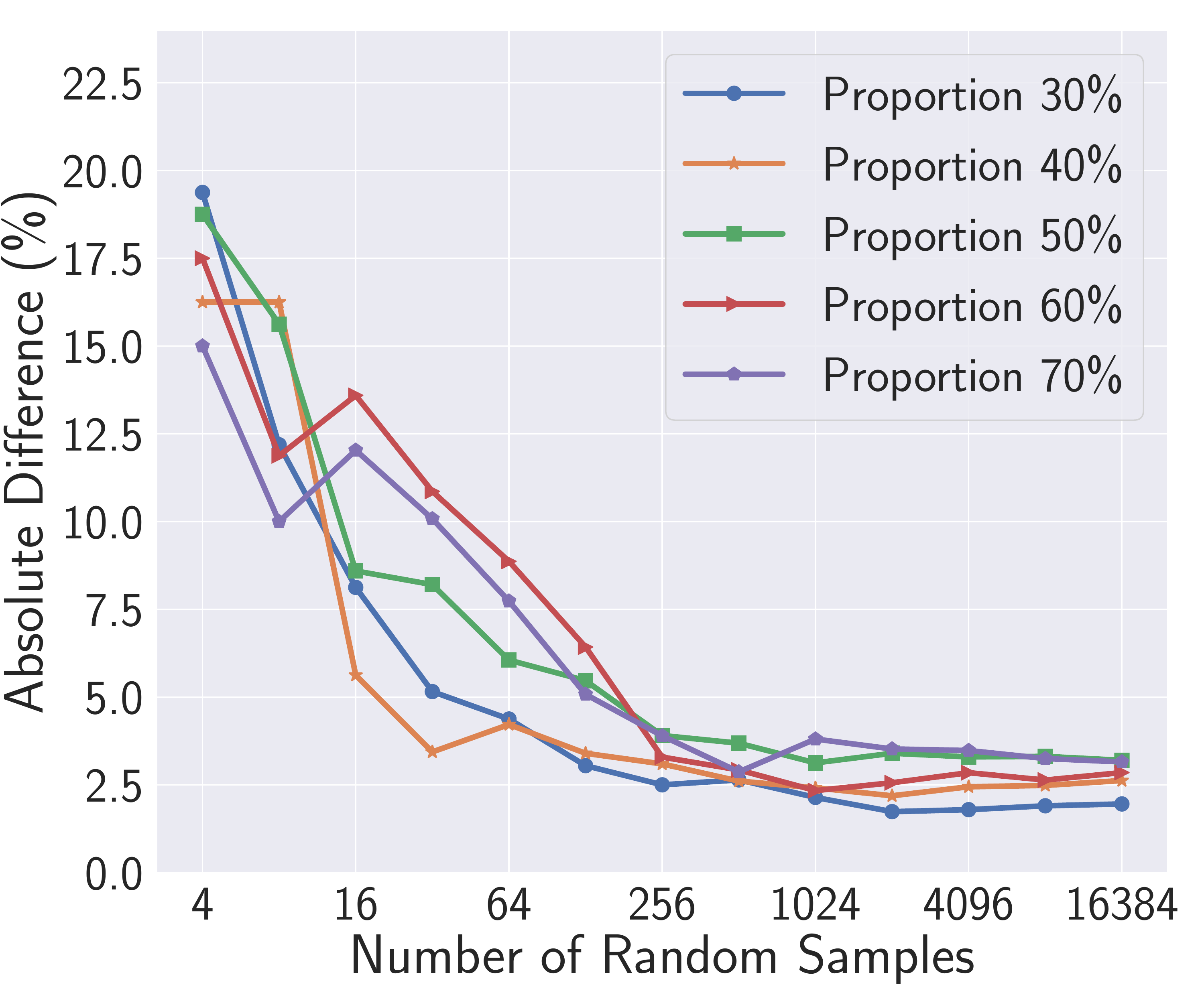}
\caption{Evaluation on $T_4$}
\label{figure:fullBBnumEUC4}
\end{subfigure}
\begin{subfigure}{0.4\columnwidth}
\includegraphics[width=\columnwidth]{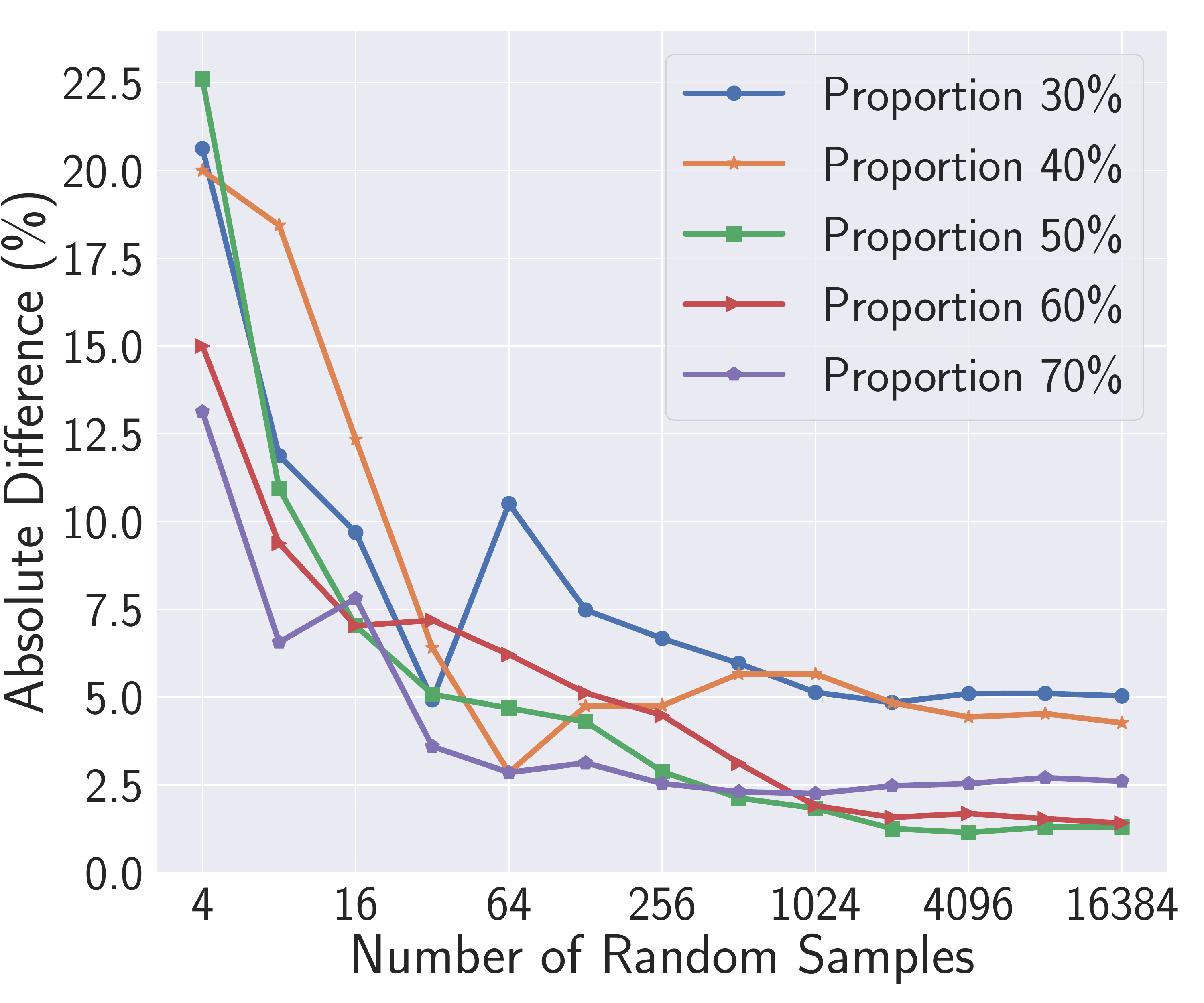}
\caption{Evaluation on $T_5$}
\label{figure:fullBBnumEUC5}
\end{subfigure}
\caption{Full black-box performance w.r.t.\ number of random samples, evaluated with the absolute difference between the inferred property and the underlying property.
Each line presents the average behavior of the target models with the same property, and how the behavior changes using the different number of random samples.}
\label{figure:fullBBnum}
\end{figure*}

We evaluate the performance of our property inference under five different settings, including the proportion of males on CelebA ($T_1$), the proportion of males on AFAD ($T_2$), the proportion of youth on AFAD ($T_3$), the proportion of 0s on MNIST ($T_4$), and the proportion of the high-income on Census Incone dataset ($T_5$).
\autoref{table:experiment_setting} lists them in detail.
For each task, we firstly split the dataset into three disjoint parts, i.e., a shadow dataset $\dataset_{\shadow}$ to draw $\dataset_{k}$, a dataset to draw $\dataset_{\target}$ with the same size as $\dataset_{\shadow}$, and a classifier dataset $\dataset_{\classify}$ to train $\classifier$.
We then split $\dataset_{\classify}$ with a proportion of 7:3 to train and test the property classifier.
Note that, all of the samples are drawn randomly but under a control on the number of images with the target analyzing attribute.
For instance, when we focus on inferring the gender distribution of the training dataset, we will control the number of females and males.
In this way, these split datasets follow the same distribution as the original datasets except the target attribute.
Moreover, the concrete size of the disjoint dataset is also presented in \autoref{table:experiment_setting}.

With the aim of inferring the distribution of the target GAN's underlying training dataset, the property of each shadow dataset $\property_{k}$ can be simply set as a range of proportions covering the target property.
In our experiments, we assume that $\property_{\target},\property_{k}\in\{30\%,40\%,50\%,60\%,70\%\}$.
In this way, we draw $\dataset_{\target}$ and $\dataset_{k}$ randomly, while controlling the size and property of each dataset.
Note that, we assume the target property $\property_{\target}$ is between 30\% and 70\%, so we simply set the property of each shadow dataset within the same range.
Moreover, we also investigate how our partial black-box attack behaves when the property of shadow models do not cover their target property, as shown in \autoref{figure:out_of_range}. 

\begin{table}[!t]
\centering
\caption{Average FID for GAN models in each task.}
\label{table:FID}
\scriptsize
\begin{tabular}{l | c | c | c | c }
\toprule
   & WGANGP $T_1$  & PGGAN $T_2$ & PGGAN $T_3$ & DCGAN $T_4$  \\
\midrule
Target GANs & 33.62 & 28.79 & 29.48 & 49.07\\
Shaodw GANs & 35.17 & 29.28 & 29.02 & 51.23\\
\bottomrule
\end{tabular}
\end{table}

For each inference task, our experiment basically goes as follows.
In the training stage, we have trained eight target models with each $\property_{\target}$, i.e., 40 target models in total, so as to examine the variation in inference accuracy.  
In the attack stage, we build up 20 shadow models for each $\property_k$, using the same GAN structures and training hyper-parameters with the target models.
Consequently, 100 shadow models are involved in the evaluation for the partial black-box attack.
The corresponding quantitative evaluation in terms of Fréchet Inception Distance (FID) metric~\cite{HRUNH17} is shown in~\autoref{table:FID} to present the quality of our shadow and target GANs.
We can find that our GAN models are in a reasonable range compared with experiments (FID ranging from 14.86 to 53.08) in~\cite{CYZF20}.
Note that we will further discuss the influence of shadow models in \autoref{section:Shadow} based on different GAN structures and underlying datasets.

% ======================================================
\subsection{Attack Evaluation}
% ======================================================

In this paper, our proposed property inference attack produces a continuous property value, instead of a discrete classification label like~\cite{GWYGB18}.
Therefore, we evaluate the effects of the attack based on the \emph{absolute difference} between the real property and the inference property.

Formally, as we focus on inferring the proportion of a certain attribute, the absolute difference of our attack can be easily calculated as $\vert\property_{\infer}-\property_{\real}\vert$, where $\property_{\infer}$ and $\property_{\real}$ range from 0\% to 100\%, represented by the percentage.
As the definition of the evaluation metric, the attack result is better when the calculated absolute difference is closer to 0\%.

\begin{figure*}[!t]
\centering
\begin{subfigure}{0.4\columnwidth}
\includegraphics[width=\columnwidth]{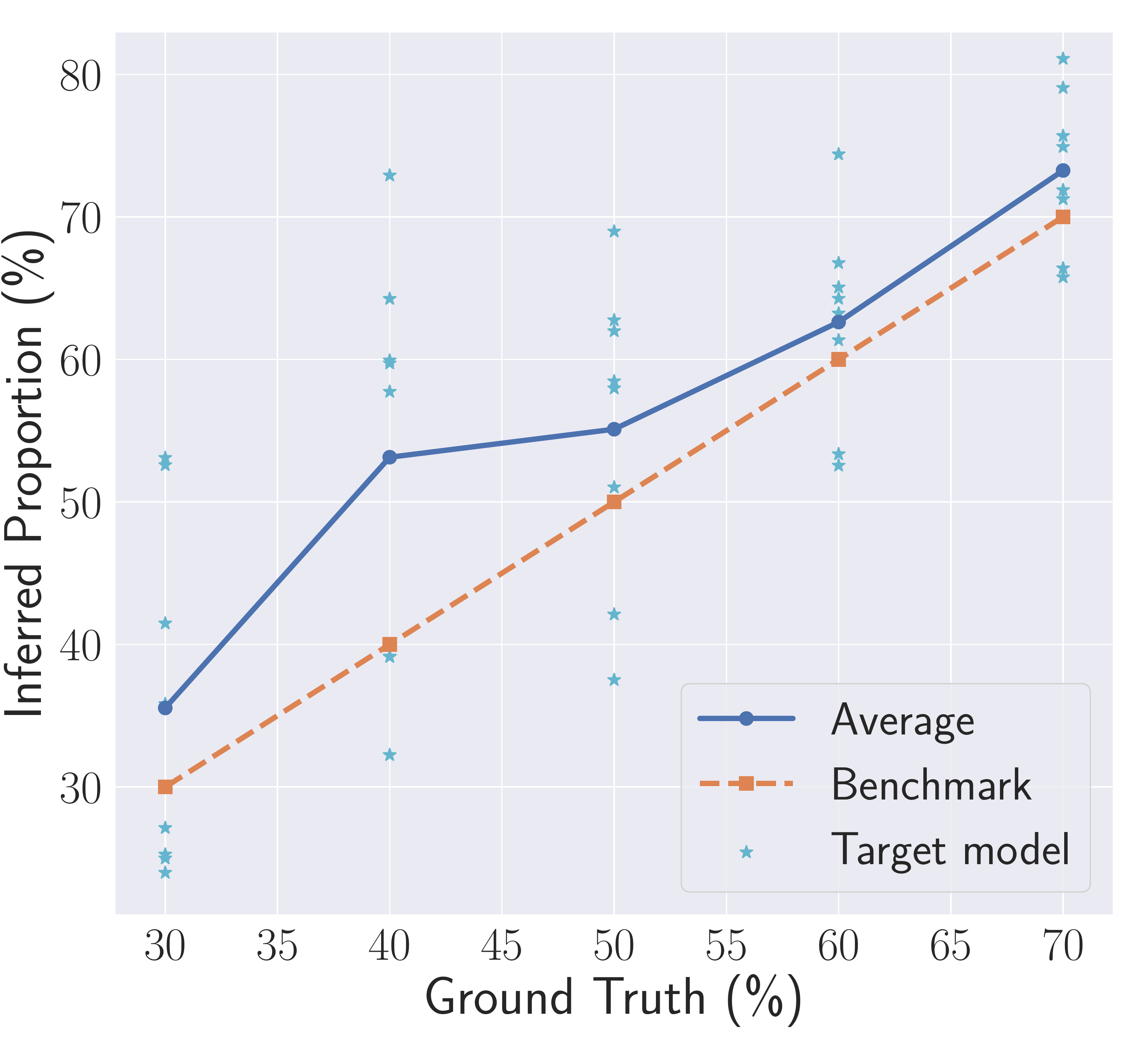}
\caption{Evaluation on $T_1$}
\label{figure:partBBsight1}
\end{subfigure}
\begin{subfigure}{0.4\columnwidth}
\includegraphics[width=\columnwidth]{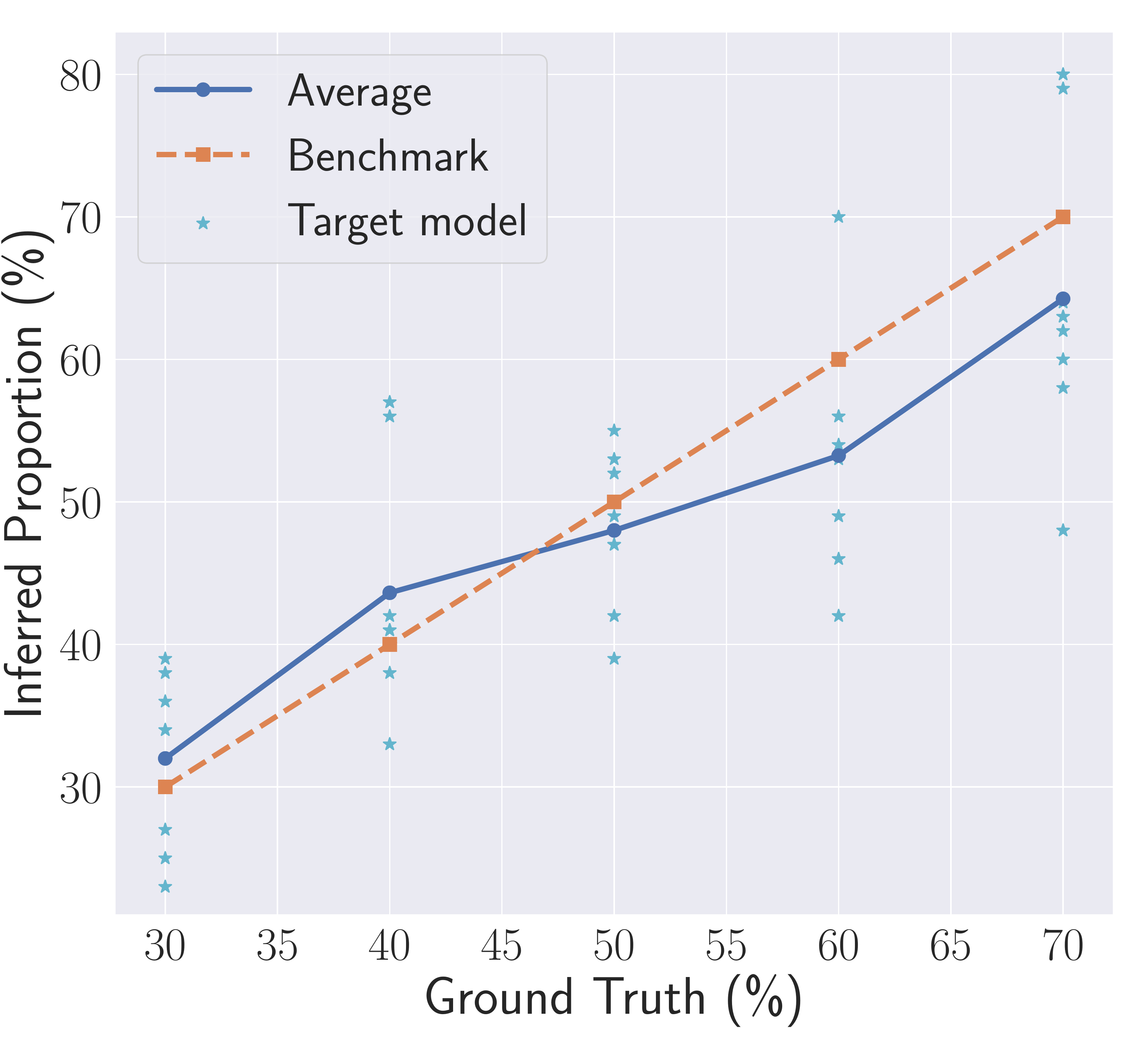}
\caption{Evaluation on $T_2$}
\label{figure:partBBsight2}
\end{subfigure}
\begin{subfigure}{0.4\columnwidth}
\includegraphics[width=\columnwidth]{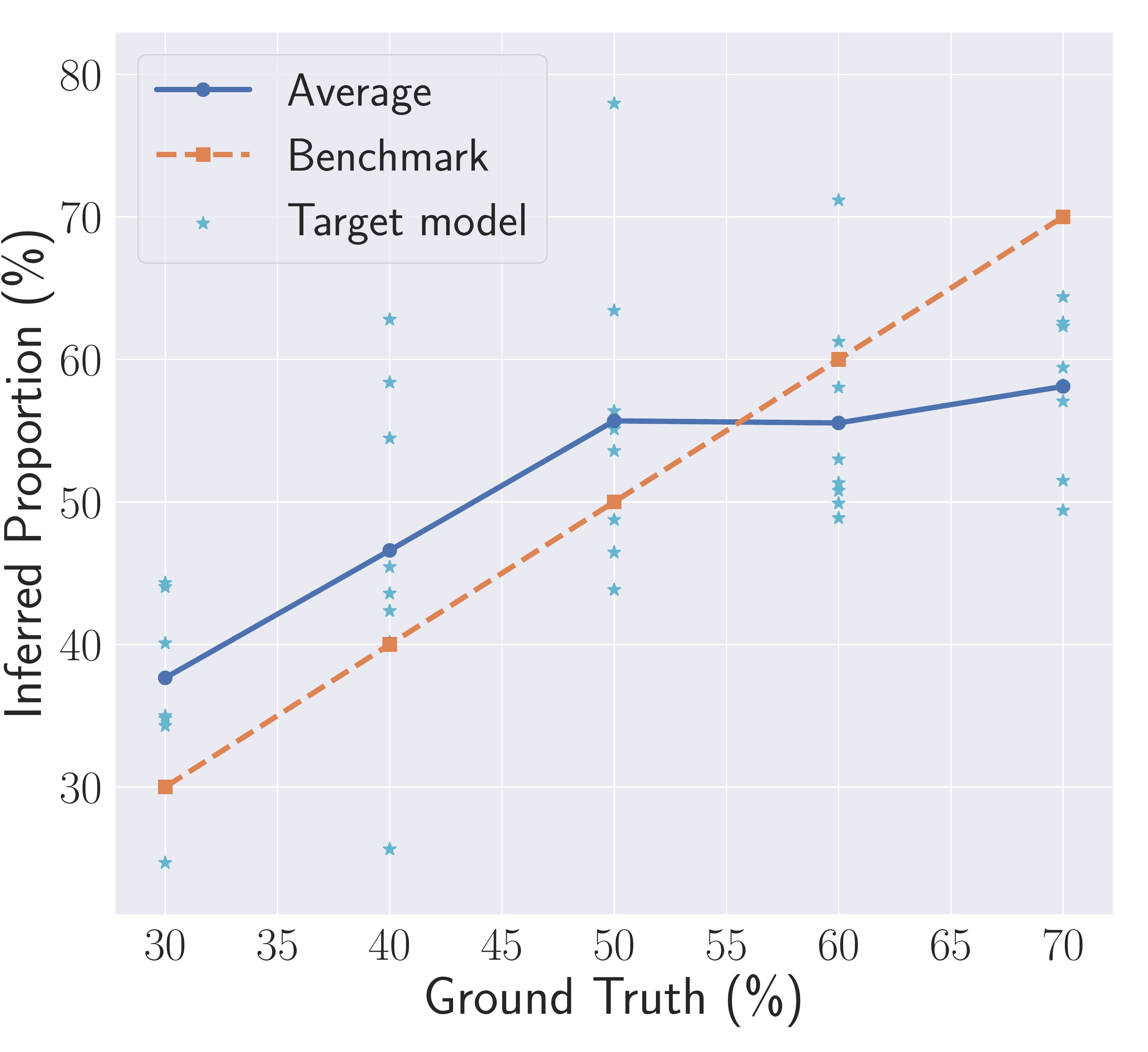}
\caption{Evaluation on $T_3$}
\label{figure:partBBsight3}
\end{subfigure}
\begin{subfigure}{0.4\columnwidth}
\includegraphics[width=\columnwidth]{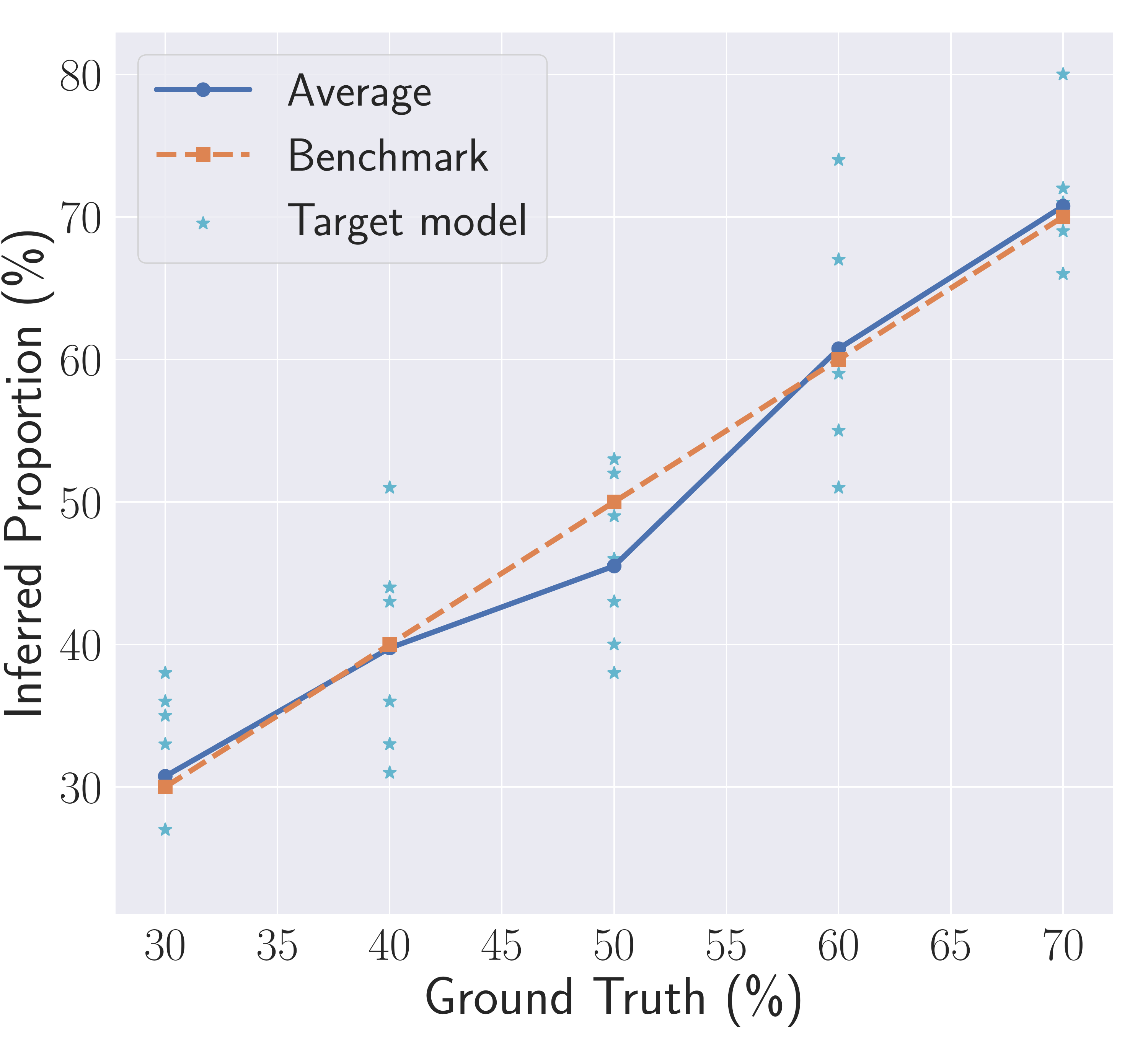}
\caption{Evaluation on $T_4$}
\label{figure:partBBsight4}
\end{subfigure}
\begin{subfigure}{0.4\columnwidth}
\includegraphics[width=\columnwidth]{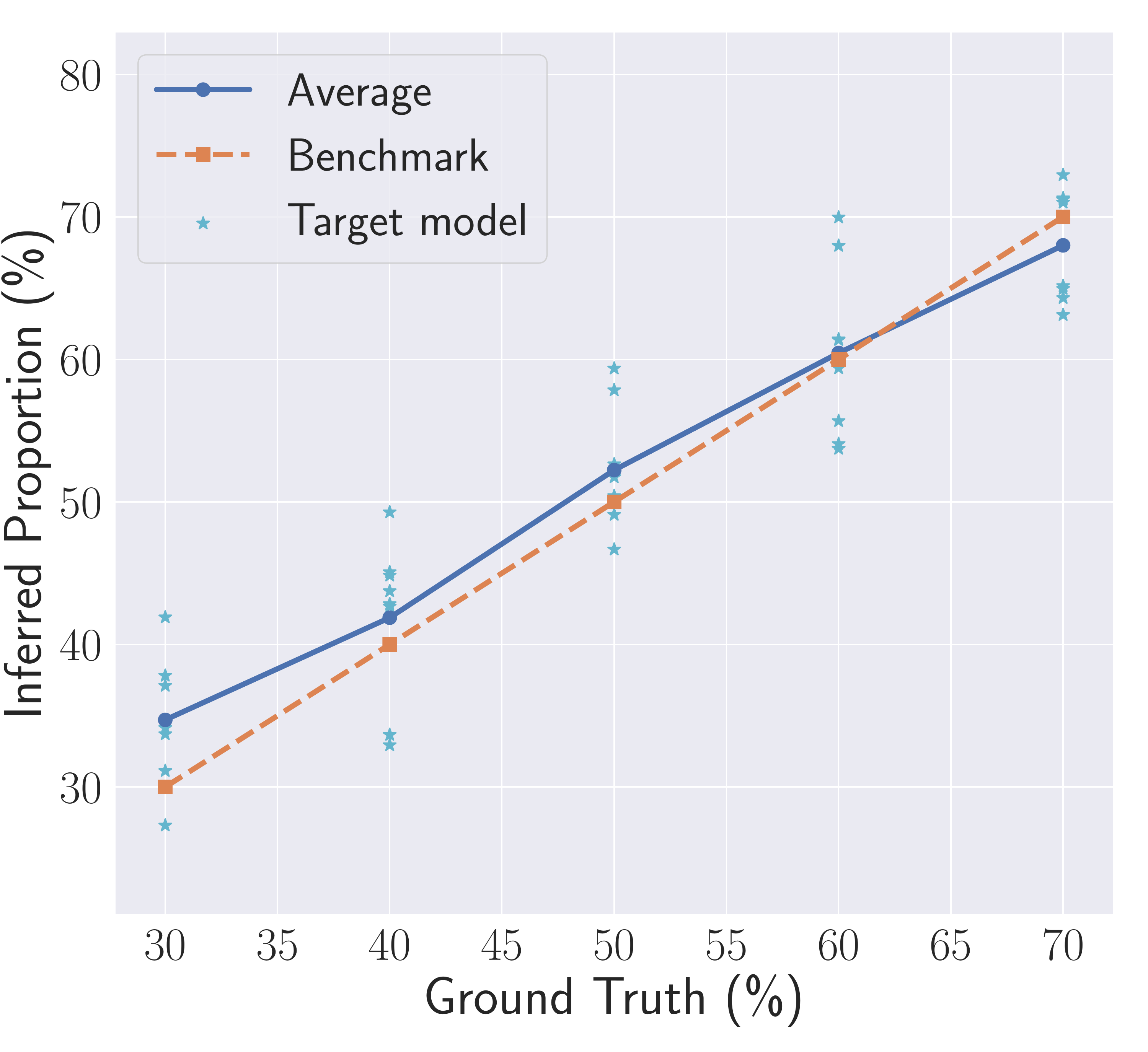}
\caption{Evaluation on $T_5$}
\label{figure:partBBsight5}
\end{subfigure}
\caption{Partial black-box attack performance.
Each point depicts a target model with corresponding underlying property (Ground Truth) and inferred property (Inferred Proportion) based on an optimized latent code set.
The average curve gives an average result for target models with the same underlying property.
The benchmark line refers to the best attack result.}
\label{figure:partBBsight}
\end{figure*}

\begin{figure*}[!t]
\centering
\begin{subfigure}{0.4\columnwidth}
\includegraphics[width=\columnwidth]{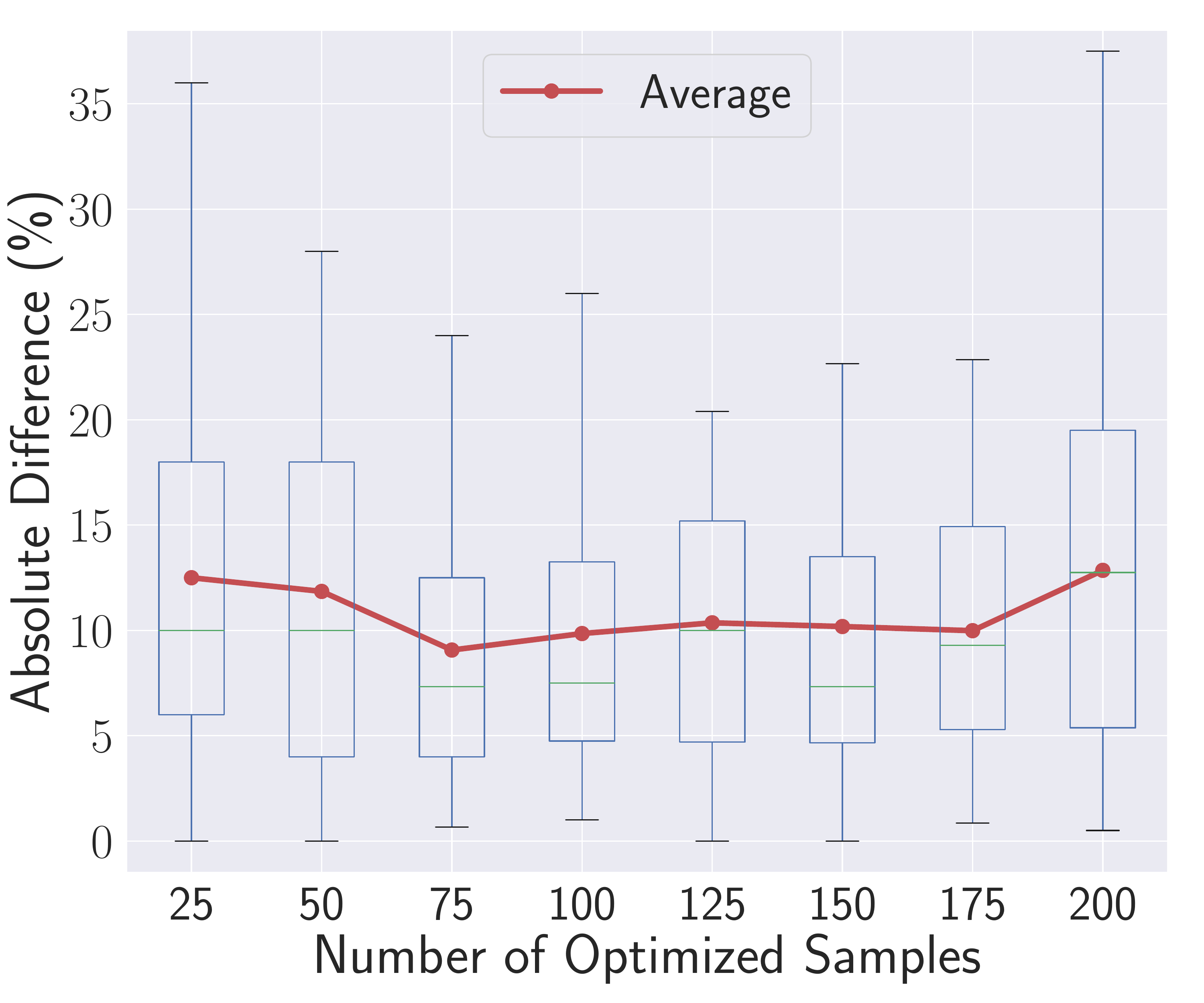}
\caption{Evaluation on $T_1$}
\label{figure:partBBnumLEUC1}
\end{subfigure}
\begin{subfigure}{0.4\columnwidth}
\includegraphics[width=\columnwidth]{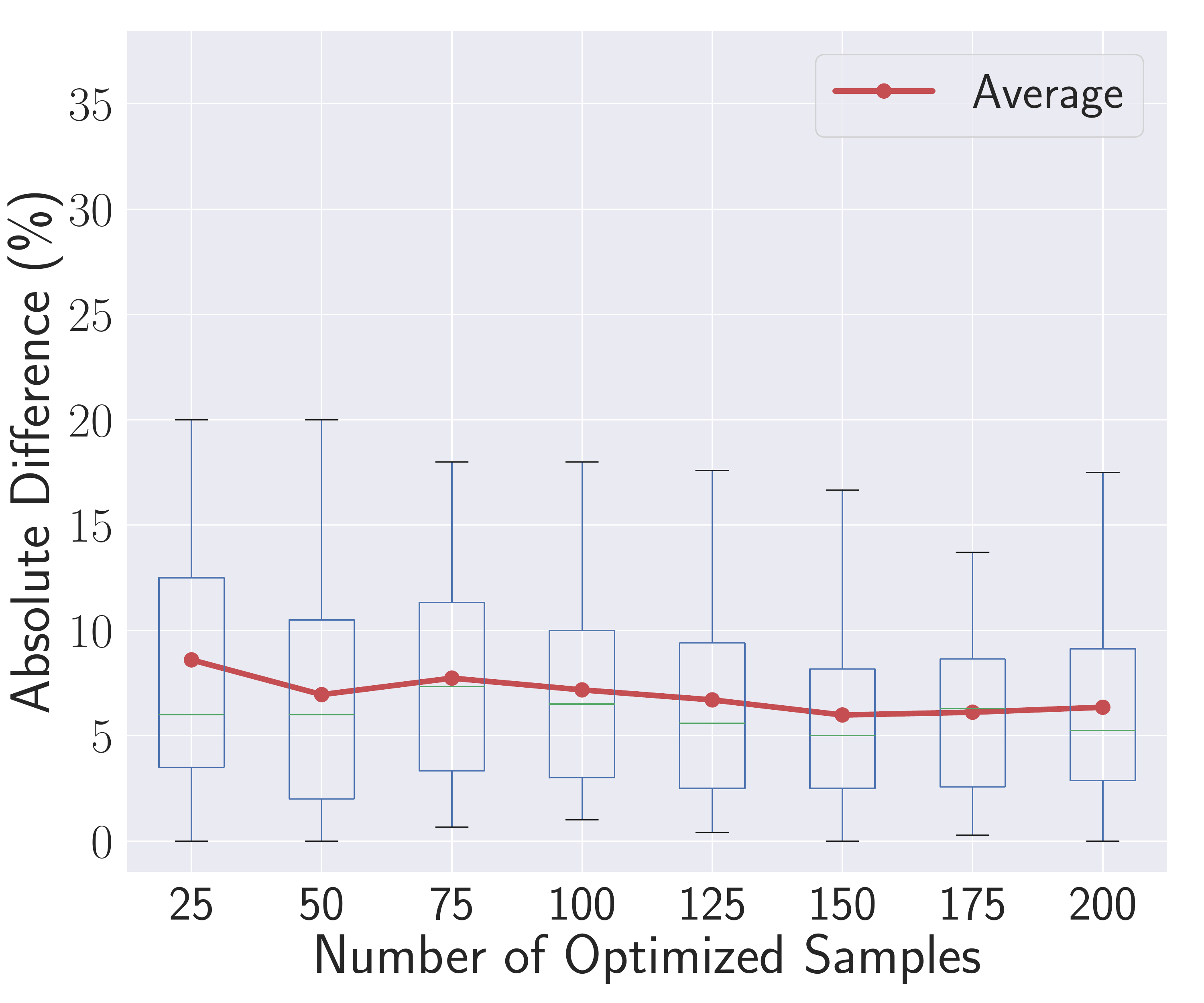}
\caption{Evaluation on $T_2$}
\label{figure:partBBnumLEUC2}
\end{subfigure}
\begin{subfigure}{0.4\columnwidth}
\includegraphics[width=\columnwidth]{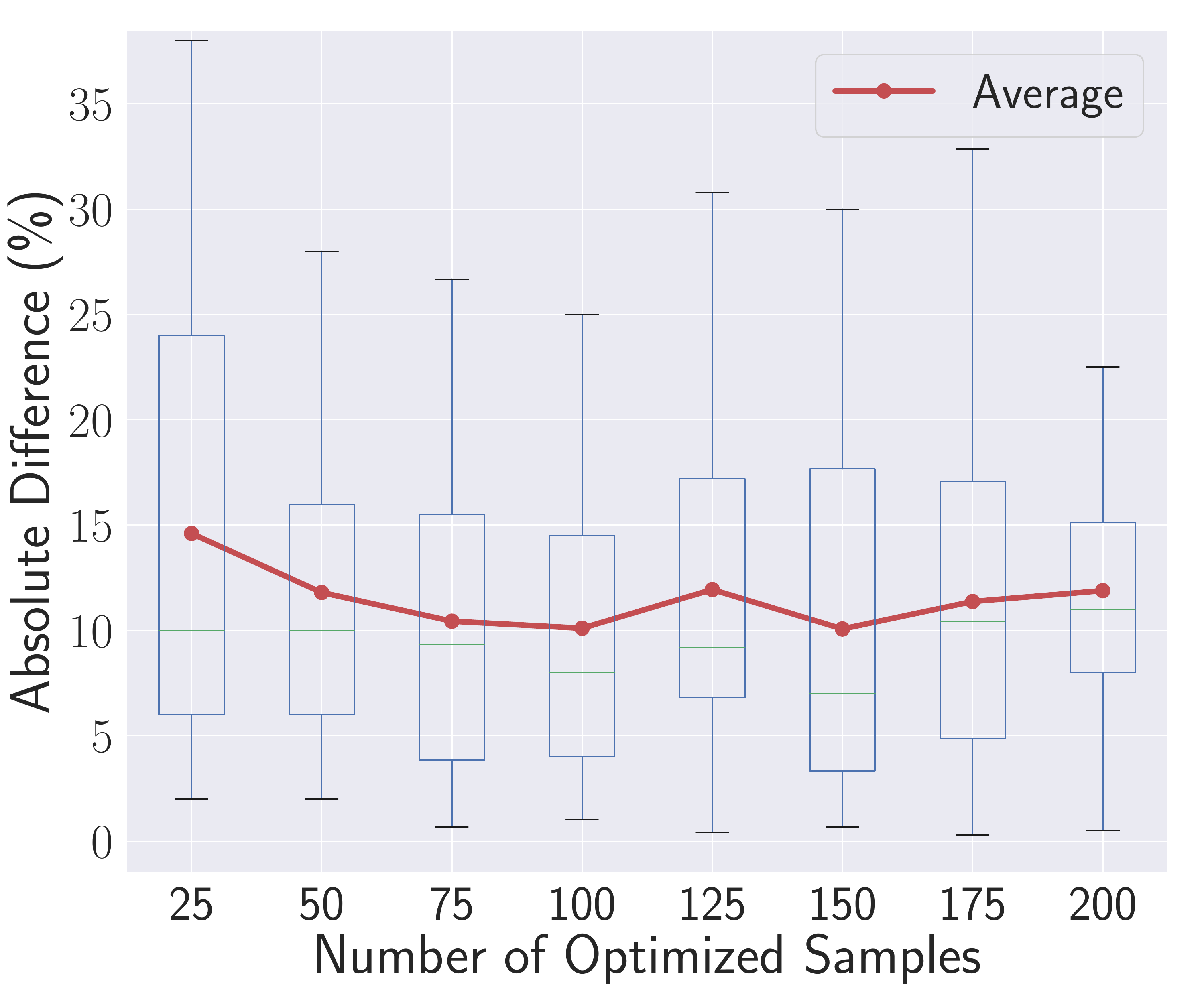}
\caption{Evaluation on $T_3$}
\label{figure:partBBnumLEUC3}
\end{subfigure}
\begin{subfigure}{0.4\columnwidth}
\includegraphics[width=\columnwidth]{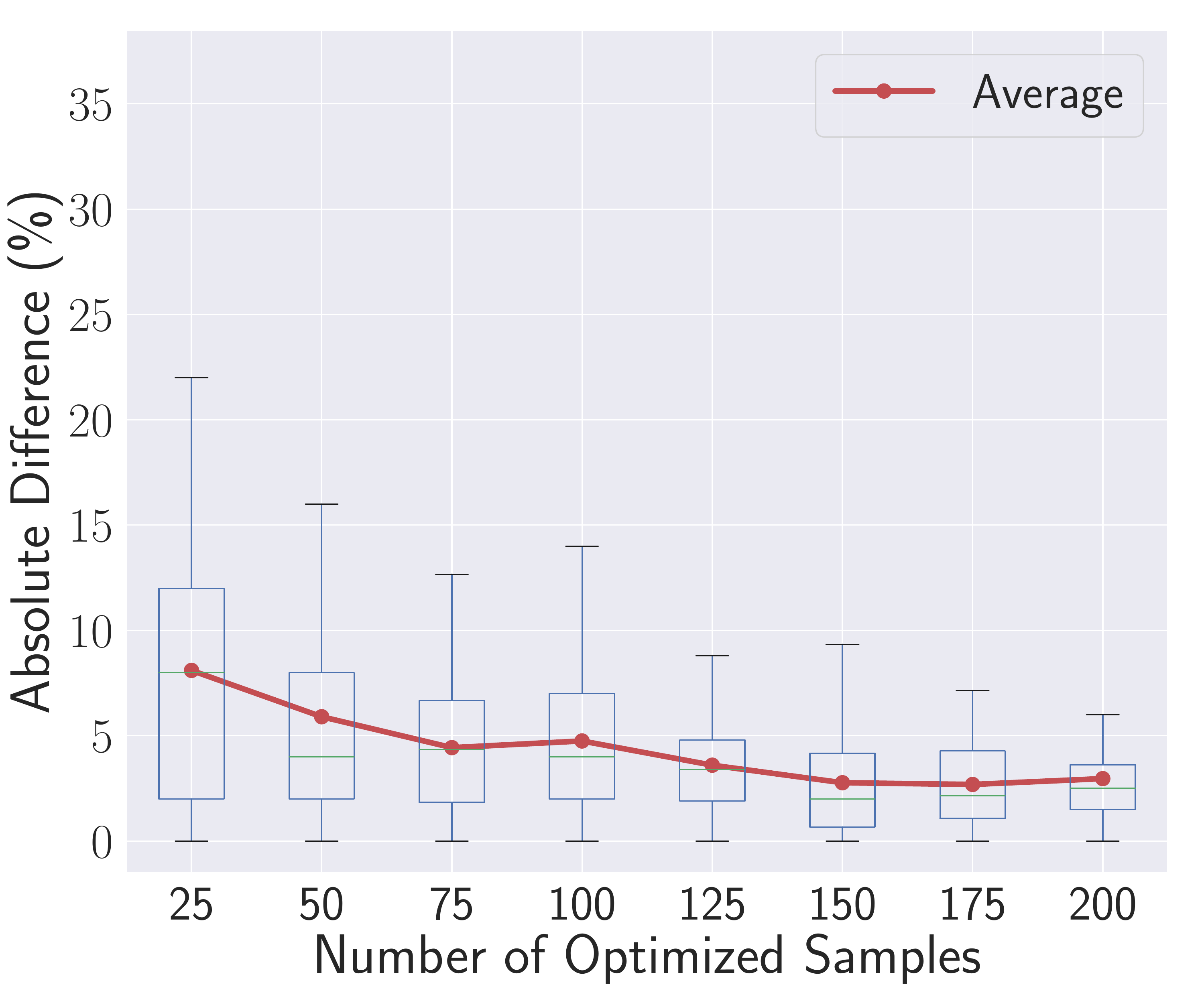}
\caption{Evaluation on $T_4$}
\label{figure:partBBnumLEUC4}
\end{subfigure}
\begin{subfigure}{0.4\columnwidth}
\includegraphics[width=\columnwidth]{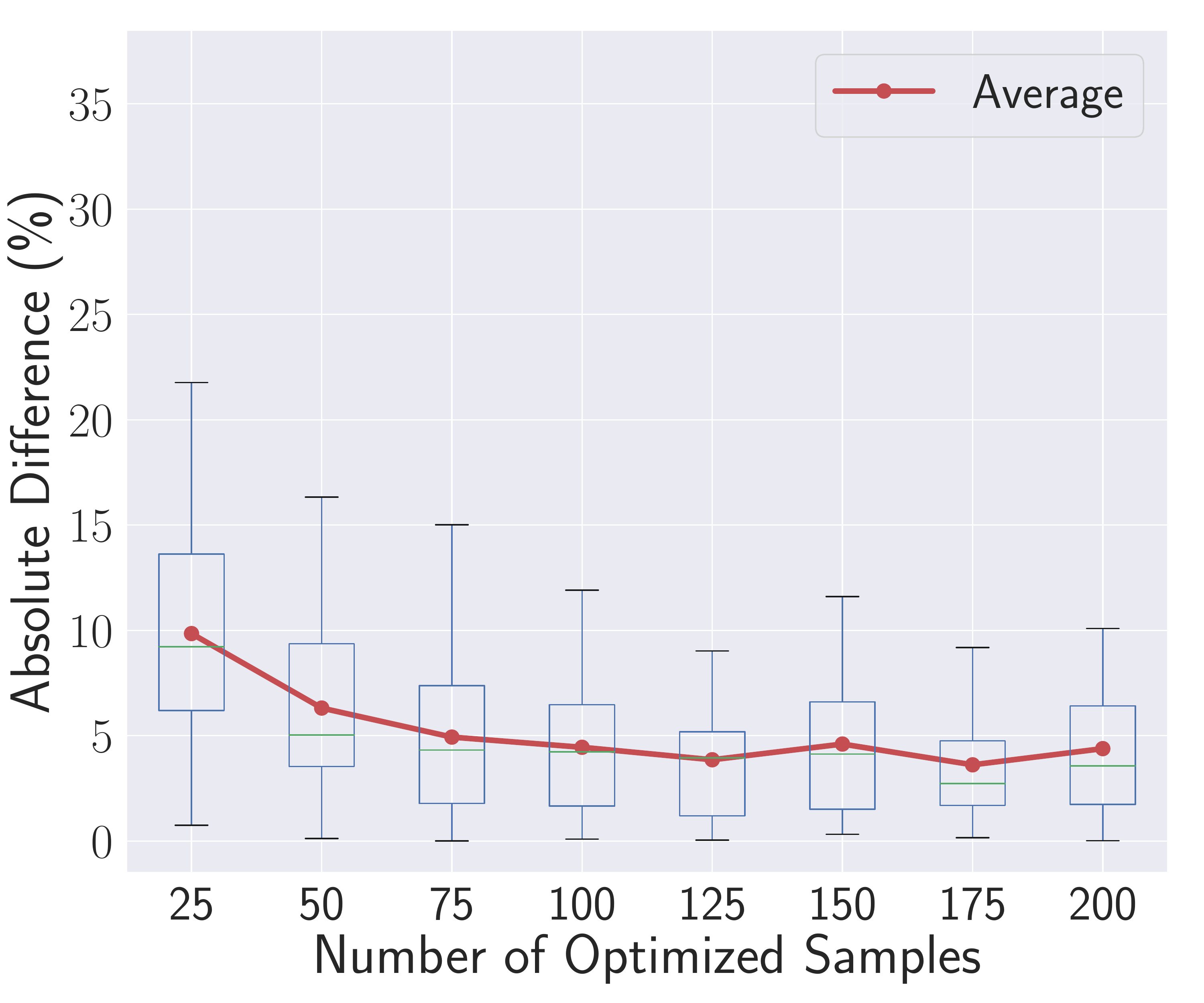}
\caption{Evaluation on $T_5$}
\label{figure:partBBnumLEUC5}
\end{subfigure}
\caption{Partial black-box performance w.r.t.\ number of optimized samples.
For each target model, we generate different numbers of optimized samples based on latent code sets and then obtain the inference results using these optimized samples.
The red curve represents the average attack performance against all target models based on different numbers of optimized samples.
We also provide box plots of our results for statistical analysis.}
\label{figure:partBBnumL}
\end{figure*}

% ======================================================
\subsection{Evaluation on Full Black-box Attack}
\label{section:evaluation_full_black-box}
% ======================================================

We first start by evaluating the full black-box method, which is the least knowledgeable setting for the adversary.

\autoref{figure:fullBBsight} shows the experimental results of our proposed full black-box attack against all target models.
The inferred property of each target model is represented as a single point, which depicts the black-box attack result based on 20K random generated samples against the corresponding underlying property (i.e., the ground truth).
We also plot the average inference result for target models sharing the same property, as well as an ideal benchmark line for comparison, on which the inferred property is exactly equal to the underlying property.
Overall, our results indicate the effectiveness of the full black-box attack, as we can clearly observe that the average inference curve corresponds closely to the benchmark line.
For instance, in \autoref{figure:fullBBsight1} focusing on target models with 30$\%$ males in the training dataset, our attack is well-behaved as the inferred proportion for each target model is very close to 30$\%$.
Moreover, we can see that in the tasks $T_1$ and $T_4$, our attack achieves quite a good attack performance, and the variances of inference results on target models are smaller than the other two.
Meanwhile, the result in the task $T_3$ is not so good, as the youth is hard to discriminate when using a local property classifier whose testing accuracy is only around 80{\%}.
How the $\classifier$ affects our attack behavior will be further explored in \autoref{section:property_classifier}.

\mypara{Number of Random Samples} 
Next, we evaluate our full black-box attack performance against an ablation study, i.e., the influence of the number of random generated samples.
We repeat our aforementioned full black-box evaluation with various random sample amounts to achieve the attack: $2^i (i = 2, 3, \cdots, 16)$.
After obtaining inferred proportions for all the target models, we average the inference results from the same underlying proportion and number of samples respectively for further analysis.
\autoref{figure:fullBBnum} plots the average full black-box attack performance against different numbers of random samples.
We can see the absolute difference of the inference result is obvious when using a relatively small amount of generated samples (less than 128).
And then the inference results become more \emph{accurate and stable} with the increasing number of random generated samples.
For instance, the green line in \autoref{figure:fullBBnumEUC2} depicts the average attack performance against target models with half male and half female training data, where the absolute difference is decreasing from 17.8\% to 3.7\% as the amount of samples grows to 1024, indicating the inference result is getting closer to the ground truth.
Moreover, focusing on $T_1$ and $T_2$, we can find the inference results on models with 70$\%$ males in the underlying training dataset are always worse than the others.
A possible explanation is that $\classifier$ has relatively higher accuracy on the female class. 
In that case, more males are misclassified as females when the proportion of males increases.

% ======================================================
\subsection{Evaluation on Partial Black-box Attack}
\label{section:evaluation_part_black-box}
% ======================================================

\begin{figure*}[!t]
\centering
\begin{subfigure}{0.5\columnwidth}
\includegraphics[width=\columnwidth]{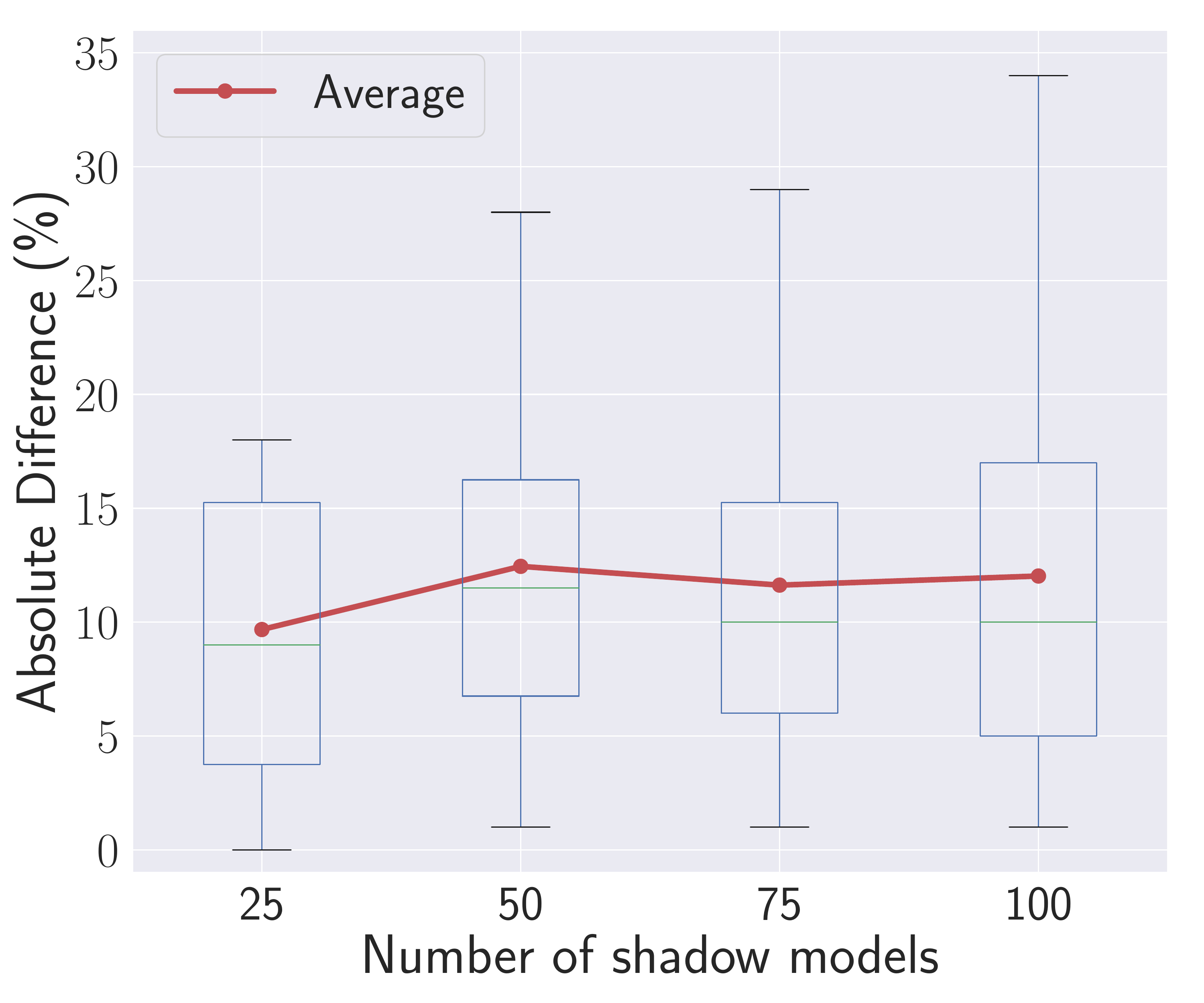}
\caption{Evaluation on $T_1$}
\label{figure:partBBnumSEUC1}
\end{subfigure}
\begin{subfigure}{0.5\columnwidth}
\includegraphics[width=\columnwidth]{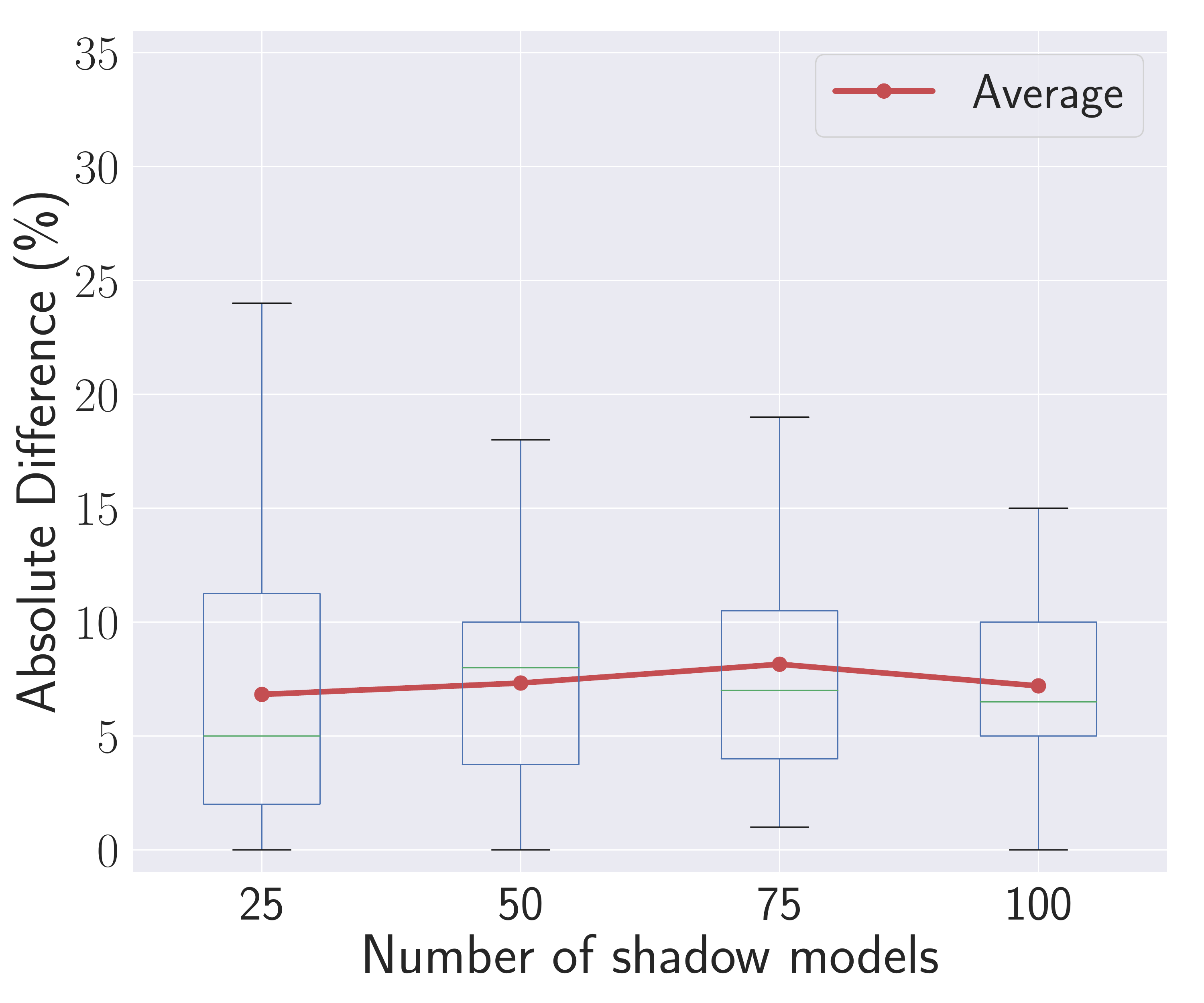}
\caption{Evaluation on $T_2$}
\label{figure:partBBnumSEUC2}
\end{subfigure}
\begin{subfigure}{0.5\columnwidth}
\includegraphics[width=\columnwidth]{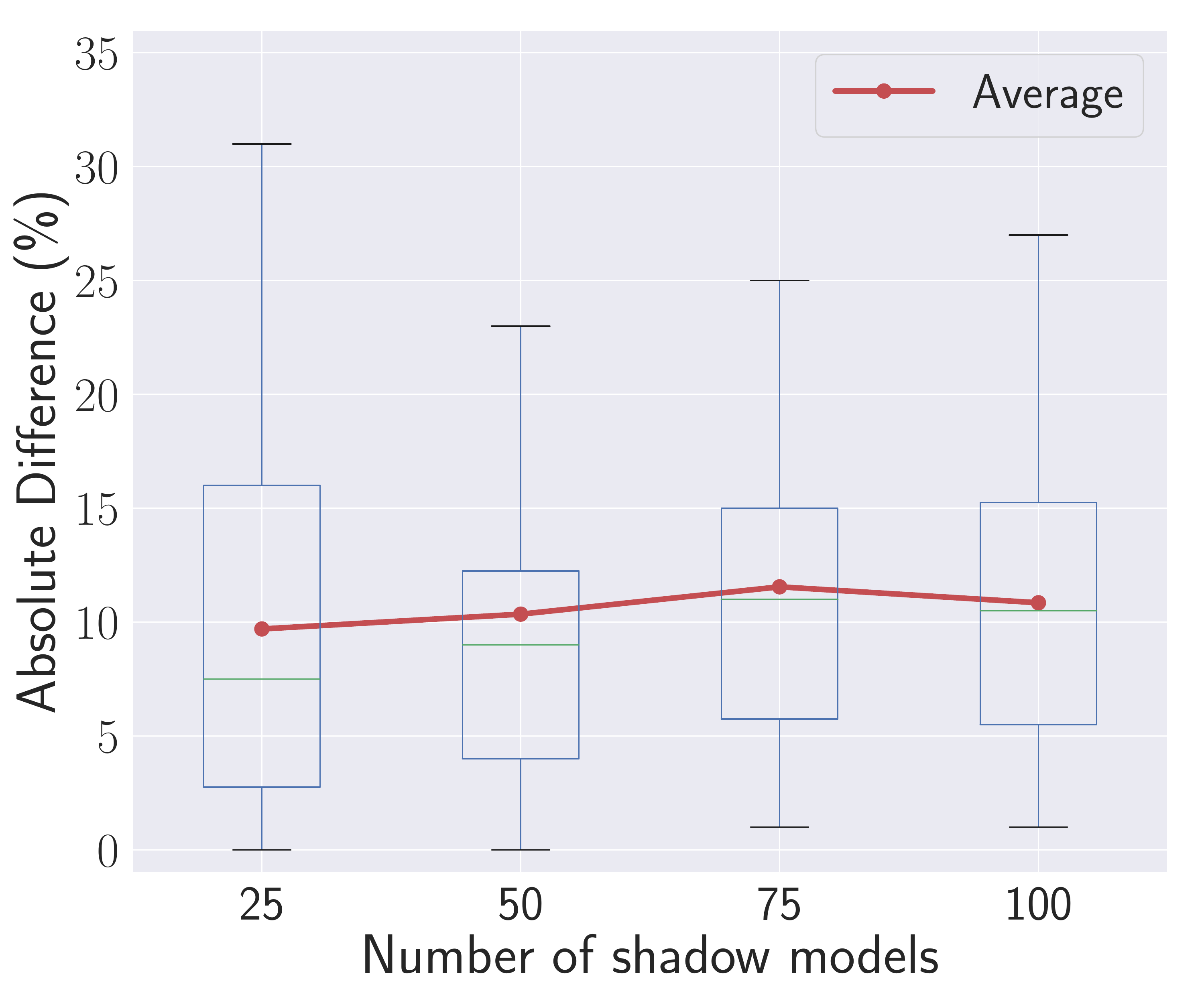}
\caption{Evaluation on $T_3$}
\label{figure:partBBnumSEUC3}
\end{subfigure}
\begin{subfigure}{0.5\columnwidth}
\includegraphics[width=\columnwidth]{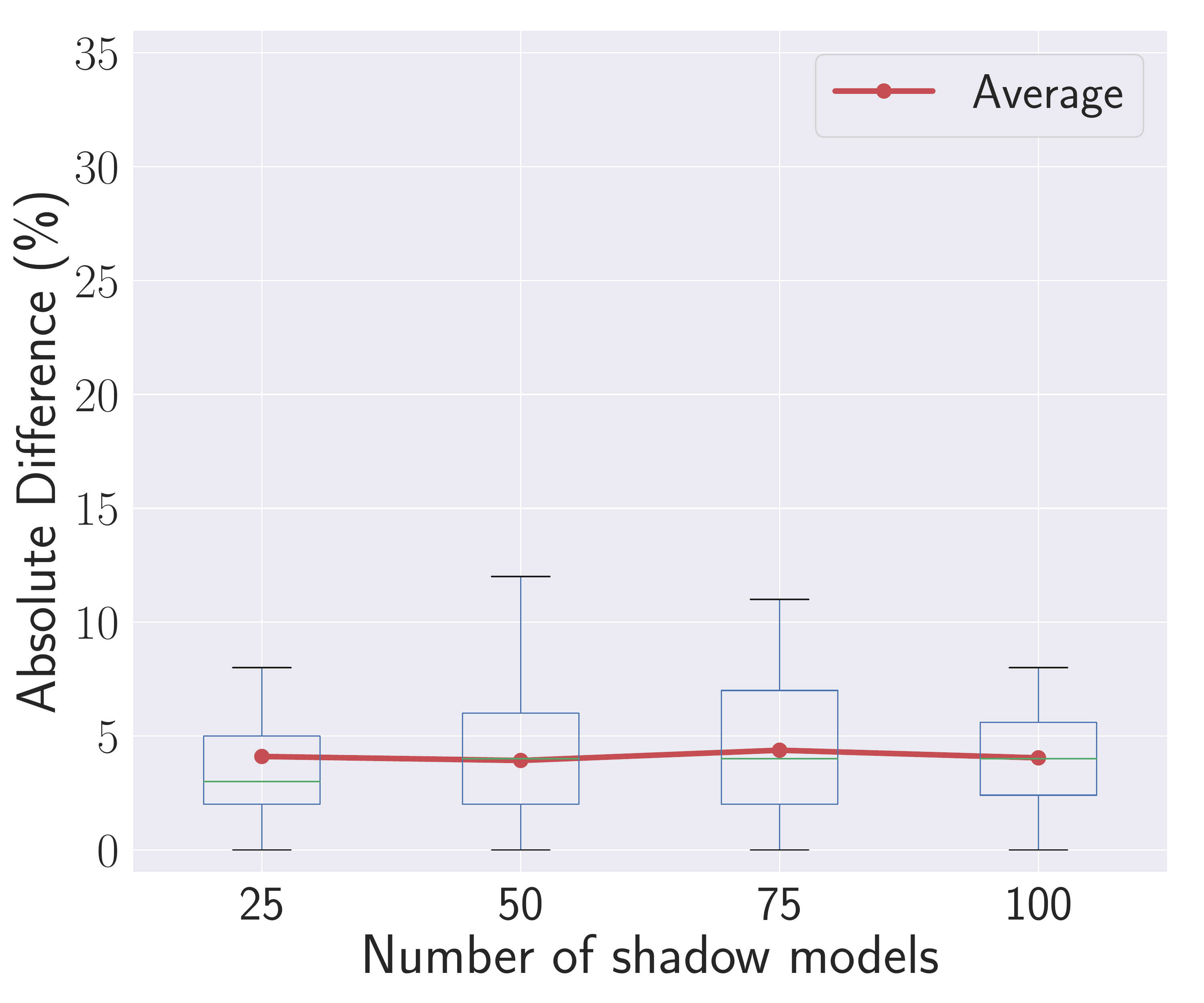}
\caption{Evaluation on $T_4$}
\label{figure:partBBnumSEUC4}
\end{subfigure}
\caption{Partial black-box performance w.r.t.\ number of shadow models.
For each target model, we generate 100 samples from the latent code sets optimized based on different numbers of shadow models and then obtain the inference results using these optimized samples.
The red curve shows the average performance of all target models through latent code sets optimized by different numbers of shadow models.
We also provide box plots of our results for statistical analysis.}
\label{figure:partBBnumS}
\end{figure*}

We then evaluate the partial black-box method, which relies on locally-trained shadow models.
\autoref{figure:partBBsight} shows the partial black-box attack results against all target models.
The inferred property of each target model is represented as a single point, which depicts the partial black-box attack result against the corresponding underlying property, using 100 optimized samples generated from an optimized latent code set.
Similar to \autoref{figure:fullBBsight}, we further plot a benchmark line as a reference to the best attack result, as well as an average line for target models with the same underlying property.
As we can see, the average inference curve lies close to the benchmark line, proving the effectiveness of our partial black-box attack.
For instance, we get an average inferred proportion of 48.4$\%$ for the target models with half males and half females in the underlying training dataset as shown in \autoref{figure:partBBsight2} (the ground truth is 50$\%$ males).
Note that we use only 100 generated samples to achieve our partial black-box attack in \autoref{figure:partBBsight}, it is reasonable that the inference result of each target GAN has a relatively large deviation. 
Similar to our results and analysis for the full black-box attack (\autoref{section:evaluation_full_black-box}), our partial black-box attack produces good results when aiming to infer the underlying distribution in tasks $T_1$, $T_2$, $T_4$ and $T_5$, while $T_3$ is the most difficult one to achieve an accurate inference.  
Moreover, we can find some obvious gaps between the benchmark line and the average curve, such as $\property_{real}=40\%$ in $T_1$ and $\property_{real}=70\%$ in $T_3$.
They are possibly caused by a significant dissimilarity between the shadow datasets and the target datasets. 
In that case, the optimization phase within the partial black-box pipeline fails to construct an effective latent code set for the target models.

\mypara{Number of Optimized Samples} 
In \autoref{section:evaluation_full_black-box}, our one key finding is that the full-black box attack performance is influenced by the number of random samples. 
Here we also investigate how the number of optimized samples impacts the accuracy of our partial black-box attack, note that the adversary uses an optimized latent code set with the corresponding size to generate these optimized samples.
\autoref{figure:partBBnumL} presents the attack performance on each task while using different numbers of optimized samples, in other words, using latent code sets with different sizes.
Fixing the number of optimized samples, we can obtain the inference results for all 40 target models.
And for each number, we take advantage of box plots to perform all the attack results against the target models.
Besides, we also plot a red curve that highlights the average performance against the target models.
In general, we can see the partial black-box attack gets more precise when increasing the number of optimized samples.
For instance, in \autoref{figure:partBBnumL} when the number of optimized samples grows from 25 to 100, the average absolute difference of all target models changes from 12.5\% to 9.8\% in $T_1$, 8.7\% to 7.2\% in $T_2$, 14.9\% to 10.1\% in $T_3$, 7.7\% to 4.9\% in $T_4$ and 9.8\% to 4.5\% in $T_5$.
However, a larger number of optimized samples also tend to cause more serious instability (e.g., the variance of inference results), scaled here by the difference between the lower and upper quartile in the box plot.
As shown in \autoref{figure:partBBnumLEUC1}, when choosing the number of optimized samples as 200 in $T_1$, the average performance increases to 12.6\%, and the attack stability increases to around 15\%.
In our remaining experiments, after trading off the attack accuracy and the optimization time cost, we set the number of optimized samples as 100, i.e., the size of the optimized latent code set is equal to 100.

\begin{figure}[!t]
\centering
\includegraphics[width=0.67\columnwidth]{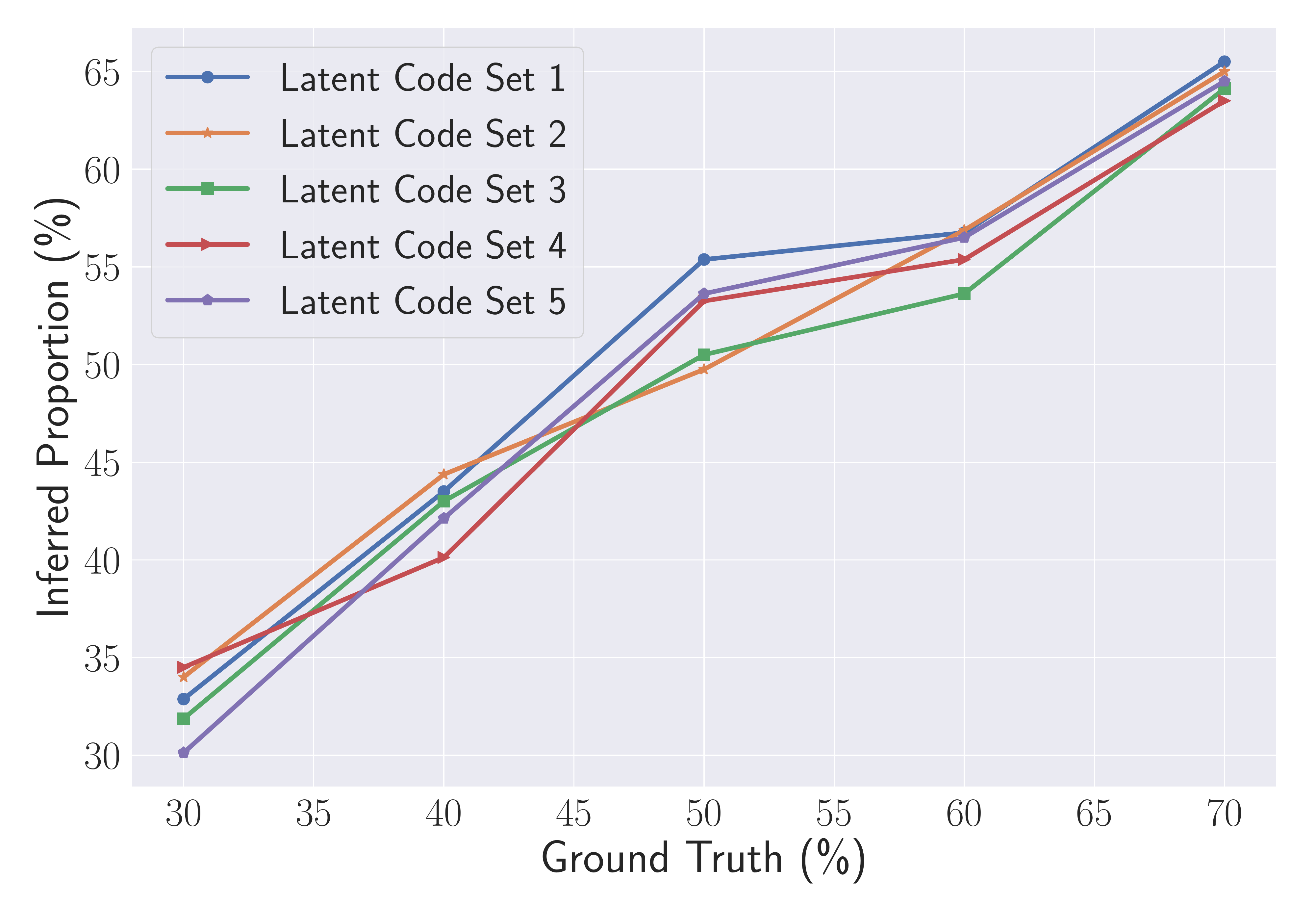}
\caption{Partial black-box performance w.r.t.\ optimization starting point.
Each line plots an average performance of the target models in $T_2$ with the same underlying property, based on 5 latent code sets optimized from different starting points.}
\label{figure:start_point}
\end{figure}

\mypara{Number of Shadow Models}
\autoref{figure:partBBnumS} shows the partial black-box attack performance on all target models against the number of local shadow models used to optimize the latent code set.
We leverage box plots to present the attack performance for all models' inference results, based on latent code sets optimized with different numbers of shadow models.
Moreover, we also plot a curve to present the average inference result against the number of shadow models.
As we can see, the average inference behavior changes slightly when using different numbers of shadow models.
For example, when increasing the number of shadow models from 50 to 100 in \autoref{figure:partBBnumSEUC3}, the average absolute difference of all target models changes slightly from 10.2\% to 10.8\% in $T_3$.
Even though the median and average attack accuracies are slightly higher when optimizing latent code set based on only 25 shadow models for most tasks, the variance of attack results is quite large, even more than 10\% in $T_3$ (scaled by the difference between the lower and upper quartile in a box plot).
It means the property inference attack suffers from severer instability when there are only a few shadow models.
Finally, considering both the accuracy and stability of the attack performance, we set the number of shadow models as 100 in our other experiments.
Moreover, our results indicate that our proposed partial black-box attack is able to achieve high inference accuracy with just a limited amount of shadow models, suggesting it is a practical and realistic threat in the real world.

\mypara{Optimization Starting Point}
Since we utilize the stochastic gradient descent (SGD) method to optimize a latent code set to implement our partial black-box attack, the optimized results may be affected by the optimization initialization, i.e., the optimization starting point.
\autoref{figure:start_point} shows the performance of our partial black-box attack on $T_2$ with five different random starting points.
Similar to \autoref{figure:fullBBsight} and \autoref{figure:partBBsight}, we exhibit the average performance for all target models.
As we can see, the five curves depicting the average inference results are close to each other.
For instance, in \autoref{figure:start_point} considering target models with 70\% males in the underlying training dataset, the average inference results range from 63.2\% to 65.2\% using five latent code sets based on different optimization starting points.
It suggests that the optimization starting point would not obviously affect our partial black-box attack performance.

\begin{figure}[!t]
\centering
\includegraphics[width=0.67\columnwidth]{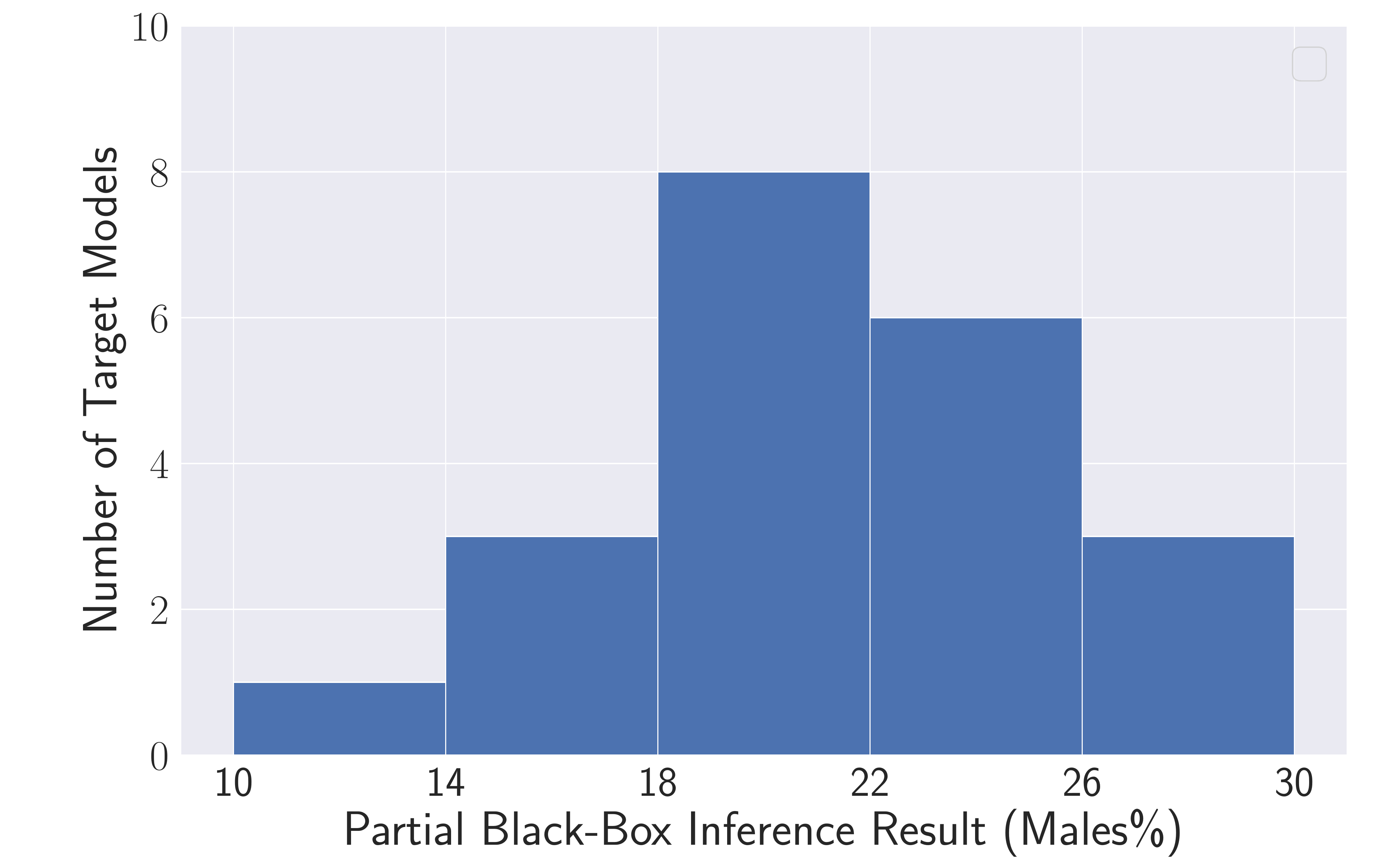}
\caption{Partial black-box performance w.r.t.\ out-of-range target property.
Each bar represents the number of target models with relative partial black-box attack results.
All 21 target GANs have 20\% males and 80\% females in the underlying dataset and follow the same setting as $T_4$.
The underlying property of the shadow models ranges from 30\% to 70\%.}
\label{figure:out_of_range}
\end{figure}

\mypara{Target Property Out of Range}
In the above experiment, we simply set the range of the property of the shadow models to cover the underlying property of the target models.
When facing an unknown distribution, the adversary can also expand the covering property range of the shadow models to 0\% to 100\%.
In this part, we examine our attack behavior when the target property is out of the range of the shadow models.
\autoref{figure:out_of_range} shows the partial black-box attack result of target GANs with 20\% males in the underlying dataset based on shadow models with 30\% to 70\% males.
As we can see in \autoref{figure:out_of_range}, there are 38\% (8/21) target GANs whose partial black-box inference error is lower than 2\%.
It indicates that our partial black-box attack still works when the target property is out of the range of the property of the shadow models.

Compared with the full black-box scenario, our proposed attack methodology on the partial black-box scenario shows more stable performance, as it is loosely related to the choice of several attack hyper-parameters, e.g., the optimization initialization and the number of shadow models.

% ======================================================
\subsection{Comparison Between Two Attacks}
% ======================================================

In this paper, we develop two property inference strategies against GANs on two practical attack scenarios: the full black-box setting and the partial black-box setting.
The key difference relies on whether the adversary is able to craft latent codes to control the target model’s output.
In the former case, we choose a more general and straightforward way by directly sending random generated latent codes to get random samples.
As it does not require any internal information of the target model (e.g., parameters, structures), this full black-box methodology is also applicable for the partial black-box adversary.
In the latter case, we propose to leverage auxiliary knowledge of the target model to help search for optimized latent codes via SGD, and then send them to the target model.
Our aforementioned experimental results show that the partial black-box adversary can achieve an accurate inference with a limited amount of latent codes.
In this subsection, we present a more comprehensive comparison between the two attacks based on their attack stability and accuracy.
As we can see in \autoref{figure:fullBBnum}, our full black-box attack can achieve a quite good attack performance based on over 256 random generated samples, 
so our comparison below will focus on using a relatively small amount of samples.

\begin{figure}[!t]
\centering
\includegraphics[width=0.66\columnwidth]{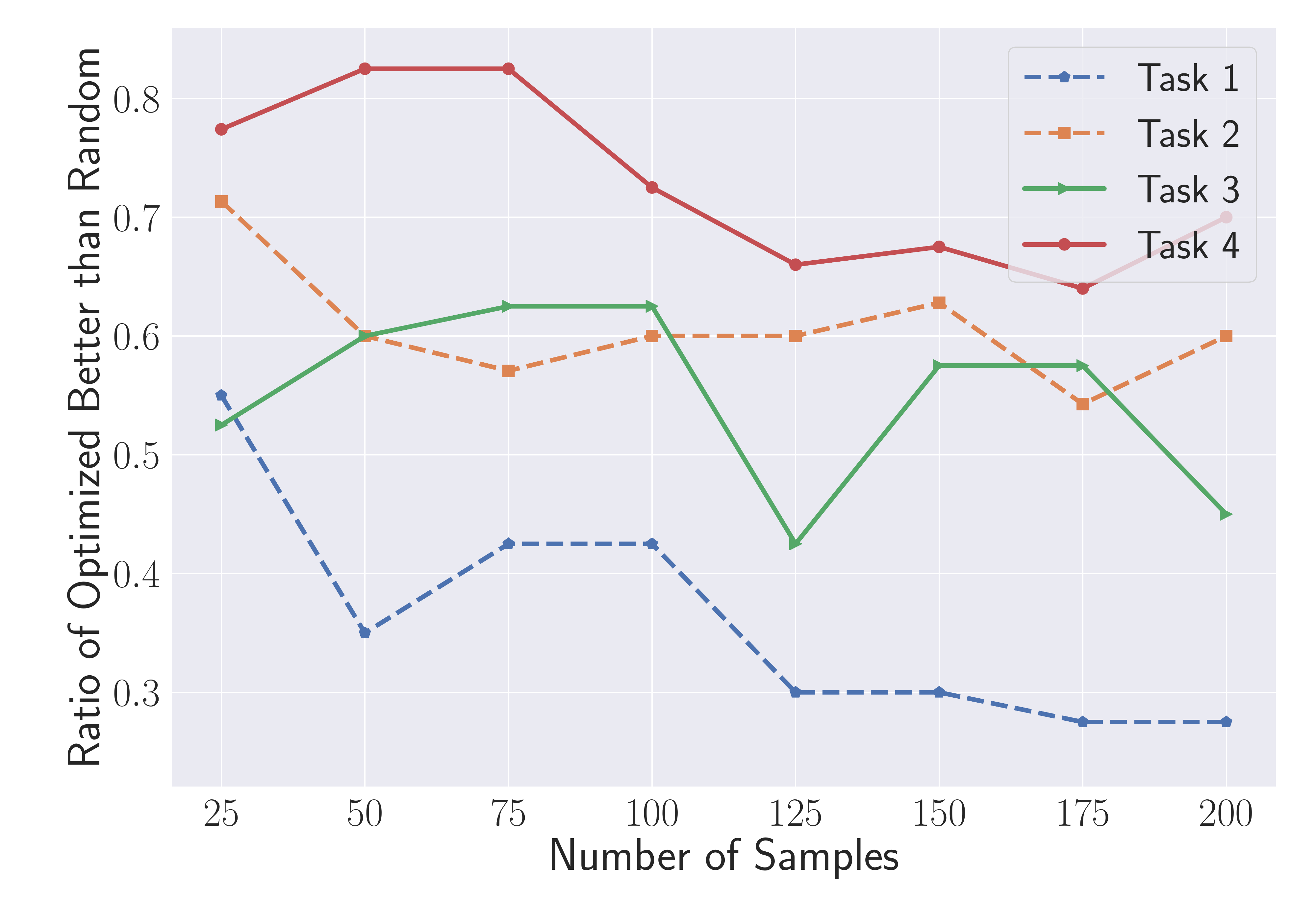}
\caption{Ratio of optimized samples behaving better than random ones.
For each task, we obtain 80 random inference results and one optimized inference result based on a limited number of generated samples for each target GAN.
Then we compare them to get the frequency of optimized samples behaving better than random ones and finally obtain the ratio considering all target models.
In this way, we show the ratio of partial black-box attack (optimized) behaving better than the full black-box attack (random) against different numbers of samples.}
\label{figure:comp2}
\end{figure}

\begin{figure*}[!t]
\centering
\begin{subfigure}{0.4\columnwidth}
\includegraphics[width=\columnwidth]{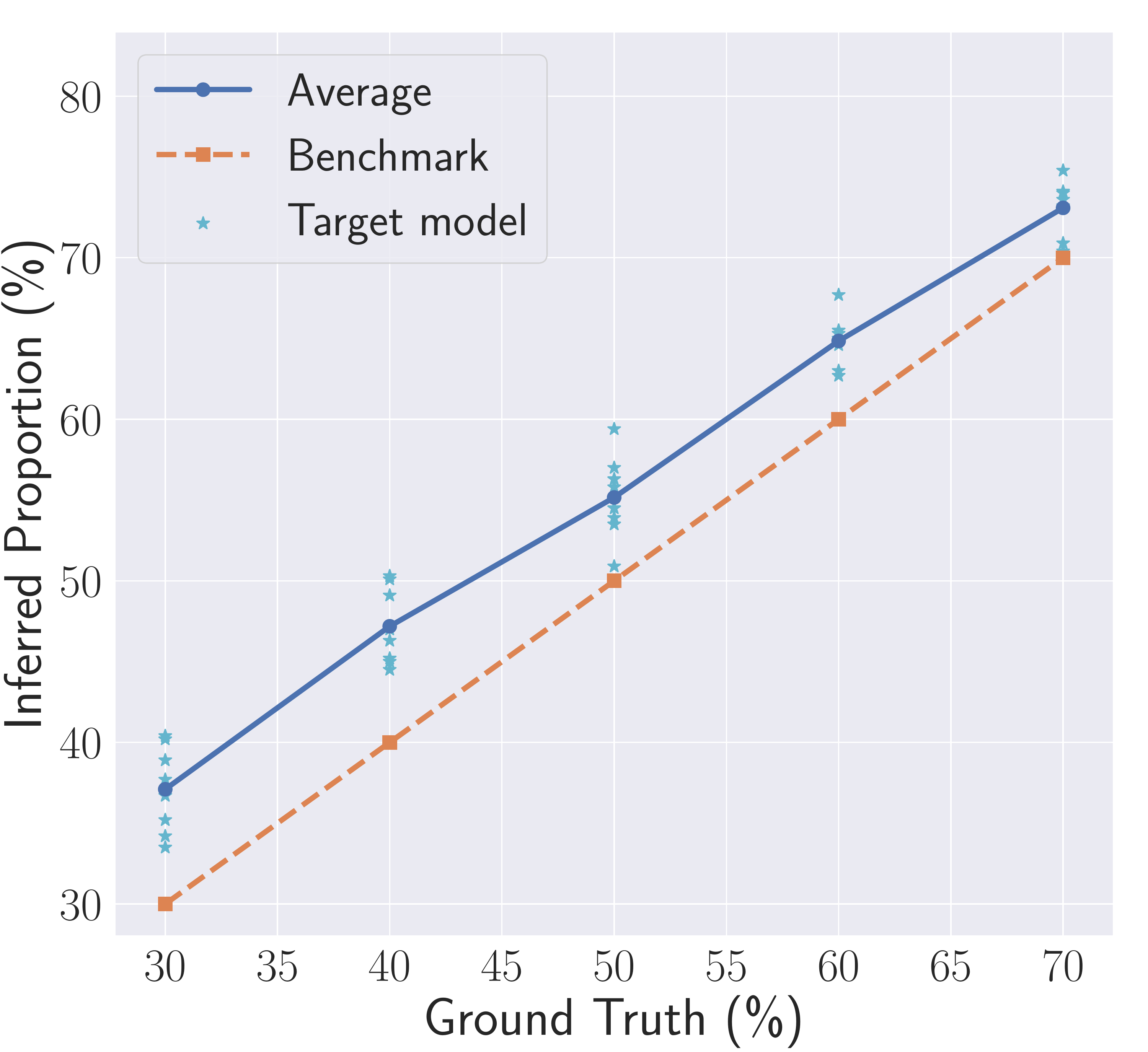}
\caption{Evaluation on $T_1$ with \\ IMDB-WIKI classifier}
\label{figure:online1}
\end{subfigure}
\begin{subfigure}{0.4\columnwidth}
\includegraphics[width=\columnwidth]{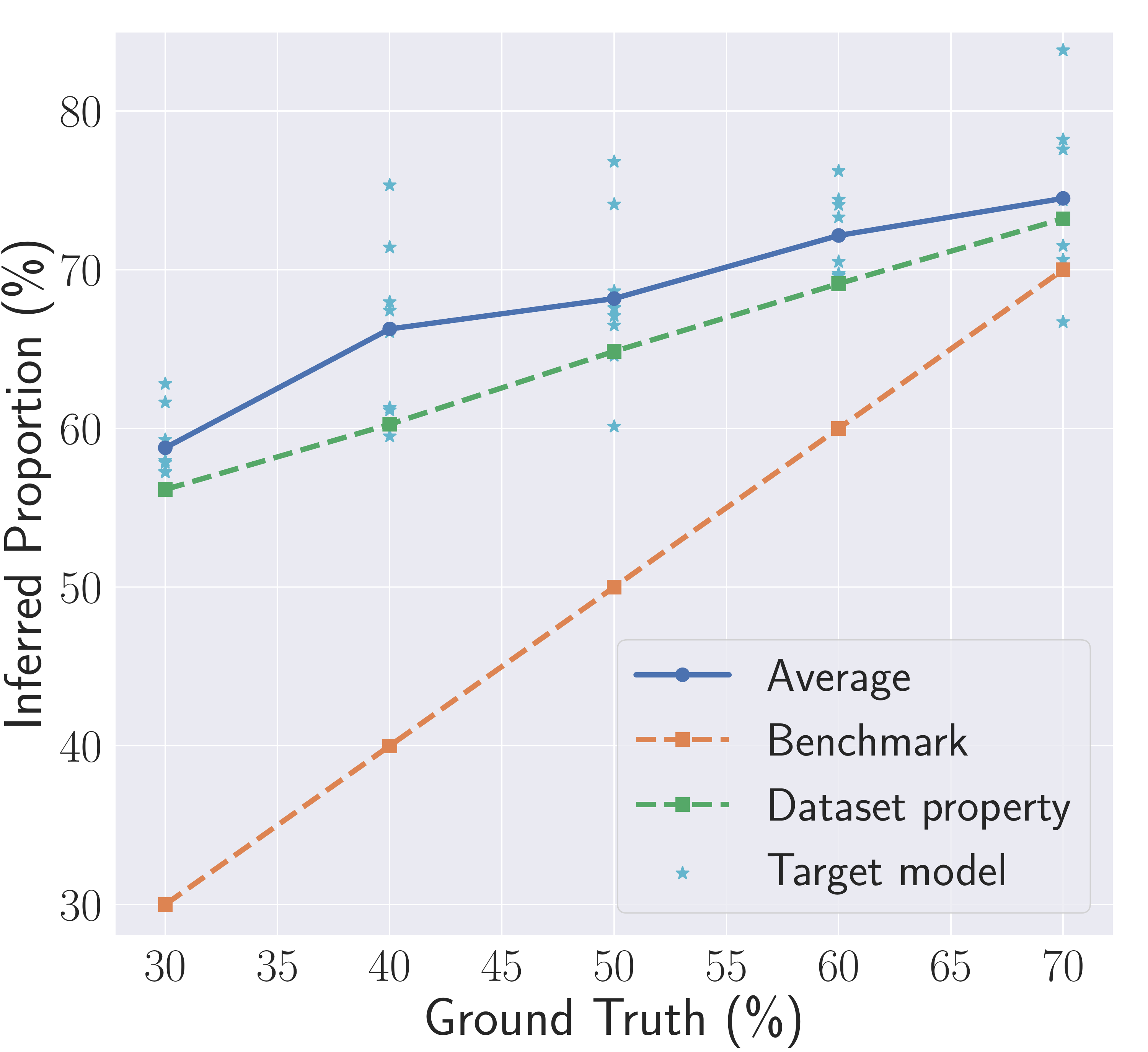}
\caption{Evaluation on $T_2$ with \\ IMDB-WIKI classifier}
\label{figure:online2}
\end{subfigure}
\begin{subfigure}{0.4\columnwidth}
\includegraphics[width=\columnwidth]{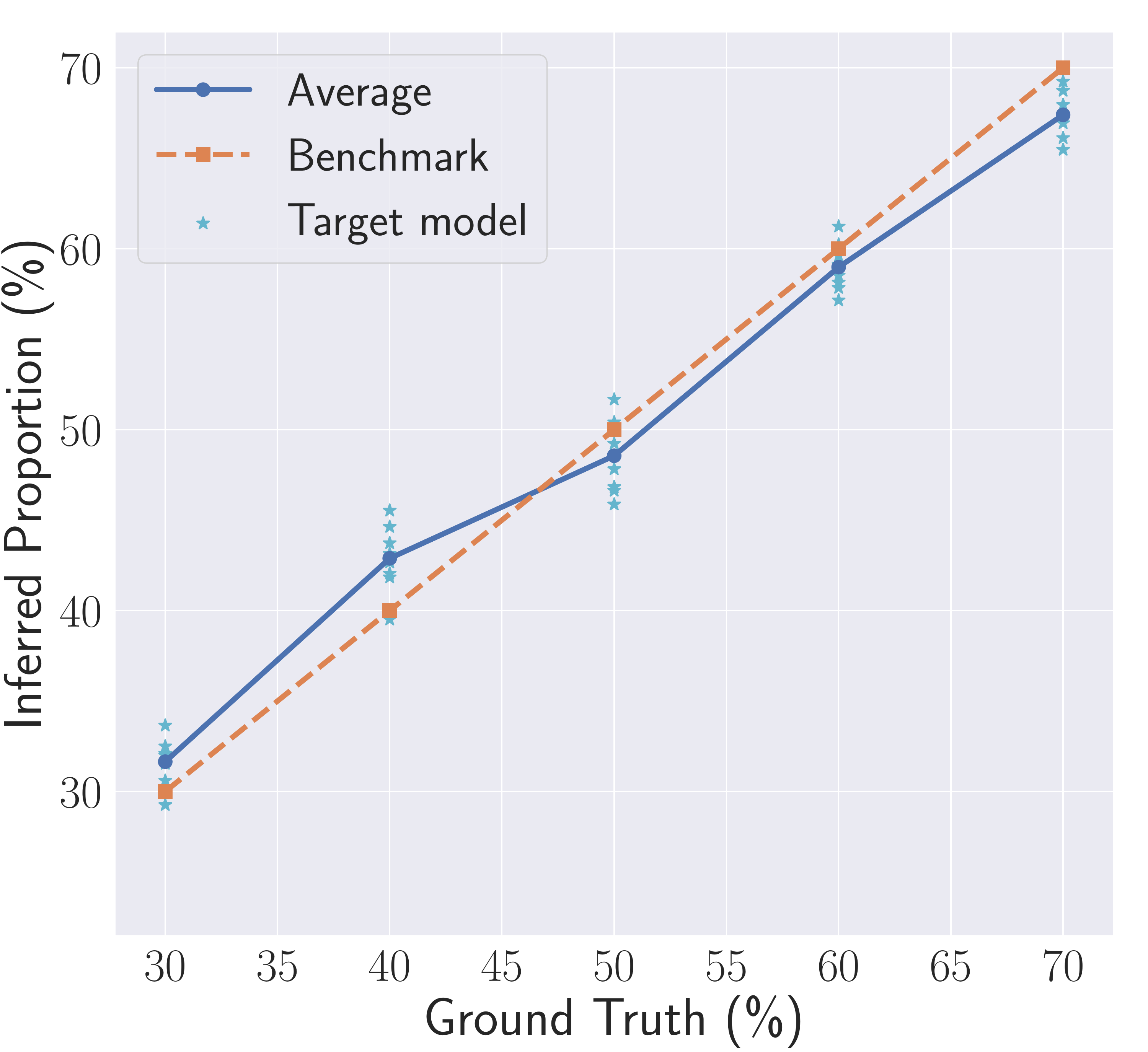}
\caption{Evaluation on $T_1$ with \\ Audience classifier}
\label{figure:online3}
\end{subfigure}
\begin{subfigure}{0.4\columnwidth}
\includegraphics[width=\columnwidth]{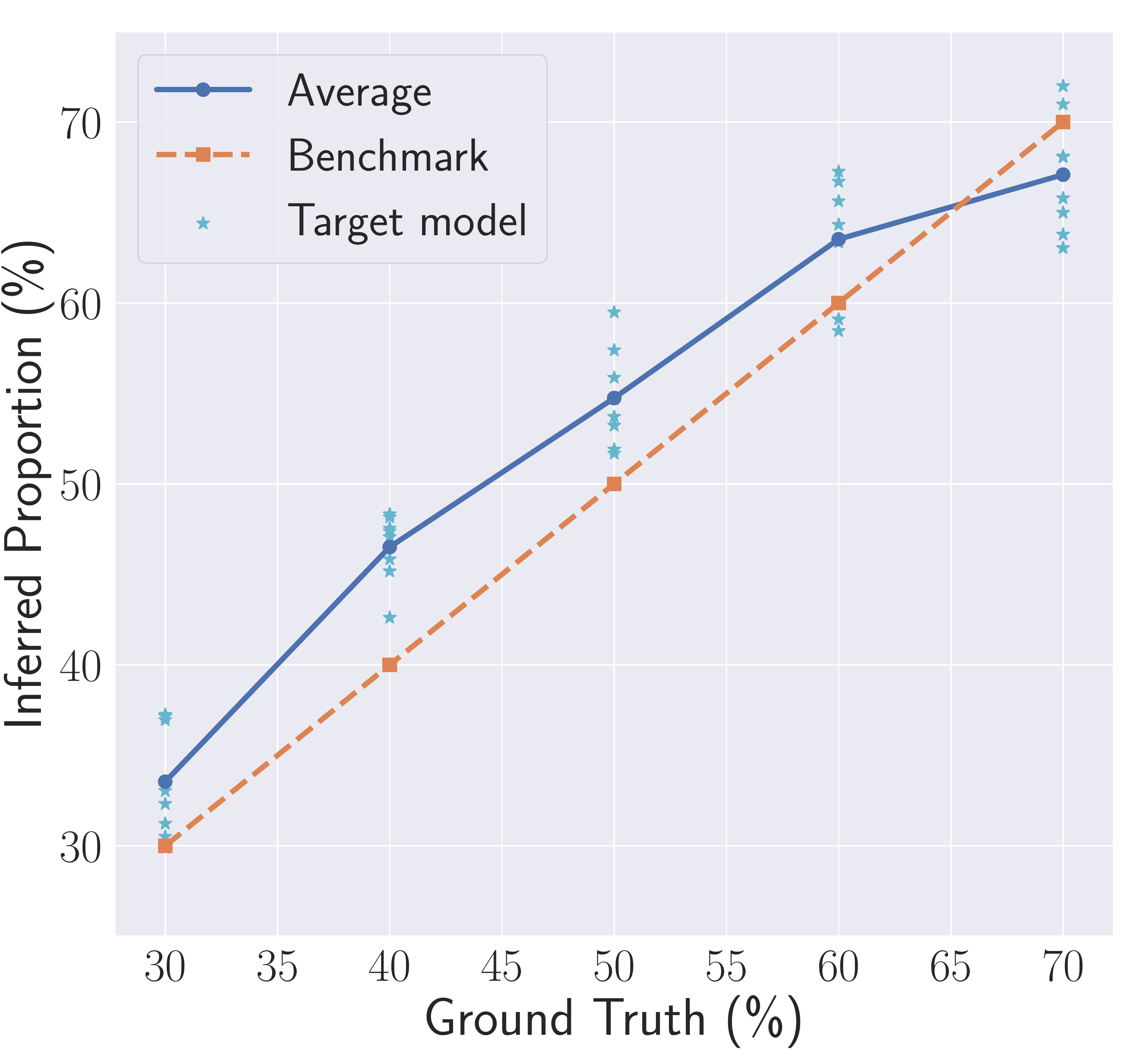}
\caption{Evaluation on $T_2$ with \\ Audience classifier}
\label{figure:online4}
\end{subfigure}
\begin{subfigure}{0.4\columnwidth}
\includegraphics[width=\columnwidth]{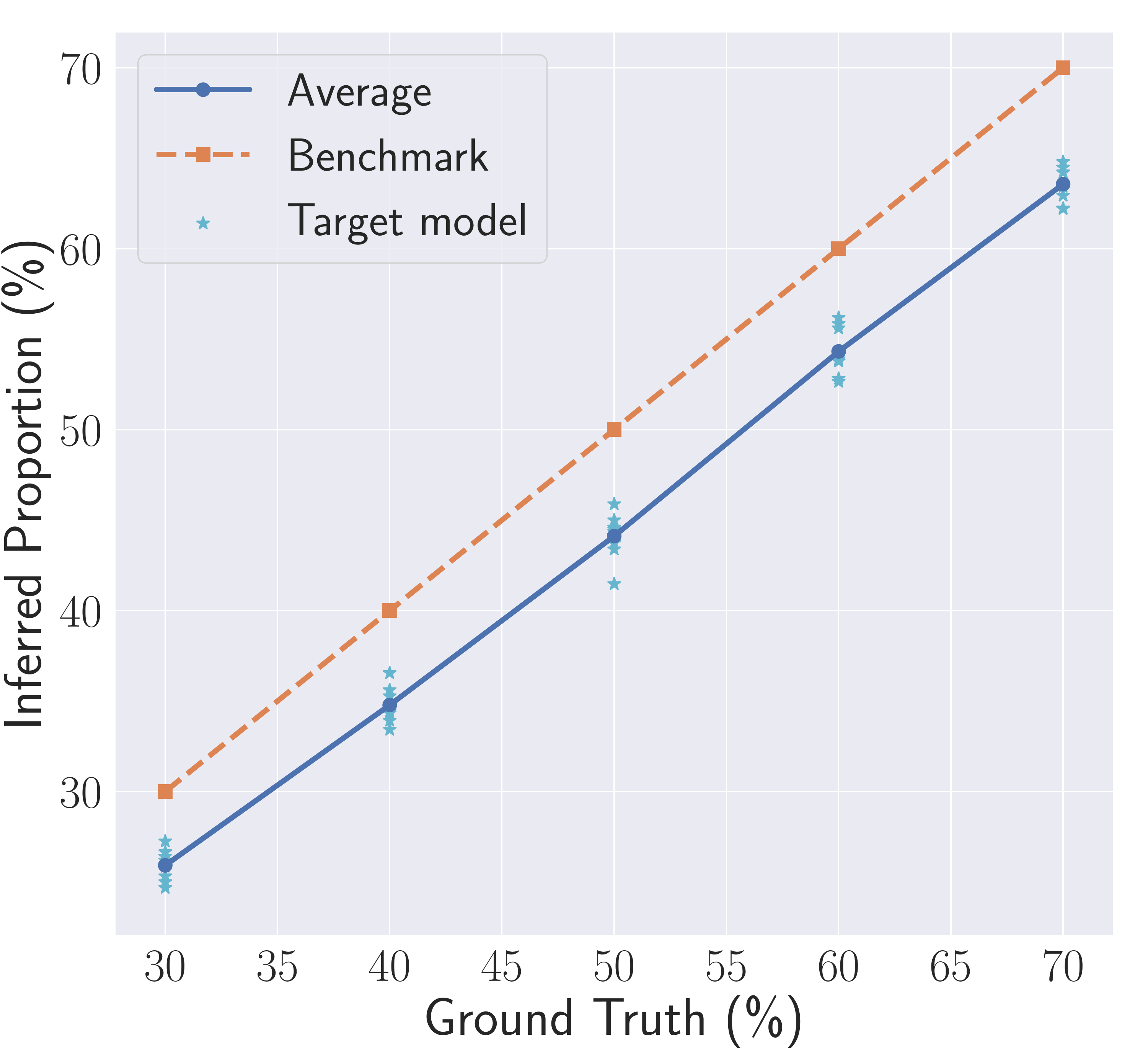}
\caption{Evaluation on $T_4$ with \\ EMNIST classifier}
\label{figure:online5}
\end{subfigure}
\caption{Full black-box performance w.r.t.\ unknown training distribution.
We adopt two off-the-shelf classifiers (based on IMDB-WIKI and Audience dataset) and a locally trained model (based on EMNIST dataset) to achieve our attacks.
Each point depicts the full black-box attack performance with corresponding underlying property and inferred property.
The blue curve plots the average performance of the target models with the same underlying property.
The orange line marks the ideal attack result.
The green line shows the dataset property reported by directly running the off-the-shelf classifier on the underlying training dataset.}
\label{figure:online}
\end{figure*}

We first give a comparison based on the \emph{attack stability} for the proposed two methodologies.
Based on the observation in \autoref{figure:fullBBnum}, we can find that the number of samples strongly affects the full black-box attack performance and the inference results are extremely unstable when using less than 256 samples.
On the other hand, the partial black-box attack methodology is not affected too much by the number of optimized samples, as the red curves in \autoref{figure:partBBnumL} are much smoother than curves in \autoref{figure:fullBBnum}.
For instance, the average absolute difference of our partial black-box attack result decreases from 8.2\% to 3.6\% in $T_4$ when the number of optimized generated samples increases from 25 to 200.
In this way, the partial black-box methodology produces a relatively stable performance when using a limited amount of samples (less than around 128 to 256).

We follow with a comparison of the \emph{attack accuracy} for the proposed two methodologies, based on the experiment established below.
For each target model and specific size of the latent code set, we perform the partial black-box attack once using the corresponding number of optimized samples.
In the meanwhile, because the full black-box attack presents a relatively unstable behavior, especially when the number of random samples is small, we repeat our full black-box attack 80 times to reduce observation randomness.
Then we compare each full black-box result with the corresponding partial black-box result.
Finally, for each tested number of samples, we respectively calculate the ratio of comparisons in which the optimized samples produce a more accurate inference with full consideration of all target models.

The statistical result is plotted in \autoref{figure:comp2}.
It shows that the partial black-box attack method is more likely to provide more accurate inferences than the full black-box attack when using a small number of samples (less than around 150).
For tasks except $T_1$, we can find that inferences with optimized samples are better than those with random samples in most cases (i.e., the ratio is exceeding 0.5).
Another observation is that, as the number of samples increases, more full black-box inferences outperform the partial black-box attack results. 

Overall, we give a conclusion for these two methodologies.

\mypara{Partial Black-box Methodology}
Our partial black-box attack achieves a better inference performance in both accuracy and stability with a limited number of optimized samples (around 150).
Note that obtaining a large number of samples from the target GAN can be possibly detected as an abnormal event, our partial black-box attack is supposed to be a more stealthy one.
Moreover, the partial black-box attack helps to reduce query charges when the adversary needs to pay for the generated samples.

\mypara{Full Black-box Methodology}
When the adversary is allowed to obtain a large number of samples, our full black-box methodology provides a more convenient way to achieve her attack, as it avoids the consumption to optimize the latent code set.
Besides, we believe that our full black-box methodology has also presented realistic threats against generative models, as it provides a more generic and easy solution without any extra knowledge of the target model. 

% ======================================================
\subsection{Evaluation on Property Classifier}
\label{section:property_classifier}
% ======================================================

Both of our full and partial black-box attack pipelines include a property classifier $\classifier$, which directly impacts the final inference result.
In this subsection, we study how the property classifier influences the property inference attack with respect to two factors: the training dataset distribution and the structure of the property classifier.  

\mypara{Training Dataset from a Different Distribution}
We firstly consider a strict situation that the adversary has no knowledge of the target model’s training dataset distribution. 
In that case, we investigate how the full black-box attack behaves when the property classifier is trained on a different dataset.
In our experiment, we adopt three property classifiers, i.e., an off-the-shelf CNN model\footnote{\url{https://data.vision.ee.ethz.ch/cvl/rrothe/imdb-wiki/}} trained on the IMDB-WIKI dataset~\cite{RTG18}, an off-the-shelf CNN model\footnote{\url{https://github.com/dpressel/rude-carnie}} trained on the Audience dataset~\cite{LH15}, and a model locally trained with the EMNIST dataset~\cite{CATS17}.
As the former two off-the-shelf models are gender classifications, we directly adopt them to achieve the full black-box attack on $T_1$ and $T_2$.
\autoref{figure:online3} and \autoref{figure:online4} shows that the property classifier based on Audience dataset has a good attack performance on both tasks.
But as shown in \autoref{figure:online2}, the IMDB-WIKI classifier results in a significant inference accuracy decline in $T_2$.
This phenomenon is possibly due to the relatively poor performance of the IMDB-WIKI classifier on the AFAD dataset.
In order to verify our guess, we run the off-the-shelf property classifier directly on the target model’s underlying training dataset and then calculate the dataset property.
The green curve in \autoref{figure:online2} depicts the result.
We can clearly see that the average attack results correspond closely to the dataset property recognized by the property classifier.
For instance, when the proportion of male training samples is 30\%, the property classifier reports a male proportion of 56\%, which is close to the average inferred proportion of 59\%. 
This phenomenon reveals that the property classifier plays a pivotal role in the inference attack.
Moreover, we also train a local classifier with the EMNIST dataset~\cite{CATS17} to achieve the full black-box attack on $T_4$.
The EMNIST dataset is a variant of the full NIST dataset and shares the same image structure and parameters as the original MNIST dataset, but it is completely disjoint with the MNIST dataset.
\autoref{figure:online5} shows that the EMNIST classifier can still help us to accomplish a relatively accurate property inference attack.
Our result shows that it is possible to achieve an accurate property inference attack even without the knowledge of the training dataset distribution, as long as the adversary owns a property classifier that has good enough accuracy on the target problem.

\begin{figure}[!t]
\centering
\includegraphics[width=0.6\columnwidth]{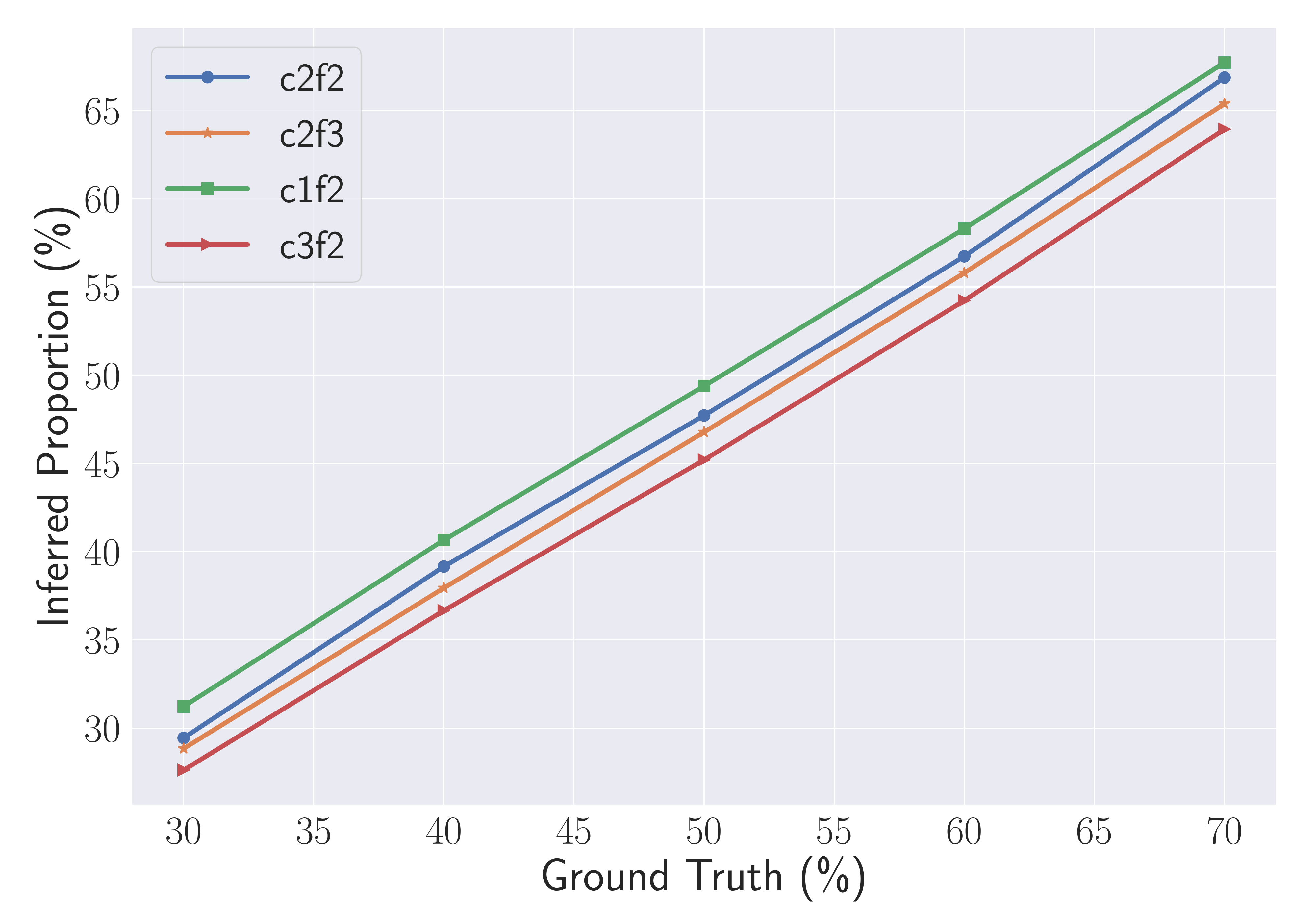}
\caption{Full black-box performance w.r.t.\ property classifier architecture.
Each line gives an average performance of target models with the same underlying property.
``c3f2'' means the classifier architecture begins with 3 convolution layers and follows with 2 fully connected layers.}
\label{figure:structure}
\end{figure}

\begin{figure}[!t]
\centering
\begin{subfigure}{0.49\columnwidth}
\includegraphics[width=\columnwidth]{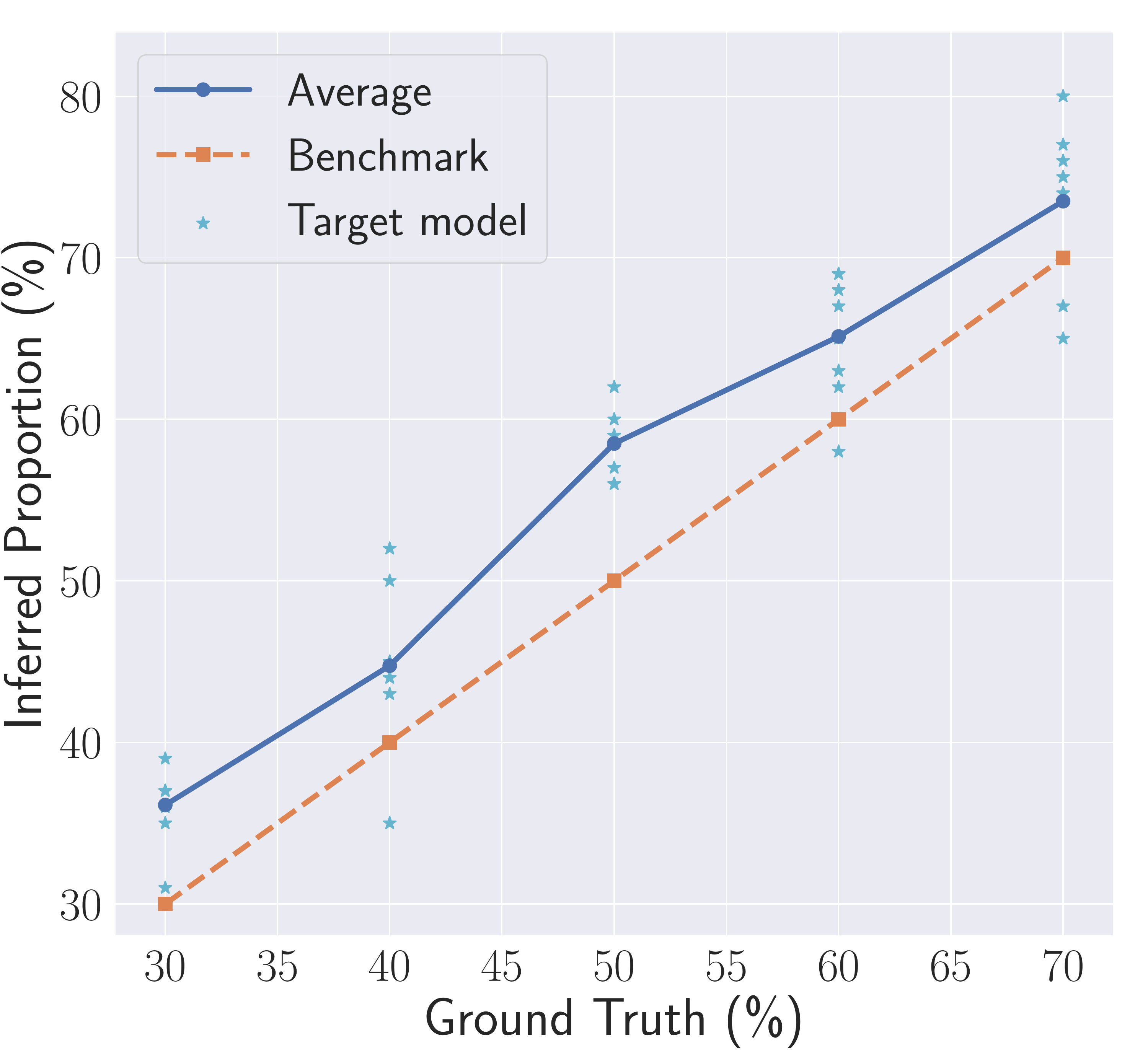}
\caption{Evaluate on $T_1$}
\label{figure:shadow1}
\end{subfigure}
\begin{subfigure}{0.49\columnwidth}
\includegraphics[width=\columnwidth]{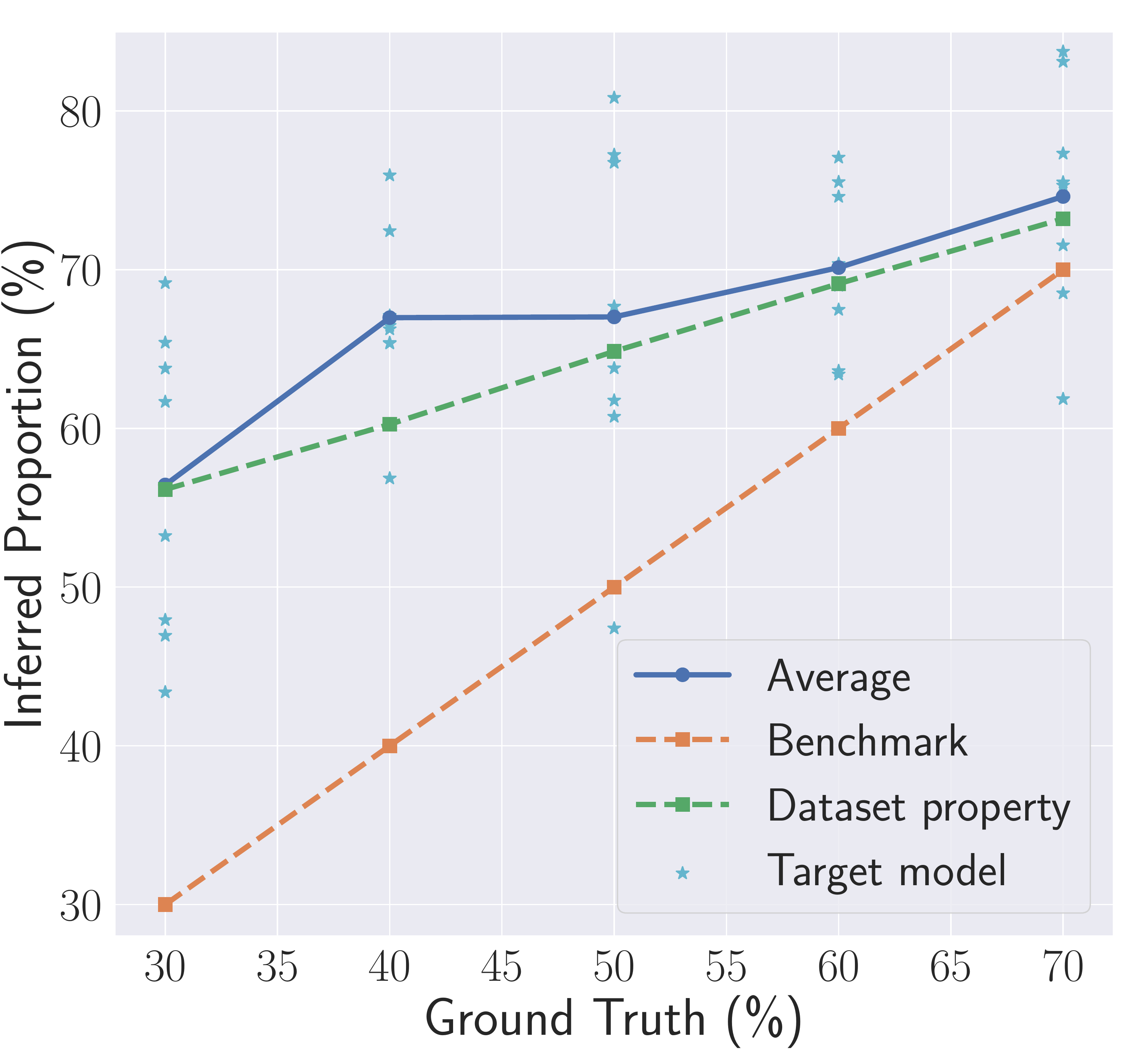}
\caption{Evaluate on $T_2$}
\label{figure:shadow2}
\end{subfigure}
\caption{Partial black-box performance w.r.t.\ unknown training distribution and structure of target GAN.
For $T_1$, we optimize the latent code set based on PGGAN with IMDB-WIKI dataset.
For $T_2$, we optimize the latent code set based on WGANGP with CelebA dataset.
For both tasks, we adopt a downloaded CNN model based on IMDB-WIKI dataset as our property classifier.
The blue curve plots the average performance of the target models with the same underlying property.
The orange line marks the best attack result.
The green line shows the training dataset property directly reported by the off-the-shelf classifier.}
\label{figure:shadow}
\end{figure}

\mypara{Classifier Architecture}
We secondly examine the attack behavior based on property classifiers with different architectures.
In this paper, our property classifier in $T_1$ is a convolutional neural network with two convolutional layers and two fully connected layers (shortened as c2f2 in this paper).
Additionally, we adopt three other structures which have different numbers of convolution layers or fully connected layers (i.e., c2f3, c1f2, and c3f2).
As shown in \autoref{figure:structure}, the average full black-box attack results in $T_1$ are very close to each other when using property classifiers with different structures.
For instance, the average inference results range from 26\% to 31\% when the underlying training dataset contains 30\% males.
As a result, our attack methodologies are only sightly influenced by the classifier architecture.

% ======================================================
\subsection{Evaluation on Shadow Models with Less Hyper-parameters}
\label{section:Shadow}
% ======================================================

As we assume that the partial black-box adversary can train shadow models with the same structure and training hyper-parameters on the auxiliary dataset of the target GAN.
In this subsection, we investigate the behavior of our partial black-box attack based on shadow models with different structure and training dataset. 

As we still control the input latent code of the target GAN in the partial black-box methodology, we only set the size of the latent code layer in shadow models the same as the target model. 
For $T_1$, we use the PGGAN trained on the IMDB-WIKI dataset as shadow models, while the target model is WGANGP based on the CelebA dataset.
\autoref{figure:shadow1} shows the partial black-box attack result is close to the benchmark line.
As a result, the latent code set optimized with shadow models still works well in spite of that the adversary has no knowledge of the main structure and dataset of the target model.
For $T_2$, we use the WGANGP trained on the CelebA dataset as shadow models.
As shown in \autoref{figure:shadow2}, there is a certain deviation between the partial black-box attack result and the benchmark line.
But the inference result corresponds closely to the dataset property recognized by the off-the-shelf property classifier.
Moreover, the similarity between \autoref{figure:shadow} and \autoref{figure:online} proves the effectiveness of our partial black-box methodology without knowledge of the structure and underlying dataset of the target model.
And this phenomenon further certifies the needless similarity of the generated samples in shadow and target GANs to achieve our partial black-box methodology.

% ======================================================
\subsection{Evaluation on Multi-class Property}
\label{section:Multi}
% ======================================================

\begin{figure}[!t]
\centering
\includegraphics[width=0.75\columnwidth]{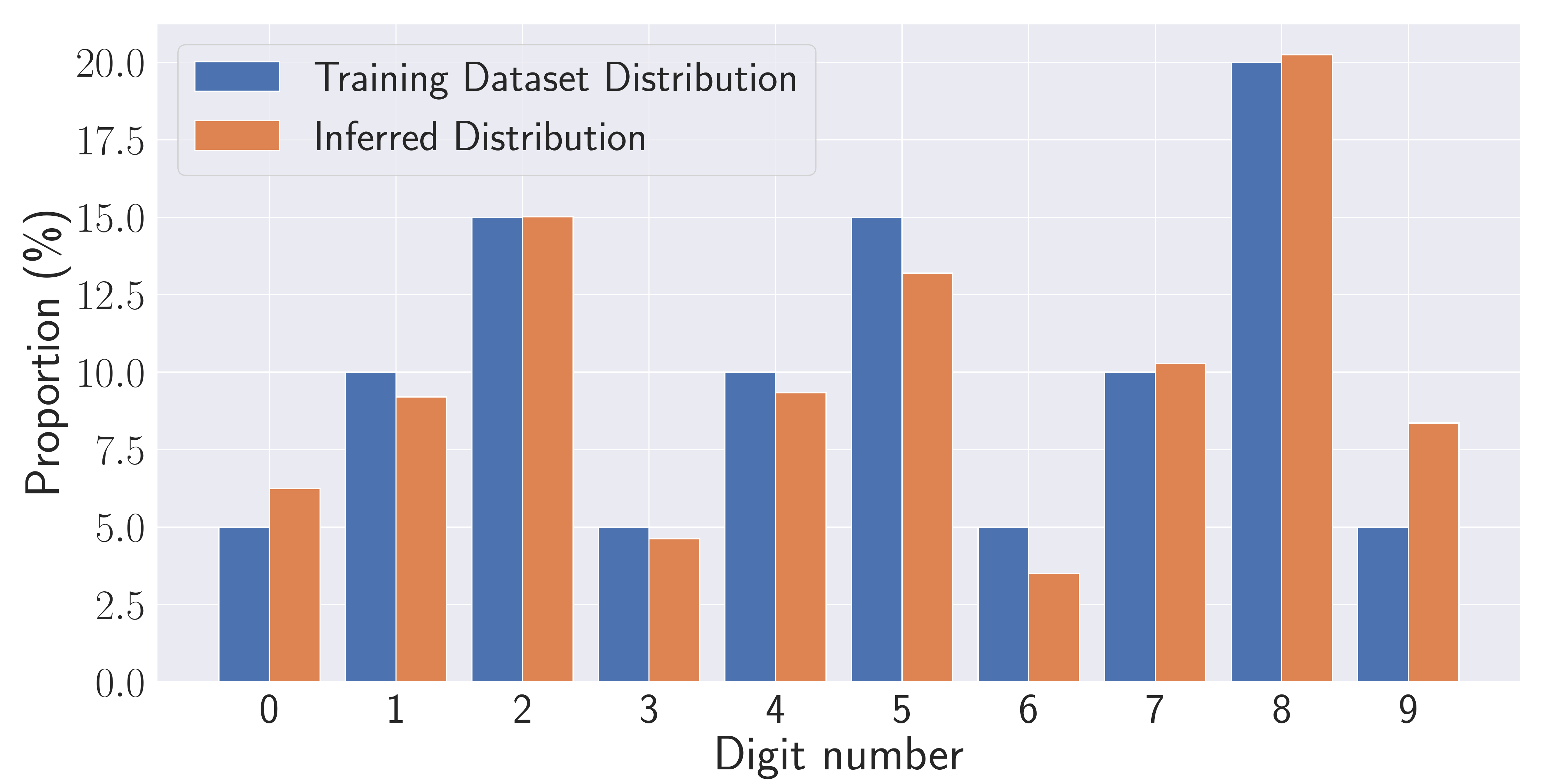}
\caption{Full black-box performance w.r.t.\ multi-class property.
The blue bar shows the distribution of each digit in the training dataset, while the orange one depicts the inferred distribution based on our full black-box attack.}
\label{figure:mnist_ori}
\end{figure}

As the five tasks shown above all focus on inferring the distribution of attributes with \emph{binary classes}, we present here an extra experiment to evaluate the performance of our attack on the property with \emph{multiple classes}, i.e., the distribution of 10 digits in the training dataset.
Facing a target DCGAN trained on MNIST with a specific distribution (including digits from 0 to 9), \autoref{figure:mnist_ori} shows how our full black-box attack behaves when inferring this multi-class property.
We can clearly observe that the inferred distribution follows a close trend with the underlying property.
Moreover, the cosine similarity is over 0.99 when considering these two distributions as vectors.
As a result, our attack gives a good performance on inferring property with not only binary classes, but also multiple classes.

\begin{figure}[!t]
\centering
\includegraphics[width=0.75\columnwidth]{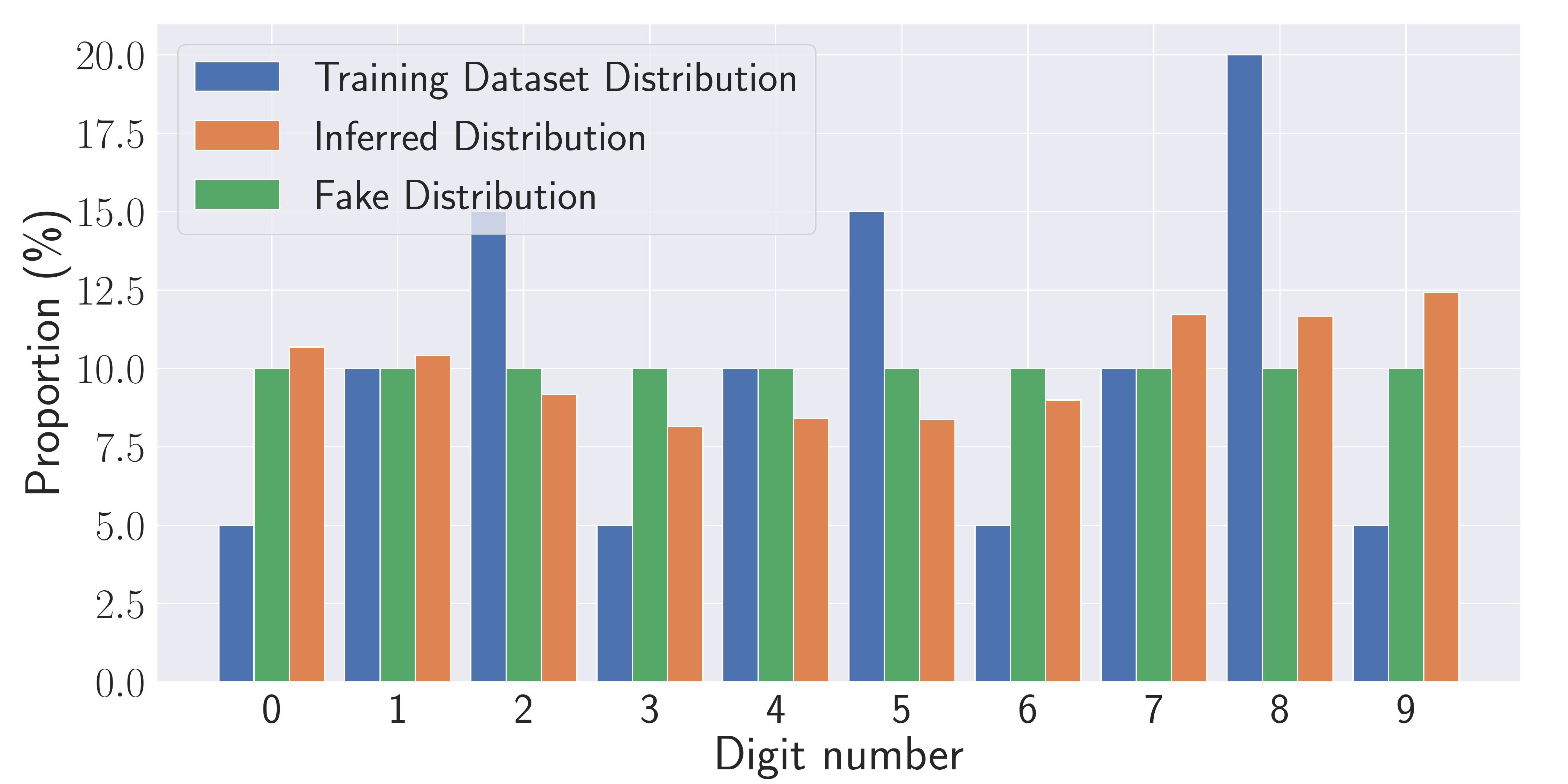}
\caption{Mitigation performance.
The blue bar shows the original distribution of each digit in the training dataset, while the orange one depicts the inferred distribution based on our full black-box attack.
The green bar shows the fake distribution after rebalancing dataset.}
\label{figure:mnist_fake}
\end{figure}

% ======================================================
\subsection{Discussion on Mitigation}
\label{section:mitigate}
% ======================================================

\mypara{Local Property Classifier}
One way to mitigate our attack is by introducing a local property classifier to pre-test the property of generated samples.
In this way, even though the real generated samples expose the underlying property of the training dataset, we can provide a subset of the generated samples to hide this distribution with the help of the local property classifier.

\mypara{Rebalancing the Training Dataset}
Another solution would be simply rebalancing the training dataset with respect to the target property by adding extra new samples to the training dataset.
\autoref{figure:mnist_fake} shows the full black-box attack performance after filling the training dataset to form a fake distribution.
We can find that the inferred property is closer to the fake distribution than the real one, and the cosine similarity reduces from 0.99 to 0.88.
This method is useful in most scenarios, but can possibly jeopardize the performance of the target GAN.

% ======================================================
\section{Enhancing Membership Inference Attack}
\label{section:MIA}
% ======================================================

So far, we have demonstrated the effectiveness of our property inference attacks against GANs.
Next, we investigate whether our property inference attacks can be used as a building block to launch other attacks.
In particular, we focus on membership inference, one of the most well-established privacy attacks against GANs~\cite{HMDC19,CYZF20}.

\mypara{Methodology}
In general, the membership inference attack intends to infer whether a target sample belongs to the underlying training dataset of a target GAN.
State-of-the-art attacks in this field proposed by Chen et al.~\cite{CYZF20} follow three steps:
\begin{itemize}
\item Use a distance metric $L(\cdot,\cdot)$ to evaluate the reconstruction error of the target sample $x$ against the target GAN $(\generator_v)$. In different scenarios, they deliver different methodologies to obtain the most similar generated sample $\mathcal{R}(x|\generator_v)$.
\item Build a shadow GAN $(\generator_r)$ to repeat the first step and get a reference reconstruction error. In this way, the calibrated reconstruction error $L_{cal}(\cdot,\cdot)$ can be calculated as:
\begin{equation}
L_{cal}(x,\mathcal{R}(x|\generator_v)) = L(x,\mathcal{R}(x|\generator_v)) - L(x,\mathcal{R}(x|\generator_r))
\end{equation}
\item Infer whether the target sample is in the training dataset based on a threshold. Formally, the attack flow works as 
\begin{equation}
\label{equation:MIA}
\mathcal{A}(x) = \mathbbm{1}[L_{cal}(x,\mathcal{R}(x|\generator_v)) < \epsilon]
\end{equation}
i.e., when the calibrated reconstruction error is smaller than a threshold, it belongs to the training dataset.
\end{itemize}

Our enhancement follows the intuition that a sample has a larger possibility to be a member when it shares the same property with the majority of samples in the target GAN's training dataset.
For instance, if the target GAN's training dataset contains more males than females (obtained by our property inference attacks), then a target male sample is more likely to be a member than a female sample.

Based on this, we add an extra item on the threshold of \autoref{equation:MIA} to enhance membership inference. 
Formally, the new membership inference attack is modified as the following.
\begin{equation}
\label{equation:enhanced_MIA}
\mathcal{A}(x) = \mathbbm{1}[L_{cal}(x,\mathcal{R}(x|\generator_v)) < \epsilon + \lambda_{p}\frac{1}{N}\sum_{i}^N \sl{f}(\property_{\it{i}})]
\end{equation}
where $\lambda_{p}$ controls the magnitude of our enhancement, $N$ refers to the number of considered attributes of the query sample, $\property_{\it{i}}$ is the proportion of the $i$th attribute in the training dataset, and $\sl{f}({\property_{\it{i}}}) = 2 {\times} {\property_{\it{i}}} - 1$.
The term $\lambda_{p}\frac{1}{N}\sum_{i}^N \sl{f}(\property_{\it{i}})$ in \autoref{equation:enhanced_MIA} helps to calibrate a target sample's membership probability with respect to the target model's training dataset's property.
When the target sample shares the same attribute as a higher proportion of the underlying training dataset ($\property_{\it{i}}>50\%$), the new threshold rises and leads to a better membership probability.

\mypara{Evaluation}
We evaluate the performance of the enhanced membership inference attack with the help of the additional knowledge of the gender distribution of samples in the training dataset, which is obtained by our property inference attack.
We set up the target GANs using 2,048 CelebA samples with a control of its underlying property (the proportion of males) and the structure is the same as $T_1$ discussed in \ref{section:target_models}.
We adopt the same full black-box membership inference attack methodology as Chen et al.~\cite{CYZF20}.
We set $\lambda_{p} = 2$ and each evaluation dataset has 2,048 members and 6,144 non-members (8,192 in total).

\begin{figure}[!t]
\centering
\includegraphics[width=0.66\columnwidth]{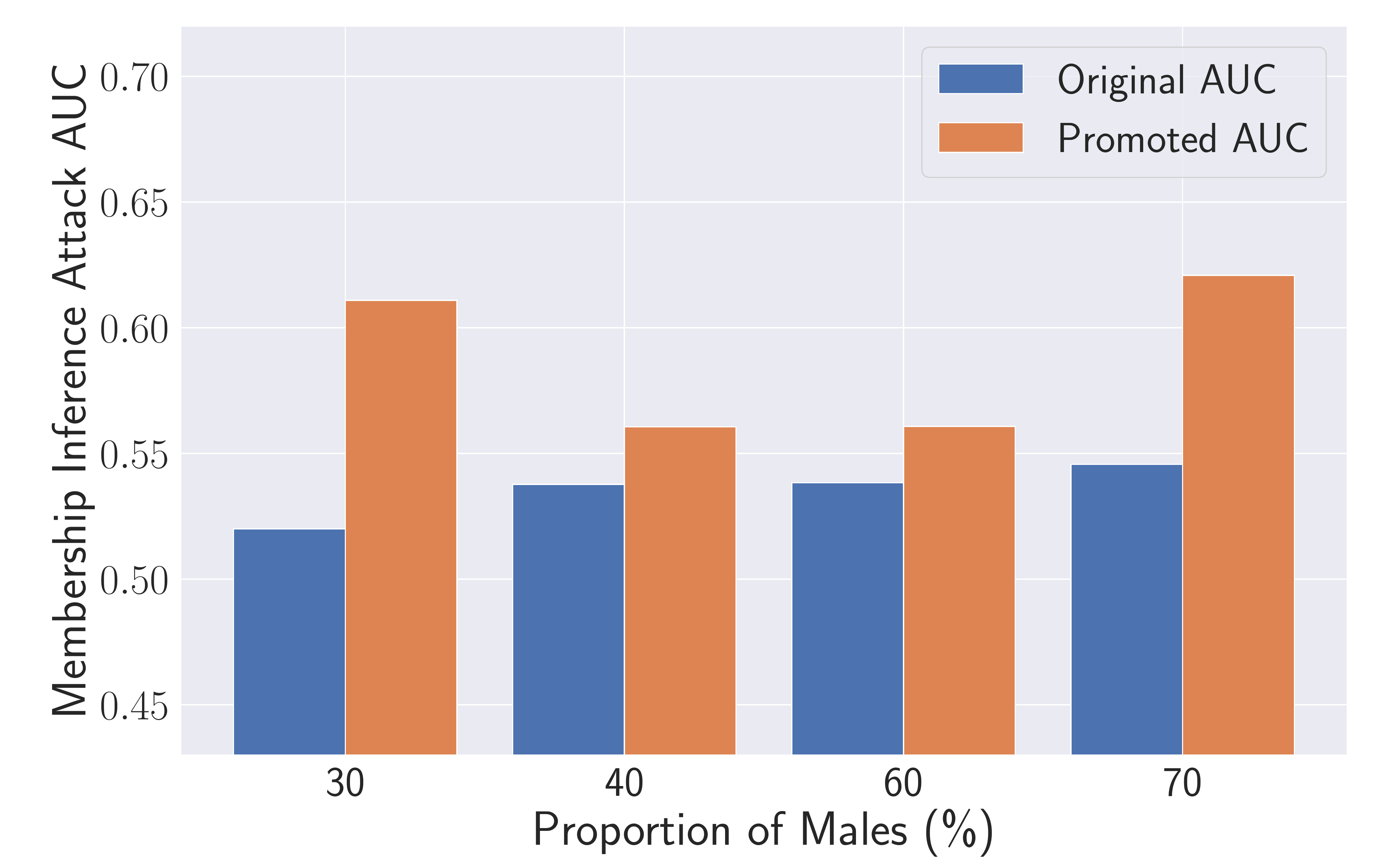}
\caption{Enhanced membership inference attack performance.}
\label{figure:MIA}
\end{figure}

As shown in \autoref{figure:MIA}, with the knowledge of the training dataset's underlying property, i.e., 30\% male samples, our enhanced membership inference's AUC (area under the ROC curve) increases from 0.52 to 0.61.
Furthermore, when the distribution of gender is more polarized, the enhancement is more pronounced.
As a result, the extra item added to the threshold in \autoref{equation:enhanced_MIA} indeed calibrates the membership probability effectively, which further demonstrates the applicability of our property inference attacks.

\begin{figure}[!t]
\centering
\includegraphics[width=0.66\columnwidth]{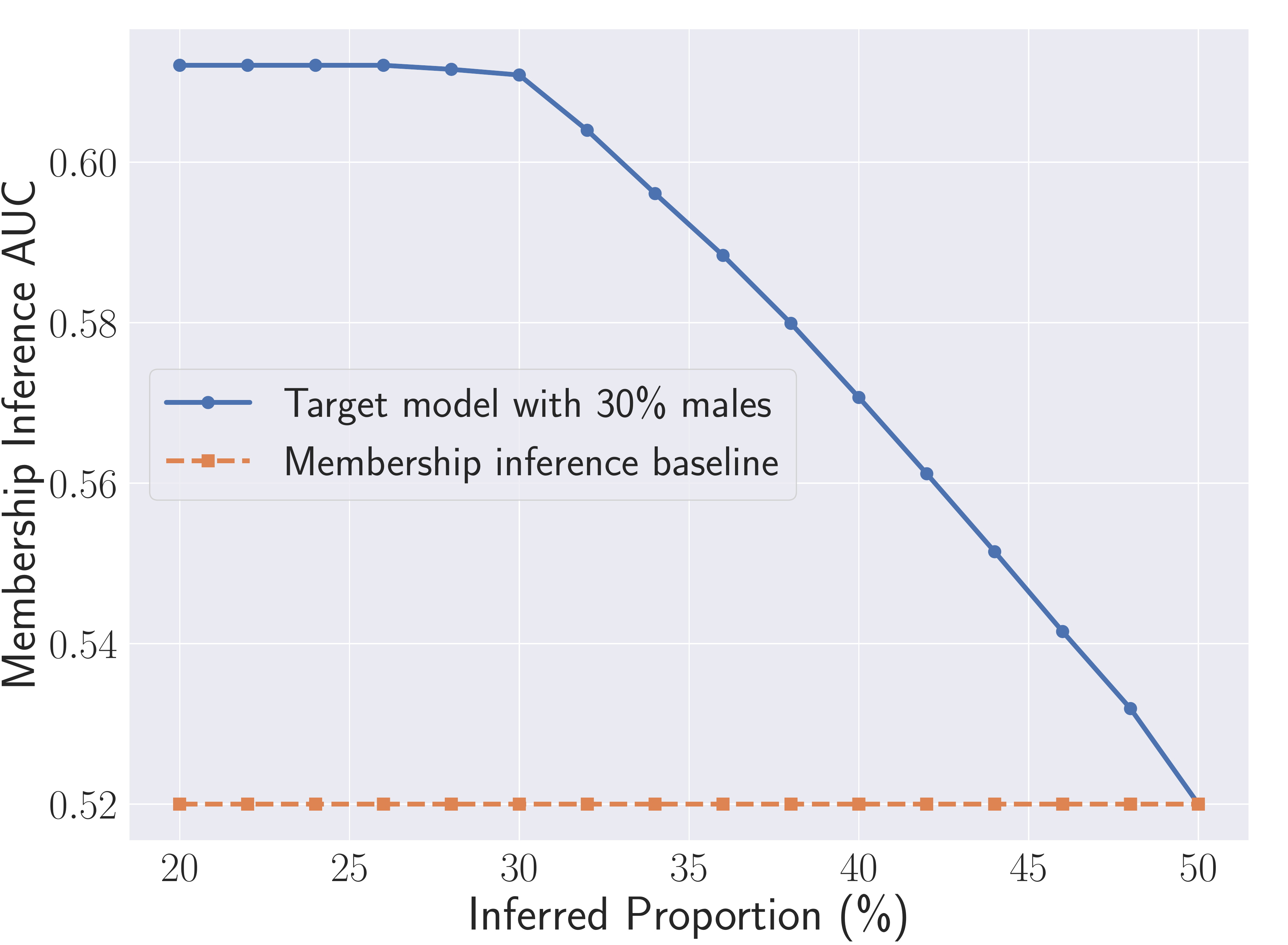}
\caption{Enhanced membership inference attack performance w.r.t inferred property with deviations.}
\label{figure:inaccurate_MIA}
\end{figure}

\mypara{Impact of Inferred Properties}
As our property inference can not deliver the exact proportion of the considered attribute in the target training dataset, we further evaluate our enhancement algorithm based on inferred proportions with deviations.
As we can see in \autoref{figure:inaccurate_MIA}, our enhancement still works on a target GAN with 30\% males when the utilized inference property is less than 50\%.
Moreover, the enhanced membership inference's AUC changes slightly when our property inference attack delivers a proportion less than the underlying property, but the AUC decreases significantly to the baseline when the inferred proportion comes closer to 50\%.
This further illustrates the applicability of our membership inference attack enhancement algorithm based on the property inference attack.

% ======================================================
\section{Related Work}
\label{section:related}
% ======================================================

\mypara{Membership Inference Attacks}
Membership inference attack tries to infer whether a sample belongs to a specific dataset.
Previous studies have demonstrated successful attacks against various targets, such as biomedical data~\cite{HSRDTMPSNC08,BBHM16,HZHBTWB19} and location data~\cite{PTC18}.

Shokri et al.~\cite{SSSS17} introduce the first membership inference attack against machine learning models.
The key idea is to leverage a set of shadow models to mimic the target model's behavior, and then train an attack model to discriminate member from non-member samples on model outputs.
Salem et al.~\cite{SZHBFB19} show that the membership inference attack can attain high accuracy even relaxing the three assumptions in~\cite{SSSS17}.
In recent years, membership inference attacks have been investigated on various other scenarios, e.g., white-box models~\cite{NSH19,LF20}, federated learning~\cite{MSCS19}, generative models~\cite{HMDC19,CYZF20}, machine unlearning~\cite{CZWBHZ21}, graph neural networks~\cite{HWWBSZ21,ONK21}, recommender systems~\cite{ZRWRCHZ21}, self-supervised models~\cite{HZ21,LJQG21}, label-only cases~\cite{CTCP21,LZ21}, etc.

Despite current research efforts on the membership inference threat for generative models, a wide range of privacy issues of generative models still remain largely unexplored.
To fill this gap, we present the first study to specify the property inference attack against GANs.
Our results show that even with limited knowledge and access to the target, it is still possible to infer sensitive properties of the training dataset accurately.

\mypara{Property Inference Attacks}
Property inference attacks aim to infer properties of the target model or the training dataset which are unintended to share by the producer.
In fact, sensitive {\it{properties}} cover a wide range of information, which would violate intellectual property if exposed.
They can be model-related, such as the model structure and activation functions; as well as data-related, such as where the data are produced or the distribution of the training data.
And our work lies in the data-related property inference attacks against GANs. 

Ganju et al.~\cite{GWYGB18} propose the first property inference attack against discriminative models, which focuses on fully connected neural networks (FCNN), while ours focuses on GANs. 
As FCNN and GANs have different types of inputs and outputs, the information that our attack exploits is different from \cite{GWYGB18}. 
As a result, our attack relies on an optimized latent code to the target model in the partial black-box case, while \cite{GWYGB18} uses the weights of FCNN as its property classifier’s input (since it is a white-box attack). 
Moreover, \cite{GWYGB18} works on the white-box scenario and only treats the inference attack as a binary prediction task, which cannot return a precise prediction of the target property.
Different from Ganju et al.~\cite{GWYGB18}, our attacks aim to predict the target property in a far more precise fashion, by modeling the attack task as a regression problem.
Furthermore, our proposed attacks work well on two more realistic scenarios: the full black-box and partial black-box setting.
Moreover, Carlini et al.~\cite{CLEKS19} demonstrate the secret leakage problem caused by unintended memorization of sequential generative models.
It focuses on recovering specific sensitive training records from sequential models, while our work targets at inferring the global data privacy of the training dataset against another important kind of generative model--GANs.

Besides the above, there also exists a wide range of study on the security and privacy risks of ML model, such as model stealing~\cite{TZJRR16,OSF19,JCBKP20,YYZTHJ20}, model inversion~\cite{FJR15}, backdoor attack~\cite{WYSLVZZ19,SWBMZ20,LMALZWZ18,CSBMSWZ21} and other attacks under specific background~\cite{SBBFZ20,MSCS19,SS19,HJBGZ21,LWHSZBCFZ22}.

% ======================================================
\section{Conclusion}
\label{section:conclusion}
% ======================================================

In this paper, we perform the first property inference attack against GANs, the goal of which is to infer the macro-level information of a target GAN's underlying training dataset.
We propose a general attack pipeline for two different attack scenarios, following the intuition that the generated samples can reflect the distribution of its underlying training dataset.
In the full black-box setting, we rely on random generated samples and a property classifier to realize our attack.
In the partial black-box setting, we introduce a novel optimization framework to reduce the number of queries with the help of shadow models.
Comprehensive experiments show the effectiveness of both the attack methodologies in a variety of setups including five property inference tasks, four datasets, and four victim GAN models.
We also compare our two methodologies to verify the advantage of the partial black-box attack when using a limited number of samples based on two factors, i.e., stability and accuracy.
We additionally show the effectiveness of our full black-box attack without \emph{any} knowledge of the target model.
Moreover, we present how to leverage our property inference attack to enhance membership inference attacks, which demonstrates the applicability of the proposed property inference method.

% ======================================================
\section*{Acknowledgement}
% ======================================================

We thank all the anonymous reviewers for their insightful suggestions and comments to improve the paper.
This work is supported by National Key R\&D Program (2020YFB1406900), National Natural Science Foundation of China (U21B2018, 61822309, 61773310, U1736205), Shaanxi Province Key Industry Innovation Program (2021ZDLGY01-02), and the Helmholtz Association within the project ``Trustworthy Federated Data Analytics'' (TFDA) (funding number ZT-I-OO1 4).

% ======================================================
\bibliographystyle{IEEEtranS}
\bibliography{normal_generated_py3}
% ======================================================

% ======================================================
\newpage
\appendix
\section{GAN models}
\label{section:append}
% ======================================================

\begin{table}[!htbp]
\centering
\caption{The structure of DCGAN in our paper.}
\label{table:DCGAN_structure}
\scriptsize
\begin{tabular}{l | c | l | c}
\toprule
Generator & Shape & Discriminator & Shape \\
\midrule
Latent codes & (100)& 
Input image & (32,32,1)\\
Reshape & (1,1,100) &  
Conv 4$\times$4 & (16,16,128)\\
Conv 4$\times$4 & (4,4,512)& 
LReL (0.2) & (16,16,128)\\
LReL (0.2) & (4,4,512)& 
Conv 4$\times$4 & (8,8,256)\\
Conv 4$\times$4 & (8,8,256)& 
LReL (0.2) & (8,8,256)\\
LReL (0.2) & (8,8,256)& 
Conv 4$\times$4 & (4,4,512)\\
Conv 4$\times$4 & (16,16,128)& 
LReL (0.2) & (4,4,512)\\ 
LReL (0.2) & (16,16,128)& 
Conv 4$\times$4 & (1,1,1)\\
Conv 4$\times$4 & (32,32,1)& 
Reshape & (1)\\
Tanh & (32,32,1)& 
Sigmoid & (1)\\
\bottomrule
\end{tabular}
\end{table}

\begin{table}[!htbp]
\centering
\caption{The structure of WGANGP in our paper.}
\label{table:WGANGP_structure}
\scriptsize
\begin{tabular}{l | c | l | c}
\toprule
Generator & Shape & Discriminator & Shape \\
\midrule
Latent codes & (100)& 
Input image & (64,64,3)\\
FC & (4*4*512) &  
Conv 5$\times$5 & (32,32,64)\\
Reshape & (4,4,512)& 
LReL (0.2) & (32,32,64)\\ 
Conv 5$\times$5 & (8,8,256)& 
Conv 5$\times$5 & (16,16,128)\\
LReL (0.2) & (8,8,256)& 
LReL (0.2) & (16,16,128)\\
Conv 5$\times$5 & (16,16,128)& 
Conv 5$\times$5 & (8,8,256)\\
LReL (0.2) & (16,16,128)& 
LReL (0.2) & (8,8,256)\\
Conv 5$\times$5 & (32,32,64)& 
Conv 5$\times$5 & (4,4,512)\\
LReL (0.2) & (32,32,64)& 
LReL (0.2) & (4,4,512)\\ 
Conv 5$\times$5 & (64,64,3)& 
Reshape & (64*64*3)\\
ReL & (64,64,3)& 
FC & (1)\\
\bottomrule
\end{tabular}
\end{table}

\begin{table}[!htbp]
\centering
\caption{The structure of PGGAN in our paper.}
\label{table:PGGAN_structure}
\scriptsize
\begin{tabular}{l@{\hskip3pt} | @{\hskip3pt}c@{\hskip3pt} | @{\hskip3pt}c@{\hskip3pt} | l@{\hskip3pt} | @{\hskip3pt}c@{\hskip3pt} | @{\hskip3pt}c}
\toprule
Generator & Shape & Act. & Discriminator & Shape & Act. \\
\midrule
Latent codes & (512) & - & Input image & (64,64,3) & - \\
FC & (4*4*512) & LReL & 
Conv 1$\times$1 & (64,64,256) & LReL \\
Reshape & (4,4,512) & - &  
Conv 3$\times$3 & (64,64,256) & LReL \\
Conv 3$\times$3 & (4,4,512) & LReL & 
Conv 3$\times$3 & (64,64,512) & LReL\\
Upsample & (8,8,512) & - & 
Downsample & (32,32,512) & - \\
Conv 3$\times$3 & (8,8,512) & LReL & 
Conv 3$\times$3 & (32,32,512) & LReL \\
Conv 3$\times$3 & (8,8,512) & LReL & 
Conv 3$\times$3 & (32,32,512) & LReL \\
Upsample & (16,16,512) & - &
Downsample & (16,16,512) & - \\
Conv 3$\times$3 & (16,16,512) & LReL &
Conv 3$\times$3 & (16,16,512) & LReL \\
Conv 3$\times$3 & (16,16,512) & LReL &
Conv 3$\times$3 & (16,16,512) & LReL \\
Upsample & (32,32,512) & - &
Downsample & (8,8,512) & - \\
Conv 3$\times$3 & (32,32,512) & LReL &
Conv 3$\times$3 & (8,8,512) & LReL \\
Conv 3$\times$3 & (32,32,512) & LReL &
Conv 3$\times$3 & (8,8,512) & LReL \\
Upsample & (32,32,512) & - &
Downsample & (4,4,512) & - \\
Conv 3$\times$3 & (64,64,256) & LReL &
Minibatch stddev & (4,4,513) & - \\
Conv 3$\times$3 & (64,64,256) & LReL &
Conv 3$\times$3 & (4,4,512) & LReL \\
Conv 1$\times$1 & (64,64,3) & linear &
Reshape & (4*4*512) & - \\
&&&FC & (512) & LReL \\
&&&FC & (1) & linear \\
\bottomrule
\end{tabular}
\end{table}

% ======================================================
\end{document}